\documentclass[amsmath,amssymb,nofootinbib]{revtex4}
\usepackage{pstricks}
\usepackage{color}
\usepackage{graphicx}
\usepackage{dcolumn}
\usepackage{bm}
\def\d{\partial}
\def\dd{\textrm{d}}
\def\g{\gamma}
\def\abar{\bar{\alpha}}
\def\om{\omega}

\begin{document}

\title{Improving the kinematics for low-x QCD evolution equations in coordinate space}
\author{Guillaume Beuf}
\email{guillaume.beuf@usc.es}
\affiliation{Departamento de F\'isica de Part\'iculas and IGFAE, Universidade de Santiago de
Compostela,\\
 E-15706 Santiago de Compostela, Spain\\}

\begin{abstract}
High-energy evolution equations, such as the BFKL, BK or JIMWLK equations, aim at resumming the high-energy (next-to-)leading logarithms appearing in QCD perturbative series. However, the standard derivations of those equations are performed in a strict high-energy limit, whereas such equations are then applied to scattering processes at large but finite energies. For that reason, there is typically a slight mismatch between the leading logs resummed by those evolution equations without finite-energy corrections and the leading logs actually present in the perturbative expansion of any observable. That mismatch is one of the sources of large corrections at NLO and NLL accuracy. In the case of the BFKL equation in momentum space, that problem is solved by including a kinematical constraint in the kernel, which is the most important finite-energy correction.
In this paper, such an improvement of kinematics is performed in mixed-space (transverse positions and $k^+$) and with a factorization scheme in the light-cone momentum $k^+$ (in a frame in which the projectile is right-moving and the target left-moving). This is the usual choice of variables and factorization scheme for the the BK equation. A kinematically improved version of the BK equation is provided, consistent at finite energies.
The results presented here are also a necessary step towards having the high energy limit of QCD (including gluon saturation) quantitatively under control beyond strict leading logarithmic accuracy.
\end{abstract}


\maketitle


\section{Introduction}\label{sec:intro}

Large logarithms arise in the perturbative expansion of observables related to QCD scattering processes,  when the total energy of the collision is much larger than all the other available scales. In order to obtain reliable theoretical results, such logarithms have to be resummed, thanks to a high-energy evolution equation. In the case of a collision between two dilute objects, the high-energy leading logarithms (LL) are in principle resummed thanks to the BFKL equation \cite{Lipatov:1976zz,Kuraev:1977fs,Balitsky:1978ic}. When one of the colliding particles is hadron or nucleus considered dense, one should instead use the JIMWLK equation \cite{Jalilian-Marian:1997jx,Jalilian-Marian:1997gr,Jalilian-Marian:1997dw,Kovner:2000pt,Weigert:2000gi,Iancu:2000hn,Iancu:2001ad,Ferreiro:2001qy} or equivalently Balitsky's hierarchy of equations \cite{Balitsky:1995ub} in order to resum the LL's, which take into account high-density effects like gluon saturation \cite{Gribov:1984tu,Mueller:1985wy,McLerran:1993ni,McLerran:1993ka,McLerran:1994vd}. These equations also apply to the case of dense-dense collisions \cite{Gelis:2008rw}, such as heavy ion collisions at high energy.
 In practice, one often uses the BK equation \cite{Balitsky:1995ub,Kovchegov:1999yj,Kovchegov:1999ua} instead, which is a mean-field truncation of Balitsky's hierarchy.

In the standard derivations of all of the aforementioned evolution equations, the high-energy limit is taken in order to simplify the kinematics. These equations are therefore valid for hypothetical collisions at infinite energy, but not necessarily for realistic collisions at large but finite energy, where finite-energy corrections may be quantitatively important. Indeed, one has to include a kinematical constraint into the BFKL equation in momentum space in order to make it self-consistent at finite energies. That kinematical constraint was first proposed as one of the ingredients to build the CCFM equation \cite{Ciafaloni:1987ur,Catani:1989sg,Catani:1989yc}, generalizing the BFKL equation. The kinematical constraint for the BFKL equation was further studied in the refs. \cite{Andersson:1995jt,Andersson:1995ju,Kwiecinski:1996td} and also included, in a different form, into the Monte Carlo code DIPSY \cite{Avsar:2005iz,Avsar:2006jy,Avsar:2007xg,Flensburg:2008ag,Flensburg:2010kq,Flensburg:2011kk,Flensburg:2012zy}.

However, the kinematical constraint has been overlooked until the BFKL equation was calculated at next-to-leading logarithmic (NLL) accuracy \cite{Fadin:1998py,Ciafaloni:1998gs}. It was then noticed that higher order corrections to the BFKL equation are typically larger than the leading order contributions, especially in the collinear limits. Those large corrections signal a breakdown of the perturbation theory resummed thanks to BFKL, and require a further resummation in the collinear regimes. In ref. \cite{Salam:1998tj}, such a collinear resummation was outlined, and it was noticed that the lack of kinematical constraint in the standard BFKL equation at LL is the main (but not unique) reason for the appearance of large NLL corrections. Hence, including the kinematical constraint into the BFKL equation corresponds to performing a significant part of the collinear resummation. Then, the full collinear resummation was performed, within various schemes, in the refs. \cite{Ciafaloni:1999yw,Altarelli:1999vw,Ciafaloni:2003rd,Altarelli:2005ni,Ciafaloni:2007gf,Altarelli:2008aj}.

For a few years, a significant effort has been devoted to the calculation of higher order corrections for high-energy processes with gluon saturation. Indeed, the NLL corrections to the BK equation have been calculated \cite{Balitsky:2008zz,Balitsky:2009xg}, as well as the NLO corrections to Deep Inelastic Scattering (DIS) structure functions \cite{Balitsky:2010ze,Beuf:2011xd} and to single inclusive hadron production in pA collisions \cite{Chirilli:2011km,Chirilli:2012jd}. The full calculation  of the JIMWLK equation and Balitsky's hierarchy at NLL accuracy is underway, and preliminary results are already available \cite{Balitsky:2013fea,Kovner:2013ona}.

Before using those higher order results in phenomenology, one should consider the issue of finite-energy corrections and collinear resummations in the presence of gluon saturation. Toy model numerical simulations \cite{Avsar:2011ds} have demonstrated that saturation effects cannot tame the large higher order corrections, so that collinear resummations have to be performed also in the case of high-energy evolution equations with gluon saturation.
Those nonlinear equations are available in mixed-space (transverse position and light-cone momentum $k^+$), whereas the kinematical constraint and the collinear resummations are known for the BFKL equation in momentum space or in Mellin space. The kinematical constraint has been investigated in mixed space only in the seminal paper \cite{Motyka:2009gi}, which nevertheless contains a few shortcomings and inaccuracies. The aim of the present paper is to revisit the issue of the mixed space version of the kinematical constraint and provide a kinematically improved version of the BK equation, self-consistent at finite energies, which corresponds to the equation \eqref{B_JIMWLK_kc_trunc}. 

This paper is organized as follows. In section \ref{sec:prelim}, after a brief presentation of the various evolution equations aiming at resumming high-energy leading logarithms (LL), several factorization schemes for that resummation are discussed. Then, the sections \ref{sec:kin_mom_space}, \ref{sec:Mellin_BFKL_BK_LL_NLL} and \ref{sec:NLO_IF_analysis} present various arguments in favor of the kinematical constraint for high-energy evolution equations. Those three sections are essentially independent of each other. More precisely, the derivation of Mueller's dipole model \cite{Mueller:1993rr,Mueller:1994jq} is revisited in section \ref{sec:kin_mom_space}, analyzing carefully the kinematics of the relevant graphs in Light-Front perturbation theory in momentum space, in an analogous way as in the ref. \cite{Motyka:2009gi} but going into more details. The section \ref{sec:Mellin_BFKL_BK_LL_NLL} reviews the Mellin space approach for the study of high-energy evolution equations in the dilute (BFKL) regime, and the knowledge about kinematical issues obtained in this way, mostly in ref. \cite{Salam:1998tj}. The section \ref{sec:NLO_IF_analysis} is devoted to the analysis of the real NLO corrections to DIS structure functions in the dipole factorization picture, as calculated in ref. \cite{Beuf:2011xd}. It is shown that those NLO corrections contain less LL contributions than the ones resummed by the standard LL evolution equations without kinematical constraint.  The section \ref{sec:kcBK}  presents the construction of a high-energy LL evolution equation in mixed-space with kinematical constraint, using on the one hand the knowledge accumulated in the previous sections and on the other hand the requirement of probability conservation along the initial-state parton cascade. The obtained equation \eqref{B_JIMWLK_kc_trunc} is the main result of the present paper. Conclusions are given in the section \ref{sec:Discussion}. Additional material is provided in appendices. The appendix \ref{App:locality_kc} presents some technical extension of the analysis within Light-Front perturbation theory performed in the section \ref{sec:kin_mom_space}. For completeness, the definition and basic properties of the Laplace transform and the Mellin representation, used various times in this paper, are recalled in the appendices \ref{App:Laplace} and \ref{App:Mellin} respectively. In the appendix \ref{App:Yfplus}, some of the calculations performed in the section \ref{sec:NLO_IF_analysis} are redone within a different prescription, for comparison.

A few remarks to the reader are in order. In this paper, the kinematical constraint is discussed thoroughly from multiple perspectives, because it is rather difficult to find the complete picture in the existing literature, where the emphasis is often on technical aspects and not on the physics. Therefore, there is partial overlap between some of the sections, and one can easily skip some parts of the paper, the first time in particular. For example, a reader not at ease with Mellin transforms can skip completely all the discussions in Mellin space, which are provided both as a cross-check and in order to make contact with the BFKL literature.
A reader already familiar with the need for the kinematical constraint can focus his attention on the sections \ref{sec:evol_variables} and \ref{sec:kcBK}.


\section{Preliminaries\label{sec:prelim}}

\subsection{LL evolution equations in mixed space in the dipole/CGC framework\label{sec:evolEqs}}

Following the idea of high-energy operator product expansion \cite{Balitsky:1995ub}, one can obtain high-energy factorization formulae for a wide class of observables, most notably in the cases of deep inelastic scattering (DIS) processes or forward particle production in hadronic collisions. Those factorization formulae typically involve the convolution of perturbatively calculable factors with the expectation value of some operators, which are products of light-like Wilson lines, evaluated in the target state.
Contrary to the case of collinear factorization, new operators appear in the high-energy factorization formulae at each perturbative order.

At leading order (LO), DIS structure functions and forward single inclusive particle production in hadron-hadron or hadron-nucleus collisions only depend on the dipole operator
\begin{equation}
 {\mathbf S}_{01} = \frac{1}{N_c} \textrm{Tr} \left(U_{\mathbf{x}_{0}}\, U_{\mathbf{x}_{1}}^\dag \right)\, ,\label{dipole_S_matrix}
\end{equation}
where $U_{\mathbf{x}_{i}}$ is the fundamental Wilson line along the $x^+$ direction\footnote{Hereafter, the frame is chosen such that the projectile (virtual photon in the DIS case) is right-moving and the target left-moving.}, at $x^-=0$ and at the transverse position $\mathbf{x}_{i}$. The expectation value of the operator \eqref{dipole_S_matrix} in the state of the target is noted $\left\langle {\mathbf S}_{01} \right\rangle_{\eta}$. Here, $\eta$ is a common regulator for the rapidity divergence of the operator and for the soft divergence of the next-to-leading order (NLO) impact factor, and play the role of a factorization scale. It will be discussed in more details in the section \ref{sec:evol_variables}.
At leading logarithmic(LL) accuracy in the high-energy limit, the $\eta$-dependence of $\left\langle {\mathbf S}_{01} \right\rangle_{\eta}$
is given by the equation \cite{Balitsky:1995ub}
\begin{eqnarray}
\partial_{\eta}  \left\langle {\mathbf S}_{01} \right\rangle_{\eta}&=&   \bar{\alpha}
\int \frac{\textrm{d}^2\mathbf{x}_{2}}{2\pi}\; \textbf{K}_{012}\: \left\langle{\mathbf S}_{02} {\mathbf S}_{21} \!-\! {\mathbf S}_{01} \right\rangle_{\eta}
\, ,\label{B_JIMWLK_dipole}
\end{eqnarray}
with the notations
\begin{eqnarray}
\abar&\equiv &\frac{N_c}{\pi}\, \alpha_s\\
\textbf{K}_{012}&\equiv &\frac{x_{01}^2}{x_{02}^2\, x_{21}^2}\: .
\end{eqnarray}
The equation \eqref{B_JIMWLK_dipole} is not closed because its right-hand side involves the new double-trace operator $\left\langle{\mathbf S}_{02} {\mathbf S}_{21} \right\rangle_{\eta}$. The evolution equation for that new operator would involve for example $\left\langle{\mathbf S}_{02} {\mathbf S}_{23} {\mathbf S}_{31} \right\rangle_{\eta}$.
Hence the equation \eqref{B_JIMWLK_dipole} is only the first in an infinite hierarchy of coupled equations, called Balitsky's hierarchy \cite{Balitsky:1995ub}.

In the Color Glass Condensate effective theory (CGC) \cite{McLerran:1993ni,McLerran:1993ka,McLerran:1994vd,Jalilian-Marian:1997jx,Jalilian-Marian:1997gr,Jalilian-Marian:1997dw,Kovner:2000pt,Weigert:2000gi,Iancu:2000hn,Iancu:2001ad,Ferreiro:2001qy,Jeon:2013zga}, valid for a dense target, the target is described by a random distribution of classical color charges, corresponding to the large-$x$ partons, and the classical gluon field radiated by those classical charges, corresponding to the low-$x$ partons. In this context, taking expectation values $\left\langle \cdots \right\rangle_{\eta}$ in the target state reduces to perform the statistical average over the distribution of classical color charges, and ${\eta}$ is also related with the cut-off separating the large-$x$ and low-$x$ partons in the target. In the CGC, the JIMWLK equation
\cite{Jalilian-Marian:1997jx,Jalilian-Marian:1997gr,Jalilian-Marian:1997dw,Kovner:2000pt,Weigert:2000gi,Iancu:2000hn,Iancu:2001ad,Ferreiro:2001qy}
is the renormalization group equation associated with the change of the cut-off between large-$x$ and low-$x$ partons.
The JIMWLK equation gives formally the LL evolution for the expectation value of any product of light-like Wilson lines, and thus  reproduces in particular Balitsky's hierarchy. The equation \eqref{B_JIMWLK_dipole} can then be called the B-JIMWLK evolution equation for $\left\langle {\mathbf S}_{01} \right\rangle_{\eta}$.

For simplicity, it is often convenient to perform the mean-field approximation
\begin{equation}
\left\langle{\mathbf S}_{02} {\mathbf S}_{21} \right\rangle_{\eta}\simeq
\left\langle{\mathbf S}_{02} \right\rangle_{\eta}\: \left\langle{\mathbf S}_{21} \right\rangle_{\eta}\, ,\label{mean-field_approx}
\end{equation}
which allows to close the equation \eqref{B_JIMWLK_dipole}, and give the Balitsky-Kovchegov (BK) equation \cite{Balitsky:1995ub,Kovchegov:1999yj,Kovchegov:1999ua}
\begin{eqnarray}
\partial_{\eta}  \left\langle {\mathbf S}_{01} \right\rangle_{\eta}&=&   \bar{\alpha}
\int \frac{\textrm{d}^2\mathbf{x}_{2}}{2\pi} \textbf{K}_{012}\: \bigg[ \left\langle{\mathbf S}_{02} \right\rangle_{\eta} \left\langle{\mathbf S}_{21} \right\rangle_{\eta} \!-\! \left\langle {\mathbf S}_{01} \right\rangle_{\eta}\bigg]
\, ,\label{BK_S}
\end{eqnarray}
which is also often written in terms of the dipole-target amplitude $\left\langle {\textbf N}_{01} \right\rangle_{\eta}=1-\left\langle {\textbf S}_{01} \right\rangle_{\eta}$ as
\begin{eqnarray}
\partial_{\eta}  \left\langle {\textbf N}_{01} \right\rangle_{\eta}&=&   \bar{\alpha}
\int \frac{\textrm{d}^2\mathbf{x}_{2}}{2\pi} \textbf{K}_{012}\: \bigg[ \left\langle{\textbf N}_{02} \right\rangle_{\eta} \!+\! \left\langle{\textbf N}_{21} \right\rangle_{\eta} \!-\! \left\langle {\textbf N}_{01} \right\rangle_{\eta}\!-\!\left\langle{\textbf N}_{02} \right\rangle_{\eta}  \left\langle{\textbf N}_{21} \right\rangle_{\eta}\bigg]
\, .\label{BK_N}
\end{eqnarray}

In the cases where the target is dilute and thus the amplitude $\left\langle {\textbf N}_{01} \right\rangle_{\eta}$ is much smaller than $1$, it is legitimate to linearize the BK equation \eqref{BK_N}, which then reduces to the dipole form \cite{Mueller:1993rr,Mueller:1994jq} of the BFKL equation \cite{Lipatov:1976zz,Kuraev:1977fs,Balitsky:1978ic}
\begin{eqnarray}
\partial_{\eta}  \left\langle {\textbf N}_{01} \right\rangle_{\eta}&=&   \bar{\alpha}
\int \frac{\textrm{d}^2\mathbf{x}_{2}}{2\pi} \textbf{K}_{012}\: \bigg[ \left\langle{\textbf N}_{02} \right\rangle_{\eta} \!+\! \left\langle{\textbf N}_{21} \right\rangle_{\eta} \!-\! \left\langle {\textbf N}_{01} \right\rangle_{\eta}\bigg]
\, .\label{BFKL_dipole}
\end{eqnarray}


\subsection{High-energy factorization schemes and evolution variables\label{sec:evol_variables}}

There are many ways to regulate the rapidity divergence of the light-like Wilson line operators just discussed, and each way is associated to a particular definition of the cut-off variable $\eta$. Nevertheless, at LL accuracy, it is always possible to write the evolution equations in $\eta$ from the previous section. By contrast, different choices of regularization for the rapidity divergence, or equivalently of high-energy factorization scheme, generically lead to different evolution equations at next-to-leading logarithmic accuracy (NLL) and beyond (see \emph{e.g.} refs. \cite{Balitsky:2008zz} and \cite{Balitsky:2009xg}).

One possible way to regularize the light-like Wilson line operators is to make them time-like, slightly changing their slope. The variable $\eta$ is then related to the new slope of the Wilson line. That method is commonly used in the case of the TMD-factorization \cite{Collins_TMD_book} and was used in the case of high-energy factorization for example in the original derivation of Balitsky's hierarchy \cite{Balitsky:1995ub}.

Another possibility is to forbid some kinematical range to the gluons included in the Wilson line operators by an explicit cut-off. For example, one allows in the Wilson lines only gluons with a $k^+$ smaller than some factorization scale $k^+_f$. That prescription was used in ref. \cite{Balitsky:2008zz} to derive the NLL corrections to the first B-JIMWLK equation \eqref{B_JIMWLK_dipole}, and it will be used in most of the rest of the present study. In that scheme, the variable
\begin{equation}
Y^+_f= \log\left(\frac{k^+_f}{k^+_{\min}}\right)   \label{Yfplus_def}
\end{equation}
is the appropriate evolution variable $\eta$ for the equations \eqref{B_JIMWLK_dipole}, \eqref{BK_S}, \eqref{BK_N} and \eqref{BFKL_dipole}, where at this stage $k^+_{\min}$ is only an arbitrary reference scale in $k^+$. Indeed, quantities such that $\left\langle {\mathbf S}_{01} \right\rangle$ cannot depend on $k^+_f$ alone, but only on a ratio of $k^+$'s, due to the required invariance under longitudinal boosts.

There are obvious variants of that regularization and factorization scheme. Indeed, one can instead include only gluons with $k^-$ larger than some factorization scale $k^-_f$ in the Wilson lines. The associated variable $\eta$ appearing in the evolution equations \eqref{B_JIMWLK_dipole}, \eqref{BK_S}, \eqref{BK_N} and \eqref{BFKL_dipole} is then
\begin{equation}
Y^-_f= \log\left(\frac{k^-_{\max}}{k^-_f}\right)   \label{Yfminus_def}\, .
\end{equation}
Yet another possibility is to include only gluons with rapidity $y=\log(k^+/k^-)/2$ smaller than a value $y_f$ in the Wilson lines. In that case, the evolution variable $\eta$ is taken to be
\begin{equation}
Y_f= y_f-y_{\min}   \label{Yf_def}\, .
\end{equation}

Finally, a last type of regularization and factorization scheme was proposed in ref.\cite{Balitsky:2009xg}, the so-called conformal dipole scheme. In QCD with massless quarks, conformal symmetry should be an anomalous symmetry, \emph{i.e.} broken only by the running of the coupling. However, the regularization schemes discussed earlier lead to an explicit breaking of conformal symmetry. Because of that, scheme-dependent non-conformal terms arise when calculating higher order perturbative corrections to the impact factors and to the high-energy evolution equations.
If the Wilson lines are regularized in a conformal way, all such non-conformal higher order terms should disappear. Unfortunately, such conformal regularization and factorization scheme is not explicitly known, and results in that scheme have been constructed only perturbatively, starting from expressions in a non-conformal scheme \cite{Balitsky:2009xg,Balitsky:2010ze}. The physical interpretation of the variable playing the role of $\eta$ in that scheme is also rather obscure. For those reasons, we will not attempt to address the case of the conformal dipole scheme, despite its mathematical attractiveness.

Let us come back to the schemes with explicit cut-off in the Wilson lines. In those cases, the factorization scale
$k^+_f$ (or $k^-_f$ or $y_f$) is covariant with respect to longitudinal boosts. It is convenient to choose that factorization scale close enough to the corresponding typical scale associated with the projectile, in order to avoid the appearence of potentially large logs in the projectile impact factor. On the other hand, for finite-energy collisions, the target is setting another typical scale in $k^+$ (or $k^-$ or $y$). It is convenient to choose the reference scale $k^+_{\min}$ (or $k^-_{\max}$ or $y_{\min}$) to be that scale provided by target. Thanks to that choice, the evolution variable $Y^+_f$ (or $Y^-_f$ or $Y_f$) is invariant under longitudinal boosts and carries the dependence on the total energy of the collision. It represents the range over which one should evolve the regularized Wilson line operators with a high-energy evolution equation of the previous section, starting from some initial condition. That initial condition is purely non-perturbative, and encodes the dynamics of the target as seen with a poor time resolution\footnote{Remember that from the point of view of the target, $x^-$ plays the role of time, not $x^+$.}.

In order to get more explicit expressions for the variables $Y^+_f$, $Y^-_f$ and $Y_f$, one has to perform some modelling of the un-evolved target. For our purposes, the following very simple model should be enough. Let the target be a collection of partons, each of them having a transverse mass $Q_0$ and carrying a fraction $x_0$ of the large component $P^-$ of the momentum of the target\footnote{In the case of a nuclear target, following the standard conventions, $P^-$ is instead the average momentum per nucleon of the target, and the momentum of the partons is still noted $x_0\, P^-$.}. Within that model\footnote{That model has been introduced in ref. \cite{Beuf:2011xd}, up to the parameter $x_0$ which has been added here for completeness.}, is it natural to choose
\begin{eqnarray}
k^-_{\max} &=& x_0\, P^-\\
k^+_{\min} &=& \frac{Q_0^2}{2\, x_0\, P^-}\label{kplus_min_def}\\
y_{\min} &=& \frac{1}{2} \log\left(\frac{Q_0^2}{2 (x_0\, P^-)^2}\right)\, .
\end{eqnarray}

Applying those ideas to the example of DIS at low $x_{Bj}$, mediated by a photon of virtuality $Q^2$ and momentum $q^+$, so that
\begin{equation}
x_{Bj} \simeq \frac{Q^2}{2\, P^-\, q^+}\, ,
\end{equation}
one finds
\begin{eqnarray}
Y^+_f &=& \log\left(\frac{x_0\, Q^2}{x_{Bj}\, Q_0^2}\right)+  \log\left(\frac{k^+_f}{q^+}\right)\label{Yfplus_final}\\
Y^-_f &=& \log\left(\frac{x_0}{x_{Bj}}\right)+  \log\left(\frac{Q^2}{2\, q^+\, k^-_f}\right)\label{Yfminus_final}\\
Y_f &=& \log\left(\frac{x_0\, Q}{x_{Bj}\, Q_0}\right)+ \log\left(\frac{Q\, e^{y_f}}{\sqrt{2}\, q^+}\right)\label{Yf_final}\, .
\end{eqnarray}
In each case, the first term is the dominant one, because the factorisation scales should be taken close enough to the scales fixed by the virtual photon
\begin{eqnarray}
k^+_f &\lesssim &  q^+\\
k^-_f &\gtrsim &\frac{Q^2}{2\, q^+}\\
e^{y_f}&\lesssim & \frac{\sqrt{2}\, q^+}{Q}\, .
\end{eqnarray}

Most of the studies in the literature have been performed at strict LL accuracy. Accordingly, no distinction between factorization schemes is usually done, and the target is evolved over a range $\log(1/x_{Bj})$ or $\log(x_0/x_{Bj})$ (with typically $x_0=0.01$). That is indeed legitimate as this order. However, it is clear from our discussion that one has to be more careful when trying to include higher order effects consistently. According to the equations \eqref{Yfplus_final}, \eqref{Yfminus_final} and \eqref{Yf_final}, the LL terms $(\abar\, Y^+_f)^n$, $(\abar\, Y^-_f)^n$ or $(\abar\, Y_f)^n$ at low $x_{Bj}$ differ from each other by terms of order NLL. Moreover, $\log(Q^2/Q^2_0)$ can be large, in practical applications to DIS.

In the following, we will often drop the $f$ subscript in $Y^+_f$, $Y^-_f$ and $Y_f$ for the variable appearing in the high-energy evolution equations, and keep the notation $Y^+_f$, $Y^-_f$ and $Y_f$  for the total range over which the Wilson line correlators should be evolved, \emph{e.g.} \eqref{Yfplus_final}, \eqref{Yfminus_final} or \eqref{Yf_final}, given the process considered, the total energy of the collision and the precise choice of factorization scale $k^+_f$ (or $k^-_f$ or $y_f$).

As a remark, note that the freedom to choose the evolution variable $Y^+_f$, $Y^-_f$ or $Y_f$ in order to specify an explicit factorization scheme is related in the traditional BFKL formalism to the freedom to choose the reference scale $s_0$ for the total energy.




\section{Kinematics of multi-gluon Fock components of photon wave-functions in momentum space\label{sec:kin_mom_space}}

Within the dipole model \cite{Mueller:1993rr,Mueller:1994jq}, one can obtain the real emission contribution to the LL high-energy evolution equations of the section \ref{sec:evolEqs} from the tree-level multi-gluon Fock components of photon wave-functions \cite{Mueller:1993rr,Kovchegov:1999yj,Kovchegov:1999ua}, which are calculable for example in light-front perturbation theory. When applied to DIS observables, those multi-gluon Fock components are building blocks for the higher-order corrections to the impact factors (see \emph{e.g.} ref. \cite{Beuf:2011xd}). The general idea is that the emission of softer and softer gluons in the photon wave-functions tends to somehow factorize from the rest of the wave-functions, and not to modify the kinematics of harder partons, so that by moving the factorization scale (typically $k^+_f$) closer to the projectile one can reinterpret the soft gluon emissions as part of the LL evolution of the target.

In that calculation, some kinematical approximations for the soft gluons are crucial, in particular in the energy denominators. Those kinematical approximations are usually done in a crude way, sufficient only at strictly LL accuracy. Performing those approximations in a fully self-consistent way leads to the kinematical constraint of refs. \cite{Ciafaloni:1987ur,Catani:1989sg,Catani:1989yc,Andersson:1995jt,Andersson:1995ju,Kwiecinski:1996td}.
In the rest of this section, that point is discussed thoroughly, extending the related discussion already available in ref. \cite{Motyka:2009gi}. The links of the kinematical constraint with the physics of the collinear and anti-collinear limits and with the ordering in formation time are also discussed for completeness.

\subsection{Kinematical approximations in the high-energy limit\label{sec:kinematical_approx}}

\begin{figure}
\setbox1\hbox to 10cm{
 \includegraphics{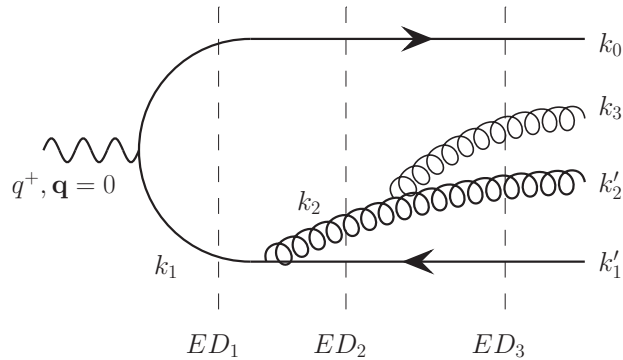}
}
\begin{center}
\hspace{-4cm}
\resizebox*{6cm}{!}{\box1}
\caption{\label{Fig:ga2qqbargg}Example of light-front perturbation theory diagram contributing to the $q\bar{q}gg$ Fock component of a photon. Energy denominators are indicated by the vertical dashed lines. The light-front time $x^+$ increases from the left to the right of the diagram, from $-\infty$ to $0$.}
\end{center}
\end{figure}


Let us consider the diagram on Fig.\ref{Fig:ga2qqbargg}, which is a typical contribution to the $q\bar{q}gg$ Fock component of the photon wave-function within light-front perturbation theory. Following the usual rules of light-front perturbation theory\cite{Kogut:1969xa,Bjorken:1970ah}, there is conservation of the transverse momentum $\mathbf{k}$ and of the $k^+$ at each vertex, but not of the light-front energy $k^-$. Instead, each of the partons is on-shell, \emph{i.e.} $2 k^+\, k^- \!-\!\mathbf{k}^2=0$ because only massless partons will be considered here. The energy denominators, which encode the energy off-shellness of each intermediate Fock state, are given by the difference between the initial $k^-$ and the total $k^-$ of the partons present in the current intermediate Fock state. The effective initial $k^-$ associated to an incoming real or virtual photon is\footnote{In the case of a photo-production reaction, the initial Fock state for the real photon wave-function is just the one-photon Fock state, with momentum $k^+_{init}=q^+$, $\mathbf{k}_{init}=\mathbf{q}=0$, and thus $k^-_{init}=q^-=0$. In the case of deep inelastic scattering in the one-photon exchange approximation, the photon is always strictly on mass shell in light-front perturbation theory. However, the correct initial state is not the one-photon Fock state but the one-lepton Fock state. Nevertheless, as explained in the appendix A.3 of Ref.\cite{Beuf:2011xd}, one can effectively start from a virtual photon initial state with $k^-_{init}=-Q^2/(2\, q^+)$, which reproduces the contribution of both the initial lepton and the final scattered lepton to each of the energy denominators. $Q^2$ is defined from the initial and final leptons 4-momentums ${k_l}^{\mu}$ and ${k_{l'}}^{\mu}$ as $Q^2=-({k_l}^{\mu}\!-\!{k_{l'}}^{\mu})({k_l}_{\mu}\!-\!{k_{l'}}_{\mu})$.} $k^-_{init}=-Q^2/(2\, q^+)$, where $q^+$ is the light-front momentum of the photon.

For the diagram on Fig.\ref{Fig:ga2qqbargg}, one has the obvious momentum conservation relations $\mathbf{k}_0=-\mathbf{k}_1$, ${k_0}^+=q^+\!-\!{k_1}^+$, $\mathbf{k}_1'=\mathbf{k}_1\!-\!\mathbf{k}_2$, ${k_1'}^+={k_1}^+\!-\!{k_2}^+$, $\mathbf{k}_2'=\mathbf{k}_2\!-\!\mathbf{k}_3$ and ${k_2'}^+={k_2}^+\!-\!{k_3}^+$. Hence, the energy denominators $ED_1$, $ED_2$ and $ED_3$ write
\begin{eqnarray}
ED_1&=& -\frac{Q^2}{2\, q^+}\!-\!{k_0}^-\!-\!{k_1}^- \;\;=\;\; -\frac{Q^2}{2\, q^+}\!-\!\frac{q^+\:{\mathbf{k}_1}^2}{2\, {k_1}^+ (q^+\!-\!{k_1}^+)}\nonumber\\
ED_2&=& -\frac{Q^2}{2\, q^+}\!-\!{k_0}^-\!-\!{k_1'}^-\!-\!{k_2}^- \;\;=\;\; -\frac{Q^2}{2\, q^+}\!-\!\frac{{\mathbf{k}_1}^2}{2\, (q^+\!-\!{k_1}^+)}\!-\!\frac{(\mathbf{k}_1\!-\!\mathbf{k}_2)^2}{2\, ({k_1}^+\!-\!{k_2}^+)}\!-\!\frac{{\mathbf{k}_2}^2}{2\, {k_2}^+} \nonumber\\
ED_3&=& -\frac{Q^2}{2\, q^+}\!-\!{k_0}^-\!-\!{k_1'}^- \!-\!{k_2'}^-\!-\!{k_3}^-\;\;=\;\; -\frac{Q^2}{2\, q^+}\!-\!\frac{{\mathbf{k}_1}^2}{2\, (q^+\!-\!{k_1}^+)}\!-\!\frac{(\mathbf{k}_1\!-\!\mathbf{k}_2)^2}{2\, ({k_1}^+\!-\!{k_2}^+)}\!-\!\frac{(\mathbf{k}_2\!-\!\mathbf{k}_3)^2}{2\, ({k_2}^+\!-\!{k_3}^+)}\!-\!\frac{{\mathbf{k}_3}^2}{2\, {k_3}^+}\, .\label{ED_exact}
\end{eqnarray}

One of the most crucial approximations in the derivation of the BFKL equation in the dipole model \cite{Mueller:1993rr} is that in the case of softer and softer gluon emissions, energy denominators should be dominated by the contribution of the last emitted gluon\footnote{Notice that it is the momentum of the gluon just at its emission, like ${\mathbf{k}_2}$, which appears in the approximation \eqref{ED_approx}, not the momentum at the end of the initial-state parton cascade, like ${\mathbf{k}_2'}$, which can be quite different, see Fig.\ref{Fig:ga2qqbargg}.}
\begin{eqnarray}
ED_2&\simeq & -{k_2}^- \;\;=\;\;-\frac{{\mathbf{k}_2}^2}{2\, {k_2}^+} \nonumber\\
ED_3&\simeq & -{k_3}^-\;\;=\;\;-\frac{{\mathbf{k}_3}^2}{2\, {k_3}^+}\, .\label{ED_approx}
\end{eqnarray}
It allows in the end to factorize the emission of each additional softer gluon.
One usually justifies that approximation by taking the gluons strongly ordered in $k^+$
\begin{equation}
q^+ > {k_1}^+,\: q^+\!-\!{k_1}^+ \gg {k_2}^+ \gg {k_2}^+ \gg \cdots \;  ,\label{kplus_ordering}
\end{equation}
  and by assuming that all the transverse momentums are of the same order
\begin{equation}
 Q^2 \simeq {\mathbf{k}_1}^2 \simeq {\mathbf{k}_2}^2 \simeq {\mathbf{k}_3}^2 \simeq \cdots  \; .\label{kt_equal}
\end{equation}
Because of the strong $k^+$ ordering, one gets in the end a LL high-energy evolution equation \eqref{B_JIMWLK_dipole}, \eqref{BK_S} or \eqref{BFKL_dipole} in the factorization scheme with cut-off in $k^+$, and thus with $Y^+=\log(k^+/k^+_{\min})$ playing the role of the evolution variable $\eta$. However, the kernel of the equation contains typically an unrestricted integration over $\mathbf{k}$, when written in momentum space. And thus there are contributions beyond the assumption \eqref{kt_equal} which violate the approximation \eqref{ED_approx} in some parametrically small part of the integration range. Due to this small inconsistency at LL accuracy, pathologically large corrections arise at NLL accuracy and beyond.

The other derivations of high-energy evolution equations such as BFKL, BK or JIMWLK always rely on some kinematical approximation equivalent to \eqref{ED_approx}. Usually, either the kinematics \eqref{kplus_ordering} and \eqref{kt_equal} is assumed, or the $k^+$ ordering \eqref{kplus_ordering} is replaced by the $k^-$ ordering
\begin{equation}
|ED_1|= \frac{Q^2}{2\, q^+}+\frac{q^+\:{\mathbf{k}_1}^2}{2\, {k_1}^+ (q^+\!-\!{k_1}^+)} \ll {k_2}^- \ll  {k_3}^- \ll\cdots \; ,\label{kminus_ordering}
\end{equation}
or by an ordering in rapidity $y=\log(k^+/k^-)/2$. Those choices provide LL evolution equations in the factorization scheme where respectively $Y^-=\log(k^-_{\max}/k^-)$ or $Y=y-y_{\min}$ plays the r\^ole of evolution variable.
When assuming \eqref{kt_equal}, those three different possible ordering become equivalent. That is why one obtains in any of those factorizations schemes a LL evolution equation with the same kernel. However, in each case, the transverse integration in the kernel is unrestricted, and contain a regime where kinematical approximations done in the derivation of the evolution equation are violated. Hence, the standard version of any high-energy evolution equation is not fully self-consistent. This problem generates the largest corrections at higher orders in the evolution equation and in the impact factor of observables sensitive to high-energy logs.
Moreover, at NLL accuracy, the kernel starts to depend on the choice of evolution variable $Y^+$, $Y^-$ or $Y$, or equivalently on the factorization scheme.

In order to address those issues, let us examine more carefully the energy denominators \eqref{ED_exact}.
When assuming the strong $k^+$ ordering \eqref{kplus_ordering} only, and nothing about the transverse momentums, one has the simplification
\begin{eqnarray}
ED_2&\simeq & -\frac{Q^2}{2\, q^+}\!-\!\frac{{\mathbf{k}_1}^2}{2\, (q^+\!-\!{k_1}^+)}\!-\!\frac{(\mathbf{k}_1\!-\!\mathbf{k}_2)^2}{2\, {k_1}^+}\!-\!\frac{{\mathbf{k}_2}^2}{2\, {k_2}^+} \;\;\simeq \;\; ED_1\!-\!\frac{{\mathbf{k}_2}^2}{2\, {k_2}^+} \label{ED2_approx_kplus_ord}\\
ED_3&\simeq& -\frac{Q^2}{2\, q^+}\!-\!\frac{{\mathbf{k}_1}^2}{2\, (q^+\!-\!{k_1}^+)}\!-\!\frac{(\mathbf{k}_1\!-\!\mathbf{k}_2)^2}{2\, {k_1}^+}\!-\!\frac{(\mathbf{k}_2\!-\!\mathbf{k}_3)^2}{2\, {k_2}^+}\!-\!\frac{{\mathbf{k}_3}^2}{2\, {k_3}^+}\;\;\simeq\;\; ED_1\!-\!\frac{{\mathbf{k}_2}^2}{2\, {k_2}^+}\!-\!\frac{{\mathbf{k}_3}^2}{2\, {k_3}^+}\, .\label{ED3_approx_kplus_ord}
\end{eqnarray}
One arrives at the last expression for $ED_2$ using the following reasoning. Due to the strong $k^+$ ordering, the term in ${\mathbf{k}_2}^2/{k_2}^+$ will be usually dominant, except if $\mathbf{k}_2$ is excessively small. In that case, any of the other terms can dominate $ED_2$. In particular, the term containing $(\mathbf{k}_1\!-\!\mathbf{k}_2)^2$ may be dominant only if $(\mathbf{k}_1\!-\!\mathbf{k}_2)^2$ is so much larger than ${\mathbf{k}_2}^2$ that the $k^+$ ordering is compensated. In that case, it is clear that ${\mathbf{k}_2}^2 \ll (\mathbf{k}_1\!-\!\mathbf{k}_2)^2\simeq {\mathbf{k}_1}^2$. Hence, once the $k^+$ ordering \eqref{kplus_ordering} is satisfied, the last expression for $ED_2$ in \eqref{ED2_approx_kplus_ord} is always a good approximation, whatever is the relative size of $Q^2$, ${\mathbf{k}_1}^2$ and ${\mathbf{k}_2}^2$. One obtains the last expression for  $ED_3$ in \eqref{ED3_approx_kplus_ord} following the same method.

Those results generalize to any light-front perturbation theory tree-level diagram contributing to the quark, anti-quark plus $N$ gluons Fock component of a photon wave-function: if gluons have a strongly decreasing $k^+$ from the first to the last emitted gluon in light-front time $x^+$, then the energy denominator $ED_{n+1}$ following to the $n$-th gluon emission is always well approximated by
\begin{eqnarray}
ED_{n+1}&\simeq& ED_1\!-\!\frac{{\mathbf{k}_2}^2}{2\, {k_2}^+}\!- \cdots -\!\frac{{\mathbf{k}_{n+1}}^2}{2\, {k_{n+1}}^+} \;\;= \;\; ED_1\!-\!{k_2}^-\!- \cdots -\!{k_{n+1}}^- \, .\label{EDn_approx_kplus_ord}
\end{eqnarray}
Hence, assuming the strong $k^+$ ordering \eqref{kplus_ordering}, one gets the approximation
\begin{eqnarray}
ED_{n+1}&\simeq& - \frac{{\mathbf{k}_{n+1}}^2}{2\, {k_{n+1}}^+} \;\;= \;\; - {k_{n+1}}^- \, .\label{EDn_approx}
\end{eqnarray}
for each energy denominator if the $k^-$ ordering \eqref{kminus_ordering} is also satisfied.
It is also possible but more cumbersome to show that if only the $k^-$ ordering \eqref{kminus_ordering} is assumed, the approximation \eqref{EDn_approx} is valid precisely when the $k^+$ ordering \eqref{kplus_ordering} is satisfied.

The approximation \eqref{EDn_approx}, which is necessary to factorize each gluon emission from the previous ones and thus to get a high-energy evolution equation like BFKL or BK, is then valid if both the $k^+$ ordering \eqref{kplus_ordering} and the $k^-$ ordering \eqref{kminus_ordering} are simultaneously satisfied. By contrast, the assumption \eqref{kt_equal} is both misleading and meaningless due to the transverse integration in the kernel of the high-energy evolution equations.

The $k^+$ ordering and the $k^-$ ordering together imply the rapidity $y$ ordering, which is intermediate between the two. When writing down a high-energy evolution equation, the choice of evolution variable $Y^+$, $Y^-$ or $Y$ makes the ordering in the corresponding variable ($k^+$, $k^-$ or $y$) automatic. The general idea behind the kinematical constraint \cite{Ciafaloni:1987ur,Catani:1989sg,Catani:1989yc,Andersson:1995jt,Andersson:1995ju,Kwiecinski:1996td} is that one should add a theta function in the kernel of the BFKL (or BK) equation, in order to impose the $k^+$ or $k^-$ (or both) ordering not already guarantied by the choice of evolution variable. This can be viewed either as an all order resummation of the largest corrections arising at NLL and beyond when calculated in the standard way, or as a improvement of the LL evolution equation, making it kinematically self-consistent.

In the appendix \ref{App:locality_kc}, the analysis of the kinematics in a dipole cascade within light-front perturbation theory is performed in a more refined way. There, it is shown that, to LL accuracy, the $k^+$ and $k^-$ orderings \eqref{kplus_ordering} and \eqref{kminus_ordering} are local instead of global, \emph{i.e.} the $k^+$ and $k^-$ of each gluon are constrained only by the $k^+$'s and $k^-$'s of the two partons forming the color dipole emitting that gluon, and not by the $k^+$'s and $k^-$'s of partons present in other branches of the cascade, contrary to statements made in ref. \cite{Motyka:2009gi}. That locality of the $k^+$ and $k^-$ orderings is crucial in order to be able to write a kinematical constrained versions of the BFKL and BK equation.


\subsection{DLL limits\label{sec:DLL_mom} and the failure of the standard high-energy evolution equations to reproduce both of them}

The DGLAP evolution of the photon corresponds to the ordering
\begin{equation}
 Q^2 \ll {\mathbf{k}_1}^2 \ll {\mathbf{k}_2}^2 \ll {\mathbf{k}_3}^2 \ll \cdots  \; ,\label{anti_kt_ord}
\end{equation}
while keeping all the $k^+$'s parametrically of the same order. When using the obtained photon wave-function to calculate photoproduction or DIS observables, this regime is the anti-collinear regime, relevant mainly for the so-called \emph{resolved photon} contributions. When taking the $k^+$ ordering \eqref{kplus_ordering} in addition to the ${\mathbf{k}}^2$ ordering \eqref{anti_kt_ord}, one arrives at the anti-collinear double leading log (DLL) regime, which is both the low-$x$ limit of the anti-collinear DGLAP evolution and the anti-collinear limit of the low-$x$ evolution equation. In that case, \eqref{kplus_ordering} and \eqref{anti_kt_ord} together imply the approximation \eqref{EDn_approx} of the energy denominators, as well as the $k^-$ ordering \eqref{kminus_ordering}. For that reason, one can conclude that generically, low-$x$ evolution equations with $Y^+$ as evolution variable should have a smooth anti-collinear limit, indeed reproducing the low-$x$ limit of the anti-collinear DGLAP evolution.

On the other hand, the collinear regime, associated with the DGLAP evolution of the target in the case of DIS, is defined by the ordering
\begin{equation}
 Q^2 \gg {\mathbf{k}_1}^2 \gg {\mathbf{k}_2}^2 \gg {\mathbf{k}_3}^2 \gg \cdots  \; ,\label{kt_ord}
\end{equation}
while keeping all the $k^-$ parametrically of the same order. Taking the $k^-$ ordering \eqref{kminus_ordering} in addition to the ${\mathbf{k}}^2$ ordering \eqref{kt_ord} defines collinear DLL regime. In that regime, the $k^+$
ordering \eqref{kplus_ordering} and the approximation \eqref{EDn_approx} of the energy denominators are automatically verified. Hence, any high-energy evolution equation formulated with the $Y^-$ evolution variable should generically give the correct collinear DLL physics in the limit \eqref{kt_ord}.

By contrast, if one assume both the $k^+$ ordering \eqref{kplus_ordering} and the collinear ${\mathbf{k}}^2$ ordering \eqref{kt_ord}, one cannot deduce anything about the validity or not of the $k^-$ ordering \eqref{kminus_ordering} or of the approximation \eqref{EDn_approx}. Hence, if one takes a high-energy evolution equation formulated with the $Y^+$ evolution variable, one has to be careful when discussing the collinear limit \eqref{kt_ord}. If the kinematical constraint has been imposed in the kernel of the evolution equation, then the $k^-$ ordering \eqref{kminus_ordering} is by definition satisfied, so that the collinear DLL physics is correctly reproduced. On the other hand, if one considers the standard version of the high-energy evolution equation, \emph{i.e.} without kinematical constraint, and derived assuming both \eqref{kplus_ordering} and \eqref{kt_equal}, one cannot obtain the correct collinear DLL limit in the regime \eqref{kt_ord}, since the $k^-$ are completely unconstrained and unordered.

Of course, by symmetry, one expect similar issues in the anti-collinear limit \eqref{anti_kt_ord} (resp. in both the collinear and anti-collinear limits) when studying a high-energy evolution equation with $Y^-$ (resp. with $Y$) playing the role of evolution variable.

Hence, for any standard definition of the multi-Regge kinematics, either \eqref{kplus_ordering} and \eqref{kt_equal}, or \eqref{kminus_ordering} and \eqref{kt_equal}, or rapidity ordering and \eqref{kt_equal}, one obtains high energy evolution equations which cannot have both the correct collinear and anti-collinear DLL limits. That problem is solved when using the kinematical constraint \cite{Ciafaloni:1987ur,Catani:1989sg,Catani:1989yc,Andersson:1995jt,Andersson:1995ju,Kwiecinski:1996td}. Actually, making appropriate all-order resummations in order to ensure both the correct collinear and anti-collinear DLL limits in the BFKL equation is essentially equivalent \cite{Salam:1998tj} as imposing the kinematical constraint in the BFKL kernel, \emph{i.e.} as imposing simultaneously the $k^+$ ordering \eqref{kplus_ordering} and  the $k^-$ ordering \eqref{kminus_ordering}.


\subsection{Formation time ordering\label{sec:Form_time}}

\begin{figure}
\setbox1\hbox to 10cm{
 \includegraphics{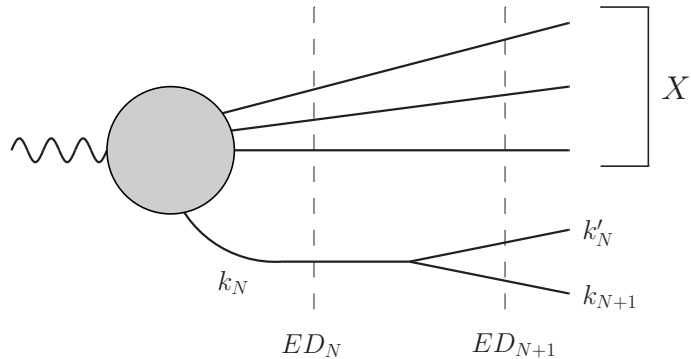}
}
\begin{center}
\hspace{-4cm}
\resizebox*{6cm}{!}{\box1}
\caption{\label{Fig:last_splitting}Tree level contribution to the wave-function of a particle in light-front perturbation theory, where only the last parton splitting is specified to be a one-to-two splitting. $X$ designates here an arbitrary Fock-state. The last two energy denominators are indicated by the vertical dashed lines. The light-front time $x^+$ increases from the left to the right of the diagram, from $-\infty$ to $0$.}
\end{center}
\end{figure}
For completeness, let us now discuss the link between energy denominators and formation time. Consider the effect of a one-to-two parton splitting at the end of a parton cascade, as shown in Fig.\ref{Fig:last_splitting}. The energy denominator $ED_{N+1}$ after that last splitting differs from the energy denominator $ED_{N}$ just before that splitting in the following way: the contribution from the parent parton is removed, and replaced by the contributions of the two daughters. Hence, following the notations in Fig.\ref{Fig:last_splitting}, one has
\begin{equation}
ED_{N+1}=ED_{N} +{k_N}^--{k_{N}'}^--{k_{N+1}}^-\label{EDenom_recur}\, .
\end{equation}
Restricting ourselves to the case of massless partons and using the momentum conservation relations
\begin{eqnarray}
{k_N}^+&=&{k_{N}'}^++{k_{N+1}}^+\\
\mathbf{k}_N&=&\mathbf{k}_{N}'+\mathbf{k}_{N+1}\, ,
\end{eqnarray}
it is elementary to rewrite the relation \eqref{EDenom_recur} as
\begin{equation}
ED_{N+1}=ED_{N}-\frac{{k_N}^+\, {\mathbf{q}_{N+1}}^2}{2\, {k_{N}'}^+\, {k_{N+1}}^+} \label{EDenom_recur_2}\, ,
\end{equation}
introducing the relative transverse momentum of the daughters with respect to the parent parton
\begin{equation}
\mathbf{q}_{N+1}=\mathbf{k}_{N+1}-\frac{{k_{N+1}}^+}{{k_N}^+}\, \mathbf{k}_{N}
= \frac{{k_{N}'}^+}{{k_N}^+}\, \mathbf{k}_{N} - \mathbf{k}_{N}'
= \frac{{k_{N}'}^+}{{k_N}^+}\, \mathbf{k}_{N+1}- \frac{{k_{N+1}}^+}{{k_N}^+}\, \mathbf{k}_{N}'   \label{Relative_transv_mon}\, .
\end{equation}

The absolute value of the second term in the right hand side of the equation \eqref{EDenom_recur_2} is exactly the inverse of the formation time (in $x^+$) associated with the considered parton splitting, \emph{i.e.} the $x^+$ interval it takes for the daughters to be at a larger transverse distance from each other than their transverse wave-length, so that they may lose their quantum coherence.
 One can deduce from the recursion relation \eqref{EDenom_recur_2} that, in the case of a parton cascade initiated from a single particle (on-shell or space-like), and involving only one-to-two splittings, each energy denominator is the opposite of the sum of the inverse formation times associated with each of the previous splittings in the cascade.

In the case of the simultaneous ordering of partons in $k^+$ and $k^-$ discussed in the section \eqref{sec:kinematical_approx}, each energy denominator is dominated by a contribution associated with the last splitting, see the equation \eqref{EDn_approx}. Hence, all cascades satisfying the simultaneous ordering in $k^+$ and $k^-$ are such that the formation times for each of the splittings are strongly decreasing as the cascade develops. And thus the formation time of the whole cascade is essentially the same as the one associated with the very first splitting. Such an ordering in formation time would not always be satisfied, when using the various usual definitions of the multi-Regge kinematics, without the kinematical constraint.

In general, QCD parton cascades at tree level can involve also one-to-three parton splittings, either from the local four-gluons vertex of QCD, or from the nonlocal vertices appearing in light-front perturbations theory. However, those vertices do not give rise to high-energy LL contributions, only to NLL ones at best, so that we can indeed ignore them safely in the present discussion.

For obvious symmetry reasons, it is tempting to guess that the simultaneous ordering in $k^+$ and $k^-$ also imply an opposite strong ordering of the formation times along $x^-$, in a frame where the same cascade seems to develop from the target instead of from the projectile. However, the $s$-channel picture used here breaks the symmetry between projectile and target, and makes it very cumbersome to check explicitly if that property is indeed true. That issue is beyond the scope of the present study.


\section{Mellin space analysis of BFKL and BK evolutions at LL and NLL accuracy\label{sec:Mellin_BFKL_BK_LL_NLL}}

Going to Mellin space allows to diagonalize the LL BFKL equation, making its study straightforward. At higher order, that representation is still very useful, although running coupling effects bring some complications \cite{Lipatov:1985uk,Chirilli:2013kca}.
Naively, it seems unlikely that such a linear transformation would help much in order to study the BK or B-JIMWLK equations, due to their nonlinearity. However, in those equations, the virtual terms are free from nonlinear contributions, and in the real terms in mixed space, the linear and nonlinear contributions are such that they have to combine into a product of dipole (or higher multipole) S-matrices, so that for example the linear real terms fully determine the nonlinear ones. The kinematical issues discussed in this study are associated with the probability density of real gluon emission. That probability density is identical in the nonlinear equations and in their BFKL linearization. Hence, those kinematical effects can be conveniently studied in the context of the Mellin representation of the BFKL equation in mixed space. No information about kinematics is lost in the linearization, because in mixed space the nonlinear terms can be reconstructed uniquely from the linearized version of the real terms.

In this section, some of the results from the seminal paper \cite{Salam:1998tj} analyzing the NLL BFKL equation are adapted to mixed space rather than momentum space, and applied in particular to the explicit result \cite{Balitsky:2008zz} for the NLL B-JIMWLK evolution of a color dipole, in the dilute target regime.

\subsection{Mellin representation of the LL BFKL evolution}

Let us choose for example $Y^+$ as evolution variable in the BFKL equation. Then, one can perform a Laplace transform with respect to $Y^+$ (see appendix \ref{App:Laplace}), which is equivalent to a Mellin transform in $k^+$. Concerning the dependence in the transverse variables, one can use a Mellin representation (see appendix \ref{App:Mellin}), which has no inverse.
For example, one writes the full Mellin representation\footnote{In order to write the formula \eqref{LaplaceMellin_rep} (and similarly \eqref{Mellin_rep}), one makes the assumption that $\left\langle {\textbf N}_{ij} \right\rangle_{Y^+}$ depend on the distance $x_{ij}$, but not on the points $\mathbf{x}_i$ and $\mathbf{x}_i$ independently. It is possible to relax that assumption by including components of higher conformal spin \cite{Lipatov:1985uk} in the Mellin representation \eqref{Mellin_rep}. However, such contributions with strictly positive conformal spin do not grow with energy, and thus are not interesting in the context of gluon saturation. Moreover for the components of conformal spin $n\neq 0$, the higher order corrections to the BFKL kernel \cite{Kotikov:2000pm} are smooth and well-behaved so that no collinear resummation is needed for them.}
\begin{equation}
\left\langle {\textbf N}_{ij} \right\rangle_{Y^+} = \int_{\om_0-i\infty}^{\om_0+i\infty} \frac{\dd \om}{2\pi i}\;
e^{\om\, Y^+}\; \hat{\textbf N}_{ij}(\om)
 = \int_{\om_0-i\infty}^{\om_0+i\infty} \frac{\dd \om}{2\pi i}\;
e^{\om\, Y^+}
\int_{1/2-i\infty}^{1/2+i\infty} \frac{\dd \g}{2\pi i}\, \left(\frac{x_{ij}^2 \, Q_0^2}{4}\right)^\g \; \hat{\cal N}(\g,\om)\label{LaplaceMellin_rep}\, ,
\end{equation}
for the solution of the mixed-space BFKL equation \eqref{BFKL_dipole} with $\eta=Y^+$.
$Q_0$ is an arbitrary momentum scale, which in practice is set to be a typical transverse momentum scale associated with the target.

The Laplace transform of the BFKL equation \eqref{BFKL_dipole} writes
\begin{eqnarray}
\om\, \hat{\textbf N}_{01}(\om) \!-\!\left\langle {\textbf N}_{01} \right\rangle_{0}&=&   \bar{\alpha}
\int \frac{\textrm{d}^2\mathbf{x}_{2}}{2\pi} \textbf{K}_{012}\: \bigg[ \hat{\textbf N}_{02}(\om) \!+\! \hat{\textbf N}_{21}(\om) \!-\!\hat{\textbf N}_{01}(\om)\bigg]
\, .\label{BFKL_Laplace}
\end{eqnarray}
Introducing the Mellin representation
\begin{equation}
\left\langle {\textbf N}_{ij} \right\rangle_{0} =\int_{1/2-i\infty}^{1/2+i\infty} \frac{\dd \g}{2\pi i}\, \left(\frac{x_{ij}^2 \, Q_0^2}{4}\right)^\g \; {\cal N}^{\, 0}(\g)\label{Mellin_rep_InitCond}\, .
\end{equation}
of the initial condition, one obtains from the equation \eqref{BFKL_Laplace}
\begin{equation}
\int_{1/2-i\infty}^{1/2+i\infty} \frac{\dd \g}{2\pi i}\, \left(\frac{x_{01}^2 \, Q_0^2}{4}\right)^\g \;  \hat{\cal N}(\g,\om)\; \Big[\om- \abar \chi(\g)\Big] =\int_{1/2-i\infty}^{1/2+i\infty} \frac{\dd \g}{2\pi i}\, \left(\frac{x_{01}^2 \, Q_0^2}{4}\right)^\g \; {\cal N}^{\, 0}(\g)\label{LaplaceMellin_rep_BFKL}\, ,
\end{equation}
where
\begin{equation}
\chi(\g)=2 \Psi(1)-\Psi(\g)-\Psi(1\!-\!\g)\label{BFKL_eigenvalue}
\end{equation}
is the characteristic function of the LL BFKL kernel at zero conformal spin. Here, $\Psi(\g)$ is the digamma function.
$\hat{\cal N}(\g,\om)$ has to be singular both in $\om$ and in $\g$, in order to provide a non-zero function $\left\langle {\textbf N}_{ij} \right\rangle_{Y^+}$ via the formula \eqref{LaplaceMellin_rep}. And due to the relation \eqref{LaplaceMellin_rep_BFKL}, the only possible contribution to the integral over $\om$ can come from a single pole of $\hat{\cal N}(\g,\om)$ at $\om= \abar \chi(\g)$. Hence, without loss of generality, one can take
\begin{equation}
\hat{\cal N}(\g,\om)= \frac{{\cal N}^{\, 0}(\g)}{\om- \abar \chi(\g)}\, ,  \label{LL_BFKL_sol_in_LaplaceMellin}
\end{equation}
and thus
\begin{eqnarray}
\left\langle {\textbf N}_{ij} \right\rangle_{Y^+}  &=&\int_{1/2-i\infty}^{1/2+i\infty} \frac{\dd \g}{2\pi i}\, \left(\frac{x_{ij}^2 \, Q_0^2}{4}\right)^\g \; \int_{\om_0-i\infty}^{\om_0+i\infty} \frac{\dd \om}{2\pi i}\; e^{\om\, Y^+}\;
\frac{ {\cal N}^{\, 0}(\g)}{\om- \abar \chi(\g)}\nonumber\\
&=&\int_{1/2-i\infty}^{1/2+i\infty} \frac{\dd \g}{2\pi i}\, \left(\frac{x_{ij}^2 \, Q_0^2}{4}\right)^\g \; {\cal N}^{\, 0}(\g) \;
e^{\abar \chi(\g)\, Y^+}
\label{LL_BFKL_sol}\, .
\end{eqnarray}
The integration over $\g$ can then be estimated using the saddle-point approximation, in either the $Y^+\rightarrow +\infty$, the $x_{ij}^2\rightarrow +\infty$ or the $x_{ij}^2\rightarrow 0$ limit.

Since $Y^+$ is used as the evolution variable, one has some control over the $k^+$ of the gluon in the parton cascades contributing to $\left\langle {\textbf N}_{ij} \right\rangle_{Y^+}$: the gluons are ordered in $k^+$ as in \eqref{kplus_ordering} when $Y^+$ is large enough.
The dipole-target amplitude $\left\langle {\textbf N}_{ij} \right\rangle_{Y^+}$ is related by some Fourier transformation to an unintegrated gluon distribution in the target (see \emph{e.g.} \cite{Dominguez:2010xd,Dominguez:2011wm} for a more details), the transverse momentum $\mathbf{k}$ of the gluon being the conjugate of the dipole vector $\mathbf{x}_{i}\!-\!\mathbf{x}_{j}$, and thus $|\mathbf{k}|\propto 1/x_{ij}$.

Hence, the regime $x_{ij}^2 \gg 4/ Q_0^2$ for $\left\langle {\textbf N}_{ij} \right\rangle_{Y^+}$ corresponds to the ${\mathbf{k}}^2$ ordering \eqref{anti_kt_ord}, \emph{i.e.} the anti-collinear regime. In that regime, the saddle point for the integration \eqref{LL_BFKL_sol} is dominated by the first singularity in $\g$ on the left of the line $\textrm{Re}(\g)=1/2$. For physically relevant initial conditions, one does not expect ${\cal N}^{\, 0}(\g)$ to have a singularity in the strip $0<\textrm{Re}(\g)\leq 1/2$, so that the dominant singularity is the one of $\chi(\g)$ in $\g=0$. Indeed,
\begin{equation}
\chi(\g)=\frac{1}{\g} + {\cal O}\left(\g^2\right)  \qquad \qquad \textrm{for } \g\rightarrow 0\, .\label{anticoll_DLL_chi}
\end{equation}
The limit $x_{ij}^2 \gg 4/ Q_0^2$ with large $Y^+$ corresponds to the anti-collinear DLL regime discussed in the section \ref{sec:DLL_mom}, associated with the simultaneous orderings \eqref{anti_kt_ord} and \eqref{kplus_ordering} of the parton cascades. From the relations \eqref{LL_BFKL_sol_in_LaplaceMellin} and \eqref{anticoll_DLL_chi}, one sees that
$\left\langle {\textbf N}_{ij} \right\rangle_{Y^+}$ is driven by a single pole at $\om=\abar/\g$ in that regime.

On the other hand, the generic solution to the DGLAP evolution\footnote{For simplicity only the mostly gluonic eigenstate of the DGLAP evolution in the singlet sector is considered here, since that eigenstate is dominant in the DLL regime.} for $\left\langle {\textbf N}_{ij} \right\rangle_{Y^+}$ in the anti-collinear regime should write
\begin{eqnarray}
\left\langle {\textbf N}_{ij} \right\rangle_{Y^+}  &=&\int_{1/2-i\infty}^{1/2+i\infty} \frac{\dd \g}{2\pi i}\, \left(\frac{x_{ij}^2 \, Q_0^2}{4}\right)^\g \; \int_{\om_0-i\infty}^{\om_0+i\infty} \frac{\dd \om}{2\pi i}\;
e^{\om\, Y^+}\; \frac{{\hat{N}}_{\, 0}(\om) }{\g- \abar \tilde{P}(\om,\abar)}
\label{AC_DGLAP_sol}\, ,
\end{eqnarray}
where ${\hat{N}}_{\, 0}(\om)$ is the Laplace transform of the initial condition for that evolution.
The DLL limit of that DGLAP solution is associated to $Y^+\rightarrow +\infty$, and thus to $\om\rightarrow 0$ because  $\tilde{P}(\om,\abar)$ and ${\hat{N}}_{\, 0}(\om)$ should not have singularities for $\textrm{Re}(\om)>0$.
To leading order in $\abar$, the DGLAP anomalous dimension writes
\begin{equation}
\abar\, \tilde{P}(\om,\abar)= \abar\, \tilde{P}(\om,0)+ {\cal O}(\abar^2)\, \quad \textrm{with} \quad \tilde{P}(\om,0)=\frac{1}{\om}-\left(\frac{11}{12}+\frac{N_f}{6\, N_c^3}\right) + {\cal O}(\om) \quad \textrm{for} \quad \om\rightarrow 0
\, ,\label{DGLAP_Mellin}
\end{equation}
so that the $\left\langle {\textbf N}_{ij} \right\rangle_{Y^+}$ is driven by a single pole at $\g=\abar/\om$ in the anti-collinear DLL regime, in agreement with the result just obtained from the LL BFKL equation. This confirms the fact that the anti-collinear DLL regime is correctly included in the LL BFKL equation with $Y^+$ as evolution variable, as already argued in the section \ref{sec:DLL_mom}.

The LL BFKL characteristic function $\chi(\g)$ has the symmetry $\chi(1\!-\!\g)=\chi(\g)$, and thus its first singularity on the right of the line $\textrm{Re}(\g)=1/2$ is the single pole
\begin{equation}
\chi(\g)=\frac{1}{1\!-\!\g} + {\cal O}\left((1\!-\!\g)^2\right)  \qquad \qquad \textrm{for } \g\rightarrow 1\, .\label{fake_coll_DLL_pole}
\end{equation}
That pole is driving the behavior of $\left\langle {\textbf N}_{ij} \right\rangle_{Y^+}$ in the limit $x_{ij}^2 \ll 4/ Q_0^2$, which is the position space analog of the transverse momentum ordering \eqref{kt_ord}. However, the evolution variable is $Y^+$, which can only impose the $k^+$ ordering \eqref{kplus_ordering}. Hence,
that pole at $\g=1$ does not correspond to the correct collinear DLL limit, but rather to the regime with the simultaneous orderings \eqref{kplus_ordering} and \eqref{kt_ord}, and with the $k^-$'s unconstrained.

As discussed in the section \ref{sec:DLL_mom}, due to the choice of high-energy factorization scheme with $Y^+$ as evolution variable, one expects the anti-collinear DLL limit to be correctly reproduced but not the collinear DLL limit. In the Mellin representation \eqref{LaplaceMellin_rep}, this failure thus shows up as unphysical singularities arising at $\g=1$.

\subsection{The collinear regime in Mellin representation\label{sec:coll_Mellin}}

Following the momentum-space discussion of the section \ref{sec:DLL_mom}, one should keep track of the variable $Y^-$ instead of $Y^+$ when studying the collinear regime, in which the transverse scales are harder and harder when going from the target to the projectile. Hence, it is natural for that purpose to choose a factorization scheme in which $Y^-$ plays the r\^ole of evolution variable. A change of factorization scheme can modify the NLL kernel of the BFKL or BK evolution but not the LL kernel. Hence, the generic solution of the LL BFKL evolution in such a factorization  scheme writes
\begin{eqnarray}
\left\langle {\textbf N}_{ij} \right\rangle_{Y^-}  &=&\int_{1/2-i\infty}^{1/2+i\infty} \frac{\dd \bar{\g}}{2\pi i}\, \left(\frac{x_{ij}^2 \, Q_0^2}{4}\right)^{\bar{\g}} \;  \int_{\om_0-i\infty}^{\om_0+i\infty} \frac{\dd \bar{\om}}{2\pi i}\; e^{\bar{\om}\, Y^-}\;
\frac{\bar{{\cal N}}^{\, 0}(\bar{\g})}{\bar{\om}- \abar \chi(\bar{\g})}
\label{LL_BFKL_minus_sol}\, .
\end{eqnarray}

The variables $Y^-$ and $Y^+$ are directly related to each other in momentum space. However, due to the transverse Fourier transform, it is difficult to relate the mixed-space representation in $Y^-$ and transverse position to the more usual mixed-space representation in $Y^+$ and transverse position.
In full momentum space, one has
\begin{equation}
Y^-=\log\left(\frac{k^-_{\max}}{k^-}\right)=\log\left(\frac{2 k^+}{\mathbf{k}^2}\, \frac{Q_0^2}{2 k^+_{\min}}\right)
=Y^++\log\left(\frac{Q_0^2}{\mathbf{k}^2}\right)\, .
\end{equation}
Starting from the mixed-space in $Y^+$ and transverse positions, the best approximation one has for $|\mathbf{k}|$ is $2/x_{ij}$. Hence, in that case, it is natural to approximate $Y^-$ as
\begin{equation}
Y^-\simeq Y^++\log\left(\frac{x_{ij}^2 \, Q_0^2}{4}\right)\, .\label{Yminus_approx}
\end{equation}
Following that idea, one can approximate the expression \eqref{LL_BFKL_minus_sol} within the standard mixed space as
\begin{eqnarray}
\left\langle {\textbf N}_{ij} \right\rangle_{Y^+}  &\simeq &\int_{1/2-i\infty}^{1/2+i\infty} \frac{\dd \bar{\g}}{2\pi i}\, \left(\frac{x_{ij}^2 \, Q_0^2}{4}\right)^{\bar{\g}} \;  \int_{\om_0-i\infty}^{\om_0+i\infty} \frac{\dd \bar{\om}}{2\pi i}\; e^{\bar{\om}\, \left(Y^++\log\left(\frac{x_{ij}^2 \, Q_0^2}{4}\right)\right)}\;
\frac{\bar{{\cal N}}^{\, 0}(\bar{\g})}{\bar{\om}- \abar \chi(\bar{\g})}\nonumber\\
&\simeq &\int_{1/2-i\infty}^{1/2+i\infty} \frac{\dd \bar{\g}}{2\pi i}\,   \int_{\om_0-i\infty}^{\om_0+i\infty} \frac{\dd \bar{\om}}{2\pi i}\; \left(\frac{x_{ij}^2 \, Q_0^2}{4}\right)^{\bar{\g}+\bar{\om}} \; e^{\bar{\om}\, Y^+}\;
\frac{\bar{{\cal N}}^{\, 0}(\bar{\g})}{\bar{\om}- \abar \chi(\bar{\g})}
\label{LL_BFKL_minus_sol_approx}\, .
\end{eqnarray}
Comparing the expressions \eqref{LL_BFKL_sol} and \eqref{LL_BFKL_minus_sol_approx}, one finds that the Laplace-Mellin variables $(\bar{\g},\bar{\om})$ suitable in the collinear regime are related to the Laplace-Mellin variables $(\g,\om)$ suitable in the anti-collinear regime as
\begin{equation}
\bar{\om}=\om \quad \textrm{and} \quad \bar{\g}=\g-\om\, .
\end{equation}

The collinear DLL regime is now obtained by taking the $x_{ij}^2 \ll 4/ Q_0^2$ limit, while supposing $Y^+\!+\!\log\left(\frac{x_{ij}^2 \, Q_0^2}{4}\right)$ large but finite.
Due to the behavior \eqref{fake_coll_DLL_pole}, $\left\langle {\textbf N}_{ij} \right\rangle_{Y^+}$ is driven in this regime by a single pole at $\bar{\om}=\abar / (1\!-\! \bar{\g})$. When translated in $(\g,\om)$ variables, this corresponds to
\begin{equation}
\om= \frac{\abar}{1\!-\!\g \!+\! \om}
\end{equation}
and thus to
\begin{eqnarray}
\om&=& \frac{1}{2}\, \left[\sqrt{(1\!-\!\g)^2+4\, \abar}-(1\!-\!\g) \right]\label{coll_DLL_full} \\
&=& \frac{\abar}{(1\!-\!\g)}-\frac{{\abar}^2}{(1\!-\!\g)^3}+{\cal O}\left(\frac{{\abar}^3}{(1\!-\!\g)^5}\right) \quad \textrm{for} \; \abar\rightarrow 0 \; \; \textrm{and} \; \g<1
\, .\label{coll_DLL_series}
\end{eqnarray}
Notice that the full expression \eqref{coll_DLL_full} is regular at $\g=1$, whereas, when truncated at any order, the series \eqref{coll_DLL_series} has coefficients with severe unphysical singularities at $\g=1$. A similar analysis of the collinear limit was performed in Ref. \cite{Salam:1998tj} for the case of the BFKL evolution in full momentum space.

\subsection{Spurious singularities in the NLL B-JIMWLK evolution for a dipole\label{sec:spurious_sing}}

The NLL generalization of the B-JIMWLK evolution equation \eqref{B_JIMWLK_dipole} for $\left\langle {\mathbf S}_{01} \right\rangle_{Y^+}$ has been calculated in ref. \cite{Balitsky:2008zz} in the factorization scheme with cut-off in $k^+$, and using the standard definition of the Regge limit, \emph{i.e.} assuming all the transverse scales to be of the same order. That equation has been studied in Mellin representation within the 2-gluons exchange approximation,  valid for a dilute target.
In that NLL equation, the terms associated with the one-loop renormalization of the coupling $\abar$ lead to terms with derivatives $\d_\g$ in the Mellin representation of the kernel. It is convenient to separate those contributions from the other NLL ones and resum them into the LL part of the equation, by promoting the coupling to a running coupling
\begin{equation}
\abar \mapsto
\abar(x_{ij}^2)= \frac{1}{b\: \log\left(\frac{4\, \exp({2\, \Psi(1)})}{x_{ij}^2 \, \Lambda_{QCD}^2}  \right)}\; ,\label{run_abar}
\end{equation}
with the one-loop beta function coefficient
\begin{equation}
b=\frac{11}{12}-\frac{N_f}{6\, N_c}\label{def_b}
\end{equation}
and the scale $\Lambda_{QCD}$ from the $\overline{\textrm{MS}}$ scheme at one-loop, with $N_c$ colors and $N_f$ flavors. The factor $4\, \exp({2\, \Psi(1)})$ appearing in the logarithm in the expression \eqref{run_abar} comes from the Fourier transform from momentum to transverse position space \cite{Kovchegov:2006vj}, but is not important for our purposes.

The simplest choice is the so-called parent dipole prescription, where the coupling in the LL kernel is taken to run with the parent dipole size $x_{01}$. Then, the NLL generalization of the equation \eqref{BFKL_Laplace} can be written in the form
\begin{eqnarray}
\om\, \hat{\textbf N}_{01}(\om) \!-\!\left\langle {\textbf N}_{01} \right\rangle_{0}&=&   \bar{\alpha}(x_{01}^2)
\int \frac{\textrm{d}^2\mathbf{x}_{2}}{2\pi} \textbf{K}_{012}\: \bigg[ \hat{\textbf N}_{02}(\om) \!+\! \hat{\textbf N}_{21}(\om) \!-\!\hat{\textbf N}_{01}(\om)\bigg]+  \bar{\alpha}^2(\cdots) \int  \textbf{K}^{NLL}\: \otimes  \:\hat{\textbf N}_{ij}(\om)\label{BFKL_NLL_dipole_Laplace_1}\\
&=&  \bar{\alpha}(x_{01}^2)  \int_{1/2-i\infty}^{1/2+i\infty} \frac{\dd \g}{2\pi i}\, \left(\frac{x_{01}^2 \, Q_0^2}{4}\right)^\g \; \chi(\g) \; \hat{\cal N}(\g,\om)\nonumber\\
& &  + \bar{\alpha}^2(\cdots) \int_{1/2-i\infty}^{1/2+i\infty} \frac{\dd \g}{2\pi i}\, \left(\frac{x_{01}^2 \, Q_0^2}{4}\right)^\g \; \chi_1(\g) \; \hat{\cal N}(\g,\om)
\, .\label{BFKL_NLL_dipole_Laplace}
\end{eqnarray}
In the right hand side of the equation \eqref{BFKL_NLL_dipole_Laplace_1}, the scale for the coupling in the NLL contribution is unconstrained at this order.

The characteristic function $\chi_1(\g)$ of the NLL kernel $\textbf{K}^{NLL}$ in the 2-gluons exchange approximation
has been calculated in ref. \cite{Balitsky:2008zz}, and shown to be identical\footnote{The small discrepancy in the original calculation of ref.  \cite{Balitsky:2008zz} was due to a mistake in the calculation of some integral, which has been corrected since then, see ref. \cite{Balitsky:2009xg}.} to the characteristic function of the momentum-space NLL BFKL kernel \cite{Fadin:1998py,Ciafaloni:1998gs}, up to terms depending on the choices of running coupling prescription and of factorization scheme, as expected. $\chi_1(\g)$ is regular for $\g$ between $0$ and $1$, and has multiple poles at $\g=0$ and $\g=1$, which write
\begin{equation}
\chi_1(\g)= -\frac{b}{\g^2} -\left(\frac{11}{12}+\frac{N_f}{6\, N_c^3}\right) \frac{1}{\g^2} + {\cal O}\left(\frac{1}{\g}\right)   \qquad \qquad \textrm{for } \g\rightarrow 0\label{double_poles_0}
\end{equation}
and
\begin{equation}
\chi_1(\g)= -\frac{1}{(1\!-\!\g)^3} -\left(\frac{11}{12}+\frac{N_f}{6\, N_c^3}\right) \frac{1}{(1\!-\!\g)^2} + {\cal O}\left(\frac{1}{1\!-\!\g}\right)   \qquad \qquad \textrm{for } \g\rightarrow 1\label{double_and_triple_poles_1}\, .
\end{equation}

In the anti-collinear regime, associated with $\g\rightarrow 0$, the presence of a double pole imply that NLL corrections dominate over LL terms, which contain only a simple pole, see \eqref{anticoll_DLL_chi}. Hence, the perturbative expansion of the BK (or BFKL) kernel breaks down in the anti-collinear regime. The collinear regime is driven by the first singularities on the right of $\g=1/2$, which are here at $\g=1$. At this point there is a triple pole in addition to the double pole, so that the breakdown of the perturbative expansion of the BK kernel is even more severe in the collinear regime than in the anti-collinear regime.

The presence of that triple pole at $\g=1$ together with the absence of triple pole at $\g=0$ confirm the previous analysis of the collinear and anti-collinear regimes, as this triple pole in \eqref{double_and_triple_poles_1} is precisely the second term in the expansion \eqref{coll_DLL_series}. Hence, the triple pole at $\g=1$ is an artifact of the factorization scheme based on the $Y^+$ variable, which is the appropriate variable to keep in addition to the transverse ones in the anti-collinear regime but not in the collinear regime. In Mellin space, this is associated with the mismatch between the variables $\g$ and $\bar{\g}$, as already discussed. The correct prediction of the pattern of triple poles also suggests that the approximation \eqref{Yminus_approx} of $Y^-$ is good enough for our purposes.

The aim of the present study is to propose a consistent scheme for the resummation of such large higher order corrections coming from factorization scheme issues. However, for completeness, let us discuss also the large corrections associated with the double poles at $0$ or $1$, before closing this section.

The first contribution in the expression \eqref{double_poles_0}, proportional to $b$, is obviously associated with the running of the coupling. One expects that the physically correct scale for the running coupling in the LL kernel is the hardest available, \emph{i.e.} the smallest dipole size among parent and daughters, when one is much smaller than the two others. The scale $x_{01}^2$ chosen for the coupling $\abar$ is thus correct in the collinear regime $x_{01}^2\ll x_{02}^2 \sim x_{21}^2$, but inappropriate in the anti-collinear regimes $x_{02}^2\ll x_{01}^2 \sim x_{21}^2$ and $x_{21}^2\ll x_{01}^2 \sim x_{02}^2$. That explains why a large correction proportional to $b$ appears in $\chi_1(\g)$ in the anti-collinear limit $\g\rightarrow 0$ but not in the collinear limit $\g\rightarrow 1$. For the same reason, one can also predict the appearance of poles of order $n+1$ with residues related to $b$ in the characteristic function of the N$^n$LL kernel at $\g=0$ but not at $\g=1$. In order to make the perturbative expansion more stable, one should then choose a running coupling prescription physically correct in all the limits, in order to avoid such large higher order corrections to appear. In position space, the two available running coupling prescriptions which satisfy that requirement are Balitsky's prescription \cite{Balitsky:2006wa} and the minimal dipole size prescription $\abar(\min (x_{01}^2,x_{02}^2,x_{21}^2))$. Despite sharing the same behavior in all the relevant limits, those two prescriptions are not identical and give different quantitative results \cite{Berger:2011ew}. Notice that in the Kovchegov-Weigert prescription \cite{Kovchegov:2006vj}
the scale in the coupling does not reduces to the parent dipole size in the collinear limit, so that it should induce
poles of order $n+1$ in the characteristic function of the N$^n$LL kernel at $\g=1$, making the perturbative expansion unstable in the collinear regime. 

The second term contributing to the double pole at $\g\rightarrow 0$ in the expression \eqref{double_poles_0} is inherited from the LO DGLAP anomalous dimension due to the duality between the DGLAP anomalous dimension and the BFKL characteristic function (see \emph{e.g.} \cite{Altarelli:1999vw}). In general, after removing the contributions related to running coupling, the perturbative expansion the DGLAP anomalous dimension can be written to all order as
\begin{equation}
\g=\abar\, \tilde{P}(\om,\abar)=\sum_{n=1}^{+\infty} \sum_{m=0}^{+\infty} p_{n,m}\;  \left(\frac{\abar}{\om}\right)^n\, \om^m
\, ,\label{DGLAP_Mellin_exp}
\end{equation}
where the terms with a given $n$ sum up to the N$^{n\!-\!1}$LO contribution to the DGLAP anomalous dimension, for example
\begin{equation}
\tilde{P}(\om,0)= \sum_{m=0}^{+\infty} p_{1,m}\; \om^{m\!-\!1}
\, .\label{DGLAP_Mellin_exp_LO}
\end{equation}
Similarly, the perturbative expansion of the BFKL intercept can be written to all orders as (discarding running coupling contributions)
\begin{equation}
\om=\sum_{q=1}^{+\infty} \sum_{k=0}^{+\infty} c_{q,k}\;  \left(\frac{\abar}{\g}\right)^q\, \g^k
\, ,\label{BFKL_Mellin_exp}
\end{equation}
where the terms with a given $q$ sum up to the N$^{q\!-\!1}$LO contribution to the BFKL intercept, for example at LO
\begin{equation}
\chi(\g)= \sum_{k=0}^{+\infty} c_{1,k}\; \g^{k\!-\!1}
\, .\label{BFKL_Mellin_exp_LO}
\end{equation}
The all-order expansions \eqref{DGLAP_Mellin_exp} and \eqref{BFKL_Mellin_exp} have to coincide in the anti-collinear low-$x$ regime $\g$, $\om$, $\abar\rightarrow 0$ with $\g \om \sim\abar$. Then, the coefficients $c_{q,k}$ are fully determined by the full set of coefficients $p_{n,m}$ and vice versa. In particular, the terms in the anomalous dimension which are the most singular in the low-$x$ limit ($\om\rightarrow 0$), which are of the form $p_{n,0}\;  (\abar/\om)^n$, are fully determined by the LO contributions \eqref{BFKL_Mellin_exp_LO} to the the BFKL intercept. The first terms give
\begin{eqnarray}
p_{1,0}&=& c_{1,0} =1 \label{DLL_DGLAP_coeff}\\
p_{2,0}&=& p_{1,0}\, c_{1,1}=0
\, ,\label{sing_DGLAP_coeff}
\end{eqnarray}
where the values $c_{1,0}=1$ and 
$c_{1,1}=0$
have been read off from the expansion \eqref{anticoll_DLL_chi}. Hence, the absence of spurious singularities at low-$x$ in the DGLAP evolution at NLO is due to the (\emph{a priori} accidental) cancellation of the term of order $\g^0$ in the expansion \eqref{anticoll_DLL_chi} of the LO BFKL characteristic function. Conversely, the terms in the BFKL intercept which are the most singular in the anti-collinear limit ($\g\rightarrow 0$), of the form $c_{q,0}\;  (\abar/\g)^q$, are fully determined by the LO contributions \eqref{DGLAP_Mellin_exp_LO} to the DGLAP anomalous dimension. At first order, one gets the relation \eqref{DLL_DGLAP_coeff} once again, and at the next order
\begin{eqnarray}
c_{2,0}&=& c_{1,0}\, p_{1,1}=-\left(\frac{11}{12}+\frac{N_f}{6\, N_c^3}\right)
\, ,\label{sing_BFKL_coeff}
\end{eqnarray}
where the value of the coefficient $p_{1,1}$ has been read off from the expansion \eqref{DGLAP_Mellin}. Hence, the second term in the expression \eqref{double_poles_0} is inherited from the LO DGLAP anomalous dimension as announced. Furthermore, one can predict poles of order $n+1$ in the N$^n$LL BFKL characteristic function and calculate their residues from the expansion of the LO DGLAP anomalous dimension \eqref{DGLAP_Mellin}.

The same analysis can be done in the collinear regime $\bar{\g}\rightarrow 1$ instead of $\g\rightarrow 0$. It would predict the double pole term in the expansion \eqref{double_and_triple_poles_1}, up to higher order terms due to the change of variable from $\bar{\g}$ to $\g$. Hence, the second term in the expansions \eqref{double_poles_0} and \eqref{double_and_triple_poles_1} should be resummed in principle by promoting the collinear and anti-collinear DLL terms in the LO BFKL or BK kernel to the full collinear and anti-collinear LO DGLAP. This task has been performed for the BFKL case both in Mellin space \cite{Ciafaloni:1999yw,Altarelli:1999vw,Altarelli:2005ni} and in momentum space \cite{Ciafaloni:2003rd}. In the case of the BK equation, it is the main step left for further studies towards a full resummation of the pathologically large higher order contributions to the kernel in mixed space. In that case, the main difficulty is that it seems quite difficult to see how the DGLAP evolution of the target arises in the dipole framework.


\section{Analysis of the NLO DIS impact factors\label{sec:NLO_IF_analysis}}

At low $x_{Bj}$, it is convenient to parameterize the DIS cross section by the transverse and longitudinal virtual photon cross sections $\sigma_{T,L}^{\gamma}(Q^2,x_{Bj})$, related to the usual structure functions $F_L(Q^2,x_{Bj})$ and $F_2(Q^2,x_{Bj})=F_T(Q^2,x_{Bj})+F_L(Q^2,x_{Bj})$ by
\begin{equation}
F_{T,L}(Q^2,x_{Bj})=\frac{Q^2}{(2\pi)^2\, \alpha_{em}}\; \sigma^{\gamma}_{T,L}(Q^2,x_{Bj})\, .\label{rel_FTL_sigmaTL}
\end{equation}
At low $x_{Bj}$, those photon cross sections $\sigma_{T,L}^{\gamma}$ obey at LO the dipole factorization \cite{Bjorken:1970ah,Nikolaev:1990ja}.
The real NLO corrections to the dipole factorization have been calculated in Ref.\cite{Beuf:2011xd}. The expression for $\sigma_{T,L}^{\gamma}$ at NLO writes\footnote{In Ref.\cite{Beuf:2011xd}, only the real NLO corrections have been calculated explicitly, whereas virtual NLO corrections have been inferred by using a unitarity argument. However, there was a flaw in the particular implementation of the unitarity requirement, so that the expression given in Ref.\cite{Beuf:2011xd} is not correct and the virtual NLO corrections need to be calculated explicitly. Here, those yet unknown virtual corrections are indicated by the ${\cal O}(\abar)$ term. This issue is being further studied \cite{BeufToAppear}, but does not affect the physics discussed in the present paper, which is driven by the NLO real corrections. The virtual ${\cal O}(\abar)$ term should be UV divergent, in order to cancel the divergences of the NLO real contribution for $\mathbf{x}_{2}\rightarrow \mathbf{x}_{0}$ and $\mathbf{x}_{2}\rightarrow \mathbf{x}_{1}$. The ${\cal O}(\abar)$ term should also have a soft log divergence regulated by the cut-off $k^+_{\min}$.}
\begin{eqnarray}
\sigma_{T,L}^{\gamma}(Q^2,x_{Bj})&=& \frac{4\, N_c\, \alpha_{em}}{(2\pi)^2}\sum_f e_f^2   \int \textrm{d}^2\mathbf{x}_{0} \int \textrm{d}^2\mathbf{x}_{1} \int_0^1 \textrm{d} z_1\, \Bigg\{ \bigg[\mathcal{I}_{T,L}^{LO}({x}_{01},z_1,Q^2)+{\cal O}(\abar)  \bigg]
\Big[1- \left\langle {\mathbf S}_{01} \right\rangle_{0}\Big]\nonumber\\
& & + \bar{\alpha} \int_{k^+_{\min}/q^+}^{1\!-\!z_1}\frac{\textrm{d}z_2}{z_2}\; \int \frac{\textrm{d}^2\mathbf{x}_{2}}{2\pi}\; \mathcal{I}_{T,L}^{NLO}(\mathbf{x}_{0},\mathbf{x}_{1},\mathbf{x}_{2},z_1,z_2,Q^2)\:
\frac{2\, C_F}{N_c}\: \Big[1- \left\langle {\mathbf S}_{012} \right\rangle_{0}\Big]
\Bigg\}\, ,\label{sigma_TL_z_min}
\end{eqnarray}
where ${\mathbf S}_{012}$ is the operator describing the eikonal interaction of a $q\bar{q}g$ tripole with the target, \emph{i.e.}
\begin{equation}
 {\mathbf S}_{012}\equiv \frac{1}{N_c\, C_F}\: \textrm{Tr} \left(U_{\mathbf{x}_{0}}\, t^a\, U_{\mathbf{x}_{1}}^\dag\, t^b \right)\: \tilde{U}^{ba}_{\mathbf{x}_{2}}
 =\frac{N_c}{2 C_F} \bigg[ {\mathbf S}_{02}\,{\mathbf S}_{21}-\frac{1}{N_c^2} {\mathbf S}_{01}\bigg]\, ,\label{def_tripole}
\end{equation}
$\tilde{U}^{ba}_{\mathbf{x}_{2}}$ being in the adjoint representation.
The notation $\left\langle\, \cdots \right\rangle_{0}$ indicates that the expectation values of the operators should be evaluated, at this stage, in a quasi-classical approximation, such as the MV model \cite{McLerran:1993ni,McLerran:1993ka,McLerran:1994vd} in the case of a large nuclear target. The expression \eqref{sigma_TL_z_min} is indeed valid at strict NLO accuracy, and does not yet contain the resummation of high-energy LL. The light-cone momentum $k^+_2=z_2 q^+$ of the additional gluon in the photon wave-function has been bounded by the longitudinal resolution $k^+_{\min}$ \eqref{kplus_min_def} of the target, in order to regulate the integral over $z_2$. Using the model for the target proposed in the section \ref{sec:evol_variables}, the lower cut-off in $z_2$ becomes
\begin{equation}
z_{\min}=\frac{k^+_{\min}}{q^+}=\frac{x_{Bj}\, Q_0^2}{Q^2\, x_0}\, .
\end{equation}

In the dipole factorization formula \eqref{sigma_TL_z_min}, the LO impact factors are
\begin{eqnarray}
\mathcal{I}_{L}^{LO}({x}_{01},z_1,Q^2)&=& 4 Q^2  z_1^2 (1\!-\!z_1)^2\, \textrm{K}_0^2\!\left(QX_2\right)\label{ImpFact_LO_L}\\
\mathcal{I}_{T}^{LO}({x}_{01},z_1,Q^2)&=& \big[z_1^2+(1\!-\!z_1)^2\big] z_1 (1\!-\!z_1) Q^2 \textrm{K}_1^2\!\left(QX_2\right)\label{ImpFact_LO_T}\, ,
\end{eqnarray}
whereas the longitudinal NLO impact factor is
\begin{eqnarray}
\mathcal{I}_{L}^{NLO}(\mathbf{x}_{0},\mathbf{x}_{1},\mathbf{x}_{2},z_1,z_2,Q^2)&=&4 Q^2 \, \textrm{K}_0^2\!\left(QX_3 \right) \Bigg\{
 \frac{z_1^2 (1\!-\!z_1)^2}{{x}_{20}^2}\,
 {\cal P}\!\left(\frac{z_2}{1\!-\!z_1}\right)\nonumber\\
& &+\frac{(z_1\!+\!z_2)^2 (1\!-\!z_1\!-\!z_2)^2 }{2\, {x}_{21}^2}\,
 \left[1+\left(1\!-\!\frac{z_2}{z_1\!+\!z_2}\right)^2\right]{\cal P}\!\left(\frac{z_2}{z_1\!+\!z_2}\right)\nonumber\\
& &  -2 z_1 (1\!-\!z_1) (z_1\!+\!z_2) (1\!-\!z_1\!-\!z_2) \left[1\!-\!\frac{z_2}{2(1\!-\!z_1)}\!-\!\frac{z_2}{2(z_1\!+\!z_2)}\right] \left(\frac{\mathbf{x}_{20}\cdot\mathbf{x}_{21}}{{x}_{20}^2\; {x}_{21}^2}\right)
\Bigg\}
\label{ImpFact_NLO_L}
\end{eqnarray}
and the transverse one
\begin{eqnarray}
& &\mathcal{I}_{T}^{NLO}(\mathbf{x}_{0},\mathbf{x}_{1},\mathbf{x}_{2},z_1,z_2,Q^2)= \Bigg[\frac{Q X_3\, \textrm{K}_1\!\left(Q X_3\right)}{X_3^2}\Bigg]^2
\Bigg\{z_1^2 (1\!-\!z_1)^2 \big[z_1^2+(1\!-\!z_1)^2\big]  \left(\mathbf{x}_{10}\!-\!\frac{z_2}{1\!-\!z_1} \mathbf{x}_{20}\right)^2  \frac{\,{\cal P}\!\left(\frac{z_2}{1\!-\!z_1}\right)}{{x}_{20}^2}\nonumber\\
& &+(1\!-\!z_1\!-\!z_2)^2 (z_1\!+\!z_2)^2 \big[(1\!-\!z_1\!-\!z_2)^2+(z_1\!+\!z_2)^2\big]  \left(\mathbf{x}_{01}\!-\!\frac{z_2}{z_1\!+\!z_2} \mathbf{x}_{21}\right)^2  \frac{\,{\cal P}\!\left(\frac{z_2}{z_1\!+\!z_2}\right)}{{x}_{21}^2}\nonumber\\
& &+2 z_1 (1\!-\!z_1) (1\!-\!z_1\!-\!z_2) (z_1\!+\!z_2) \Big[z_1(z_1\!+\!z_2)+(1\!-\!z_1\!-\!z_2)(1\!-\!z_1)\Big]\nonumber\\
& & \quad \times \left[1\!-\!\frac{z_2}{2(1\!-\!z_1)}\!-\!\frac{z_2}{2(z_1\!+\!z_2)}\right]
\left(\mathbf{x}_{10}\!-\!\frac{z_2}{1\!-\!z_1} \mathbf{x}_{20}\right)\!\!\cdot\!\! \left(\mathbf{x}_{01}\!-\!\frac{z_2}{z_1\!+\!z_2} \mathbf{x}_{21}\right)
\left(\frac{\mathbf{x}_{20}\cdot\mathbf{x}_{21}}{{x}_{20}^2\; {x}_{21}^2}\right)\nonumber\\
& & + \frac{z_2^2\, z_1\, (1\!-\!z_1\!-\!z_2)\, (1\!-\!2 z_1\!-\!z_2)^2}{(1\!-\!z_1) (z_1\!+\!z_2)}\; \frac{\big(\mathbf{x}_{20}\wedge\mathbf{x}_{21}\big)^2}{{x}_{20}^2\; {x}_{21}^2}\nonumber\\
& & + z_2\, z_1^2\, (1\!-\!z_1\!-\!z_2) \bigg[\frac{z_1\, (1\!-\!z_1\!-\!z_2)}{(1\!-\!z_1)}+\frac{(1\!-\!z_1)^2}{(z_1\!+\!z_2)} \bigg] \left(\mathbf{x}_{10}\!-\!\frac{z_2}{1\!-\!z_1} \mathbf{x}_{20}\right)\!\!\cdot\!\! \left(\frac{\mathbf{x}_{20}}{{x}_{20}^2}\right)\nonumber\\
& &+  z_2\, z_1\, (1\!-\!z_1\!-\!z_2)^2 \bigg[\frac{z_1\, (1\!-\!z_1\!-\!z_2)}{(z_1\!+\!z_2)}+\frac{(z_1\!+\!z_2)^2}{(1\!-\!z_1)} \bigg] \left(\mathbf{x}_{01}\!-\!\frac{z_2}{z_1\!+\!z_2} \mathbf{x}_{21}\right)\!\!\cdot\!\! \left(\frac{\mathbf{x}_{21}}{{x}_{21}^2}\right)\nonumber\\
& &+ \frac{z_2^2\, z_1^2\, (1\!-\!z_1\!-\!z_2)^2}{2} \bigg[\frac{1}{(1\!-\!z_1)^2}+\frac{1}{(z_1\!+\!z_2)^2} \bigg]
\Bigg\}
\, .\label{ImpFact_NLO_T}
\end{eqnarray}
The variables $X_2$ and $X_3$ which appear in the impact factors are defined by
\begin{eqnarray}
X_2^2&=& z_1\, (1\!-\!z_1)\, {x}_{01}^2\label{X2}\\
X_3^2&=& z_1\, (1\!-\!z_1\!-\!z_2)\, {x}_{01}^2 + z_2\, (1\!-\!z_1\!-\!z_2)\, {x}_{02}^2 + z_2\, z_1\, {x}_{21}^2\label{X3}\, ,
\end{eqnarray}
and
\begin{equation}
{\cal P}(z)=\frac{1}{2} \left[1+ \left(1\!-\!z\right)^2\right]=1-z+\frac{z^2}{2}\label{cal_P}
\end{equation}
is related to the non-regularized quark to gluon LO DGLAP splitting function as
\begin{equation}
P_{gq}(z)=2\, C_F\;\frac{{\cal P}(z)}{z}\label{Pgq_splitting}\, .
\end{equation}


\subsection{Mixed-space analysis of the real NLO DIS impact factors}


\subsubsection{DIS impact factors and the formation time of intermediate Fock states\label{sec:Form_time_mixed_space}}

As argued in the section II.C.2 of Ref.\cite{Beuf:2011xd}, the factors in the LO and NLO DIS impact factors \eqref{ImpFact_LO_L}, \eqref{ImpFact_LO_T}, \eqref{ImpFact_NLO_L} and \eqref{ImpFact_NLO_T} containing the modified Bessel functions $\textrm{K}_0$ or $\textrm{K}_1$ have a kinematical origin. The quantities $2 q^+\, X_2^2$ and $2 q^+\, X_3^2$ are mixed-space expressions for the formation time of the Fock state ($q\bar{q}$ or $q\bar{q}g$ respectively) in the photon wave-function which is resolved by interaction with the target. On the other hand, $2 q^+/Q^2$ is the lifetime of the virtual photon. Hence,  $Q^2 X_2^2$ and $Q^2 X_3^2$ are the ratios of the formation time of the $q\bar{q}$ and $q\bar{q}g$ Fock states over the photon lifetime. The Fock states which have not enough time to form during the photon lifetime should not contribute to the photon-target cross sections. That property is guarantied by the exponential suppression at large values of $Q^2 X_2^2$ or $Q^2 X_3^2$ provided by the modified Bessel functions in the LO and NLO DIS impact factors.

In the real-photon limit $Q^2\!\!\rightarrow\!\! 0$, the longitudinal photon contribution disappears as expected, thanks to $Q^2 \, \textrm{K}_0^2\!\left(QX_n \right)\!\rightarrow\! 0$. And the only change in the transverse photon case is the switch from exponential suppression to power suppression $Q^2 \, \textrm{K}_1^2\!\left(QX_n \right)\!\rightarrow\! 1/X_n^2$ at large $X_n$, effectively allowing the contribution to $\sigma_{T}^{\gamma}(Q^2\!=\!0,x_{Bj})$ of Fock states with arbitrary large formation time.\footnote{This observation seems to suggest that in the real-photon limit, there will be sizable contributions from $q\bar{q}$ dipoles with arbitrarily large transverse separation, among other Fock states. But of course, such contributions should be suppressed at the non-perturbative level by confinement effects. Hence, this just confirms that one should not trust perturbative calculations of the photon wave-function in the case of a real or quasi-real photon.}

Physically, one expects that NLO corrections can contribute to high-energy logarithms only if there is a strong ordering in the formation time, as discussed section \ref{sec:Form_time}, with the $q\bar{q}g$ state being a short-lived fluctuation of a $q\bar{q}$ dipole. Indeed, when $X_3^2\simeq X_2^2$, the Bessel function factor in  $\mathcal{I}_{T,L}^{NLO}$ reduces to the one in $\mathcal{I}_{T,L}^{LO}$.


\subsubsection{Resummation of high-energy LL's in DIS at NLO\label{sec:std_subtr_LL}}

When calculating an observable at fixed order in perturbation theory in the high-energy limit, one obtains large logs at each order, starting at NLO. By definition, high-energy evolution equations like those discussed in the section \ref{sec:evolEqs} should allow us to resum the leading high-energy logs appearing in the higher order perturbative corrections into the non-perturbative objects appearing at lower orders. In the case of DIS cross sections \eqref{sigma_TL_z_min}, only the dipole amplitude $\left\langle {\mathbf S}_{01} \right\rangle_{0}$ appear at LO. After performing the LL resummation, it should be replaced in the LO contribution by the dipole amplitude $\left\langle {\mathbf S}_{01} \right\rangle_{Y_f^+}$ evolved over the range $Y_f^+=\log(k_f^+/k^+_{\min})=\log(z_f/z_{\min})$, where $k_f^+$ is the chosen factorization scale. In practice, one substitutes in the LO contribution the expression
\begin{equation}
\left\langle {\mathbf S}_{01} \right\rangle_{0}=\left\langle {\mathbf S}_{01} \right\rangle_{Y_f^+}+\delta\!\left\langle {\mathbf S}_{01} \right\rangle_{Y_f^+}\, ,
\end{equation}
where $\delta\!\left\langle {\mathbf S}_{01} \right\rangle_{Y_f^+}$ is a counter-term allowing to avoid double counting of LL's. When using the standard version of the dipole B-JIMWLK evolution equation \eqref{B_JIMWLK_dipole} at LL with the factorization scheme in $k^+$, the counter-term writes
\begin{equation}
\delta\!\left\langle {\mathbf S}_{01} \right\rangle_{Y_f^+}\bigg|_{\textrm{LL}}= -\, \bar{\alpha} \int_{0}^{Y_f^+} \!\!\!\!\dd Y_2^+
\int \frac{\textrm{d}^2\mathbf{x}_{2}}{2\pi}\; \textbf{K}_{012}\: \left\langle{\mathbf S}_{02} {\mathbf S}_{21} \!-\! {\mathbf S}_{01} \right\rangle_{Y_2^+}\, .
\end{equation}
That counter-term is supposed to remove the LL contributions from the fixed-order NLO corrections. It can be split uniquely into a term associated with real corrections (or $q\bar{q}g$ Fock states) and one associated with virtual corrections (or $q\bar{q}$ Fock states) as
\begin{equation}
\delta\!\left\langle {\mathbf S}_{01} \right\rangle_{Y_f^+}\bigg|_{\textrm{LL}}=\delta\!\left\langle {\mathbf S}_{01} \right\rangle_{Y_f^+}\bigg|_{\textrm{LL, real}}+\delta\!\left\langle {\mathbf S}_{01} \right\rangle_{Y_f^+}\bigg|_{\textrm{LL, virt}}\, ,
\end{equation}
with
\begin{eqnarray}
\delta\!\left\langle {\mathbf S}_{01} \right\rangle_{Y_f^+}\bigg|_{\textrm{LL, real}}&=&
\bar{\alpha} \int_{0}^{Y_f^+} \!\!\!\!\dd Y_2^+
\int \frac{\textrm{d}^2\mathbf{x}_{2}}{2\pi}\; \textbf{K}_{012}\: \frac{2\, C_F}{N_c}\: \Big[1- \left\langle {\mathbf S}_{012} \right\rangle_{Y_2^+}\Big]\label{LL_ct_real}\\
\delta\!\left\langle {\mathbf S}_{01} \right\rangle_{Y_f^+}\bigg|_{\textrm{LL, virt}}&=&
-\, \bar{\alpha} \int_{0}^{Y_f^+} \!\!\!\!\dd Y_2^+
\int \frac{\textrm{d}^2\mathbf{x}_{2}}{2\pi}\; \textbf{K}_{012}\: \frac{2\, C_F}{N_c}\: \Big[1- \left\langle {\mathbf S}_{01} \right\rangle_{Y_2^+}\Big]\label{LL_ct_virt}\, .
\end{eqnarray}

At this stage, the photon-target cross-section \eqref{sigma_TL_z_min} becomes, at NLO+LL accuracy,
\begin{eqnarray}
& &\!\!\!\!\!\!\!\!\!\!\!\!\!\!
\sigma_{T,L}^{\gamma}(Q^2,x_{Bj})= \frac{4\, N_c\, \alpha_{em}}{(2\pi)^2}\sum_f e_f^2   \int \textrm{d}^2\mathbf{x}_{0} \int \textrm{d}^2\mathbf{x}_{1} \int_0^1 \textrm{d} z_1\, \Bigg\{ \mathcal{I}_{T,L}^{LO}({x}_{01},z_1,Q^2)
\bigg[1- \left\langle {\mathbf S}_{01} \right\rangle_{Y_f^+}\bigg|_{\textrm{LL}}\:\bigg]\nonumber\\
& &\qquad\qquad\qquad +{\cal O}(\abar)
\Big[1- \left\langle {\mathbf S}_{01} \right\rangle_{0}\Big]-\mathcal{I}_{T,L}^{LO}({x}_{01},z_1,Q^2) \:\: \delta\!\left\langle {\mathbf S}_{01} \right\rangle_{Y_f^+}\bigg|_{\textrm{LL, virt}} \nonumber\\
& &\!\!\!\!\!\!\!\!\!\!\!\!\!\!
 + \bar{\alpha} \int_{k^+_{\min}/q^+}^{1\!-\!z_1}\frac{\textrm{d}z_2}{z_2}\; \int \frac{\textrm{d}^2\mathbf{x}_{2}}{2\pi}\; \mathcal{I}_{T,L}^{NLO}(\mathbf{x}_{0},\mathbf{x}_{1},\mathbf{x}_{2},z_1,z_2,Q^2)\:
\frac{2\, C_F}{N_c}\: \Big[1- \left\langle {\mathbf S}_{012} \right\rangle_{0}\Big]-\mathcal{I}_{T,L}^{LO}({x}_{01},z_1,Q^2) \:\: \delta\!\left\langle {\mathbf S}_{01} \right\rangle_{Y_f^+}\bigg|_{\textrm{LL, real}}
\Bigg\} .\label{sigma_TL_NLO+LL}
\end{eqnarray}

High-energy LL's should cancel independently between the virtual terms in the second line of \eqref{sigma_TL_NLO+LL} and between real terms in the third line. In the expression \eqref{sigma_TL_NLO+LL}, the operator expectation values $\left\langle {\mathbf S}_{01} \right\rangle$ and $\left\langle {\mathbf S}_{012} \right\rangle$ are not evaluated at the same $Y^+$ in the fixed order results and in the counter-terms \eqref{LL_ct_real} and \eqref{LL_ct_virt}. However, that mismatch is an effect of order NNLO (or NLL) in the photon-target cross-section, beyond the accuracy of the present results and irrelevant when discussing the cancelation of high-energy LL's in the second and third lines of \eqref{sigma_TL_NLO+LL}.

The NLO DIS impact factors \eqref{ImpFact_NLO_L} and \eqref{ImpFact_NLO_T} satisfy the factorization property
\begin{equation}
\mathcal{I}_{T,L}^{NLO}(\mathbf{x}_{0},\mathbf{x}_{1},\mathbf{x}_{2},z_1,z_2=0, Q^2) = \textbf{K}_{012}\;\;\; \mathcal{I}_{T,L}^{LO}({x}_{01},z_1,Q^2)\label{Fact_IF_NLO_zero_z2}
\end{equation}
in the case of infinitely soft gluon.
Hence, one can rewrite the last term in the equation \eqref{sigma_TL_NLO+LL} as
\begin{eqnarray}
\mathcal{I}_{T,L}^{LO}({x}_{01},z_1,Q^2) \:\: \delta\!\left\langle {\mathbf S}_{01} \right\rangle_{Y_f^+}\bigg|_{\textrm{LL, real}}&=&\bar{\alpha} \int_{k^+_{\min}/q^+}^{z_f}\frac{\textrm{d}z_2}{z_2}\; \int \frac{\textrm{d}^2\mathbf{x}_{2}}{2\pi}\; \mathcal{I}_{T,L}^{NLO}(\mathbf{x}_{0},\mathbf{x}_{1},\mathbf{x}_{2},z_1,z_2=0,Q^2)\nonumber\\
& &\qquad\qquad\qquad\qquad\qquad\qquad \times
\frac{2\, C_F}{N_c}\: \Big[1- \left\langle {\mathbf S}_{012} \right\rangle_{Y_2^+}\Big]
\, ,\label{LL_ct_real_NLO-like}
\end{eqnarray}
making the comparison with the original real NLO contribution easier. A reasonable choice of the factorization scale $k_f^+=z_f\, q^+$ should be such that $z_{\min}\ll z_f\lesssim (1\!-\! z_1)$, so that $\log ((1\!-\! z_1)/z_f)$ is not large. Therefore, in the first term in the third line of \eqref{sigma_TL_NLO+LL}, only the interval $[z_{\min},z_f]$ can produce high-energy LL's in the $z_2$ integral. Then, the counter-term can be taken into account via a "$+$" prescription. For example, if one approximates $\left\langle {\mathbf S}_{012} \right\rangle_{Y_2^+}$ by $\left\langle {\mathbf S}_{012} \right\rangle_{0}$ (up to higher order corrections) in the counter-term \eqref{LL_ct_real_NLO-like}, the third line of \eqref{sigma_TL_NLO+LL} can be written as
\begin{eqnarray}
& &\bar{\alpha} \int_{z_f}^{1\!-\!z_1}\frac{\textrm{d}z_2}{z_2}\; \int \frac{\textrm{d}^2\mathbf{x}_{2}}{2\pi}\; \mathcal{I}_{T,L}^{NLO}(\mathbf{x}_{0},\mathbf{x}_{1},\mathbf{x}_{2},z_1,z_2,Q^2)\:
\frac{2\, C_F}{N_c}\: \Big[1- \left\langle {\mathbf S}_{012} \right\rangle_{0}\Big]\nonumber\\
&&+ \bar{\alpha}\int \frac{\textrm{d}^2\mathbf{x}_{2}}{2\pi}\; \frac{2\, C_F}{N_c} \Big[1- \left\langle {\mathbf S}_{012} \right\rangle_{0}\Big]\:  \int_{0}^{z_f}\frac{\textrm{d}z_2}{(z_2)_{+}}\; \mathcal{I}_{T,L}^{NLO}(\mathbf{x}_{0},\mathbf{x}_{1},\mathbf{x}_{2},z_1,z_2,Q^2)\:\, ,\label{NLO_DIS_with_LL_subtr}
\end{eqnarray}
with, by definition,
\begin{equation}
\frac{1}{(z)_{+}}\: f(z)= \frac{1}{z}\: \Big(f(z)-f(0)\Big)\, .
\label{plus_prescription}
\end{equation}
Indeed, the "$+$" prescription removes the log divergence from the $z_2$ integral, allowing to drop the lower cut-off $z_{\min}=k^+_{\min}/q^+$.

This treatment of the LL resummation and of the soft divergence of the NLO DIS impact factor is equivalent to the methods used in refs. \cite{Balitsky:2010ze,Beuf:2011xd}, and also in ref. \cite{Chirilli:2012jd} in the case of forward single-inclusive hadron production in pA collisions. In all those references, the absence of soft log divergence was implicitly taken as an evidence for the absence of high-energy LL's in the final expression \eqref{NLO_DIS_with_LL_subtr} of the real NLO correction after subtraction of the counter-term, implying that the LL's have been properly resumed into the $\left\langle {\mathbf S}_{01} \right\rangle_{Y_f^+}$ evolved with the standard dipole B-JIMWLK equation \eqref{B_JIMWLK_dipole}. However, as will be discussed in the next section \ref{sec:Issues_LL_subtr}, that assumption is not correct.

The scale $Y^+$ at which the operator expectation values $\left\langle {\mathbf S}_{01} \right\rangle$ and $\left\langle {\mathbf S}_{012} \right\rangle$ in the NLO corrections and in the counter-term should be evaluated is not under control at this order in perturbation theory. However, a change in that scale is expected to have a sizable effect in practice, so that this issue deserves further discussion.
It is natural to expect that some of the NLL terms contained in NNLO corrections to $\sigma_{T,L}^{\gamma}$ will contribute to the LL evolution of the operator present in the real NLO correction up to some scale $Y^+>0$, in the same way as the LL contributions contained in NLO corrections have lead to evolve $\left\langle {\mathbf S}_{01} \right\rangle$ from $Y^+=0$ up to $Y^+=Y_f^+$. Such an effect might also modify the scale $Y^+$ relevant for the counterterms and for the virtual NLO correction. Ultimately, the scale $Y^+$ at which the operator expectation values $\left\langle {\mathbf S}_{01} \right\rangle$ or $\left\langle {\mathbf S}_{012} \right\rangle$ has to be evaluated
should be the same in each NLO correction as in the corresponding counterterm, in order to insure the cancelation of the $z_2\rightarrow 0$ divergence to all orders. The two most natural values one can expect for that scale are:
 \begin{itemize}
   \item the factorization scale $Y_{f}^+$
   \item the scale $Y_2^+=\log(z_2\; q^+/k^+_{\min})=\log(z_2/z_{\min})$, as suggested by the expressions  \eqref{LL_ct_real} and \eqref{LL_ct_virt} of the counter-term.
 \end{itemize}
It will be shown in the section \ref{sec:Mellin_NLO_IF} and in the appendix \ref{App:Yfplus} that $Y_2^+$ is the only one of those possibilities which allows for a smooth transition between the high-energy regime and the collinear regime.



\subsubsection{Issues with the standard resummation of high-energy LL's\label{sec:Issues_LL_subtr}}

The standard resummation of low-x LL's from the NLO to the LO term explained in the previous section is motivated by the assumption that at sufficiently small $z_2$, $\mathcal{I}_{T,L}^{NLO}$ is well approximated by its value at $z_2=0$, which takes the factorized form \eqref{Fact_IF_NLO_zero_z2}. That assumption is indeed true in most of the phase-space, but not in all of it. For any given small but finite $z_2$, that approximation of $\mathcal{I}_{T,L}^{NLO}$ breaks down when $\mathbf{x}_{2}$ is far enough from $\mathbf{x}_{0}$ and $\mathbf{x}_{1}$ in the transverse plane. In that case, one can have in particular $X_3\gg X_2$, so that the exact $\mathcal{I}_{T,L}^{NLO}$ is exponentially smaller than its factorized approximation \eqref{Fact_IF_NLO_zero_z2}, due to the behavior of the Bessel function factor. In that regime, the counter-term \eqref{LL_ct_real_NLO-like} is much larger in absolute value than the real NLO term before subtraction, \emph{i.e.} the first term in the third line of the equation \eqref{sigma_TL_NLO+LL}, so that one is doing an over-subtraction of leading logs.

Moreover, the exponential suppression of the unsubtracted real NLO term in that kinematical regime is precisely the property discussed in the section \ref{sec:Form_time_mixed_space}: a $q\bar{q}g$ Fock state should not contribute to the photon-target cross section if its formation time is larger than the virtual photon lifetime. Due to its inability to reproduce that physically-motivated suppression of $q\bar{q}g$ Fock states with a gluon emitted a very large transverse distance, the counter-term \eqref{LL_ct_real} gives a sizable negative contribution to $\sigma_{T,L}^{\gamma}$ from a kinematical regime where nothing should happen.

Those serious issues with the standard resummation of low-x LL's and the associated counter-terms are the counterpart for the DIS impact factors in mixed space of the kinematical issues with the low-$x_{Bj}$ evolution equations discussed in the section \ref{sec:kin_mom_space}. Imposing the kinematical constraint (or its mixed space version) introduced in that section allows to solves simultaneously the kinematical problems encountered in the analysis
of the evolution equations and of the DIS impact factors, as will be shown in the rest of the present paper.


\subsubsection{Better soft gluon approximation to the real NLO DIS impact factors\label{sec:NLO_IF_approx}}

In order to understand more quantitatively the problems with the standard resummation of high-energy LL, one has to study more carefully the behavior of the impact factors $\mathcal{I}_{T,L}^{NLO}$ at low but finite $z_2$.

Assuming $z_2\ll z_1$ and $z_2\ll 1\!-\!z_1$ but nothing about $x_{10}$, $x_{20}$ and $x_{21}$, the longitudinal NLO impact factor \eqref{ImpFact_NLO_L} simplifies as
\begin{eqnarray}
\mathcal{I}_{L}^{NLO}(\mathbf{x}_{0},\mathbf{x}_{1},\mathbf{x}_{2},z_1,z_2,Q^2)
&\simeq &4 Q^2 \, \textrm{K}_0^2\!\left(QX_3\right) z_1^2 (1\!-\!z_1)^2  \Bigg\{\frac{1}{{x}_{20}^2}
+\frac{1}{{x}_{21}^2}    -2 \left(\frac{\mathbf{x}_{20}\cdot\mathbf{x}_{21}}{{x}_{20}^2\; {x}_{21}^2}\right)
\Bigg\} \qquad \textrm{for } z_2\ll z_1, 1\!-\!z_1\nonumber\\
&\simeq & \mathcal{I}_{L}^{LO}({x}_{01},z_1,Q^2) \quad \frac{\textrm{K}_0^2\!\left(QX_3\right)}{\textrm{K}_0^2\!\left(QX_2\right)} \quad \textbf{K}_{012}\, .
\end{eqnarray}
Hence, at low $z_2$, $\mathcal{I}_{L}^{NLO}$ is well approximated by its $z_2=0$ value if and only if $X_3\simeq X_2$, in agreement with the previous discussion.

Assuming $z_2\ll z_1$ and $z_2\ll 1\!-\!z_1$ only, the expression \eqref{X3} for $X_3^2$ simplifies a bit, as
\begin{eqnarray}
X_3^2&\simeq & z_1\, (1\!-\!z_1)\, {x}_{10}^2 + z_2\, (1\!-\!z_1)\, {x}_{20}^2 + z_2\, z_1\, {x}_{21}^2\label{X3_low_z2_bis}\, .
\end{eqnarray}
Generically, the first term tends to dominates the expression, so that $X_3$ reduces to $X_2$. However, this is not true anymore when ${x}_{20}^2$ or ${x}_{21}^2$ is so much larger than ${x}_{10}^2$ that the smallness of $z_2$ is compensated. In this regime, one has necessarily ${x}_{10}^2\ll {x}_{20}^2 \simeq {x}_{21}^2$, which allows to simplify further the expression \eqref{X3_low_z2_bis}. Hence, assuming $z_2\ll z_1$ and $z_2\ll 1\!-\!z_1$ only, the expressions
\begin{eqnarray}
X_3^2&\simeq & z_1\, (1\!-\!z_1)\, {x}_{10}^2 + z_2\, {x}_{20}^2\nonumber\\
 &\simeq & z_1\, (1\!-\!z_1)\, {x}_{10}^2+ z_2\, {x}_{21}^2\label{X3_low_z2}
\end{eqnarray}
always provide correct approximations of $X_3^2$, no matter what are the relative transverse distances ${x}_{10}$, ${x}_{20}$ and ${x}_{21}$. The approximation \eqref{X3_low_z2} is the mixed space analog of the approximation \eqref{ED2_approx_kplus_ord} of the energy denominators in momentum space.

For the impact factor $\mathcal{I}_{L}^{NLO}$, the situation is thus the following. At low $z_2$, \emph{i.e.} $z_2\ll z_1\lesssim 1$ and $z_2\ll 1\!-\!z_1\lesssim 1$, one should split the integration range for $\mathbf{x}_{2}$ into two domains:
\begin{itemize}
  \item For $z_1\, (1\!-\!z_1)\, {x}_{10}^2\gg z_2\, {x}_{20}^2$ and/or $z_1\, (1\!-\!z_1)\, {x}_{10}^2\gg z_2\, {x}_{21}^2$, $\mathcal{I}_{L}^{NLO}$ is well approximated by its factorized $z_2=0$ value \eqref{Fact_IF_NLO_zero_z2}
  \item For $z_1\, (1\!-\!z_1)\, {x}_{10}^2\lesssim z_2\, {x}_{20}^2 \simeq z_2\, {x}_{21}^2$, the expression \eqref{Fact_IF_NLO_zero_z2} is a bad approximation of $\mathcal{I}_{L}^{NLO}$, so that the standard resummation of high-energy LL is not correct in this regime.
\end{itemize}

The analysis of the transverse impact factor $\mathcal{I}_{T}^{NLO}$ is more cumbersome not only due to its more complicated expression \eqref{ImpFact_NLO_T}, but also because transverse recoil effects start to matter. Assuming $z_2\ll z_1$ and $z_2\ll 1\!-\!z_1$ only, $\mathcal{I}_{T}^{NLO}$ reduces to
\begin{eqnarray}
& &\mathcal{I}_{T}^{NLO}(\mathbf{x}_{0},\mathbf{x}_{1},\mathbf{x}_{2},z_1,z_2,Q^2)\simeq \frac{Q^2 \, \textrm{K}_1^2\!\left(Q X_3\right)}{X_3^2}
\Bigg\{z_1^2 (1\!-\!z_1)^2 \big[z_1^2+(1\!-\!z_1)^2\big]\nonumber\\
& &\times\bigg[  \left(\mathbf{x}_{10}\!-\!\frac{z_2}{1\!-\!z_1} \mathbf{x}_{20}\right)^2  \frac{1}{{x}_{20}^2}+ \left(\mathbf{x}_{10}\!+\!\frac{z_2}{z_1} \mathbf{x}_{21}\right)^2  \frac{1}{{x}_{21}^2}-2 \left(\mathbf{x}_{10}\!-\!\frac{z_2}{1\!-\!z_1} \mathbf{x}_{20}\right)\!\!\cdot\!\! \left(\mathbf{x}_{10}\!+\!\frac{z_2}{z_1} \mathbf{x}_{21}\right)
\left(\frac{\mathbf{x}_{20}\cdot\mathbf{x}_{21}}{{x}_{20}^2\; {x}_{21}^2}\right)\bigg]\nonumber\\
& & + z_2\, z_1\, (1\!-\!z_1) \big[z_1^2+(1\!-\!z_1)^2\big] \bigg[ \left(\mathbf{x}_{10}\!-\!\frac{z_2}{1\!-\!z_1} \mathbf{x}_{20}\right)\!\!\cdot\!\! \left(\frac{\mathbf{x}_{20}}{{x}_{20}^2}\right)
 -\left(\mathbf{x}_{10}\!+\!\frac{z_2}{z_1} \mathbf{x}_{21}\right)\!\!\cdot\!\! \left(\frac{\mathbf{x}_{21}}{{x}_{21}^2}\right)\bigg]\nonumber\\
& &+ \frac{z_2^2}{2}\big[z_1^2+(1\!-\!z_1)^2\big]+z_2^2\, (1\!-\!2 z_1)^2\; \frac{\big(\mathbf{x}_{20}\wedge\mathbf{x}_{21}\big)^2}{{x}_{20}^2\; {x}_{21}^2}
\Bigg\}
\, .\label{ImpFact_NLO_T_low_z_2_generic}
\end{eqnarray}
As discussed in the section II.C.1 of Ref.\cite{Beuf:2011xd}, $x_{10}$ is the transverse distance between the quark and the anti-quark at the time $x^+=0$ when the $q\bar{q}g$ state crosses the target. However, $x_{10}$ does not necessarily reflect the size of the parent dipole before emission of the gluon, due to transverse recoil effects. Taking those recoil effects, the relevant size of the parent dipole is either
\begin{equation}
\left|\mathbf{x}_{10}\!-\!\frac{z_2}{1\!-\!z_1} \mathbf{x}_{20}\right| \quad \textrm{or} \quad \left|\mathbf{x}_{10}\!+\!\frac{z_2}{z_1\!+\!z_2} \mathbf{x}_{21}\right|\simeq \left|\mathbf{x}_{10}\!+\!\frac{z_2}{z_1} \mathbf{x}_{21}\right|\, ,
\end{equation}
depending if the gluon is emitted from the quark or the anti-quark.

Let us first consider the regime
\begin{equation}
{x}_{10}\gg\frac{z_2}{1\!-\!z_1} {x}_{20} \quad \textrm{and} \quad  {x}_{10}\gg\frac{z_2}{z_1} {x}_{21}\, ,\label{no_recoil_regime}
\end{equation}
while still assuming $z_2\ll z_1$ and $z_2\ll 1\!-\!z_1$. Then, the transverse recoil effects are negligible, and the contributions in the third and fourth line of the expression \eqref{ImpFact_NLO_T_low_z_2_generic} as well.
Note that the first term in the fourth line and the ones in the third line are due to instantaneous interactions in light-front perturbation theory \cite{Beuf:2011xd}. The transverse impact factor $\mathcal{I}_{T}^{NLO}$ thus reduces in that regime to
\begin{eqnarray}
\mathcal{I}_{T}^{NLO}(\mathbf{x}_{0},\mathbf{x}_{1},\mathbf{x}_{2},z_1,z_2,Q^2)&\simeq & \frac{Q^2\, \textrm{K}_1^2\!\left(Q X_3\right)}{X_3^2}
\Bigg\{z_1^2 (1\!-\!z_1)^2 \big[z_1^2+(1\!-\!z_1)^2\big] x_{10}^2 \bigg[\frac{1}{{x}_{20}^2}+\frac{1}{{x}_{21}^2}-2
\left(\frac{\mathbf{x}_{20}\cdot\mathbf{x}_{21}}{{x}_{20}^2\; {x}_{21}^2}\right)\bigg]
\Bigg\}\nonumber\\
&\simeq & \mathcal{I}_{T}^{LO}({x}_{01},z_1,Q^2) \quad \frac{X_2^2}{X_3^2}\; \frac{\textrm{K}_1^2\!\left(QX_3\right)}{\textrm{K}_1^2\!\left(QX_2\right)} \quad \textbf{K}_{012}
\, .\label{ImpFact_NLO_T_low_z_2_recoil-less}
\end{eqnarray}
It is important to notice that the inequalities \eqref{no_recoil_regime}, at low $z_2$, do not provide any information about the relative size of $X_3$ and $X_2$. Both $X_3\simeq X_2$ and $X_3\gg X_2$ are still possible, so that the situation is analogous to the $\mathcal{I}_{L}^{NLO}$ case.

Hence, for $z_2\ll z_1\lesssim 1$ and $z_2\ll 1\!-\!z_1\lesssim 1$, in the case of $\mathcal{I}_{T}^{NLO}$, one should split the integration range for $\mathbf{x}_{2}$ in three domains:
\begin{itemize}
  \item For $z_1\, (1\!-\!z_1)\, {x}_{10}^2\gg z_2\, {x}_{20}^2$ and/or $z_1\, (1\!-\!z_1)\, {x}_{10}^2\gg z_2\, {x}_{21}^2$, $\mathcal{I}_{T}^{NLO}$ is well approximated by its factorized $z_2=0$ value \eqref{Fact_IF_NLO_zero_z2}
  \item For\footnote{The factors $z_1$ or $1\!-\!z_1$, typically not too small, have been dropped for simplicity.} ${x}_{10}^2/z_2 \lesssim {x}_{20}^2 \simeq {x}_{21}^2 \ll {x}_{10}^2/z_2^2$, the expression \eqref{Fact_IF_NLO_zero_z2} is a bad approximation of $\mathcal{I}_{T}^{NLO}$ and should be replaced by \eqref{ImpFact_NLO_T_low_z_2_recoil-less}
  \item For  $ {x}_{10}^2/z_2^2\lesssim {x}_{20}^2 \simeq {x}_{21}^2$, the recoil effects become important, and the instantaneous interaction contributions are of the same order as the other ones.
      Deep into that regime, for $ {x}_{10}^2/z_2^2\ll {x}_{20}^2 \simeq {x}_{21}^2$, a correct approximation of $\mathcal{I}_{T}^{NLO}$ is
\end{itemize}
\begin{eqnarray}
\mathcal{I}_{T}^{NLO}(\mathbf{x}_{0},\mathbf{x}_{1},\mathbf{x}_{2},z_1,z_2,Q^2)&\simeq & \mathcal{I}_{T}^{LO}({x}_{01},z_1,Q^2) \quad \frac{\textrm{K}_1^2\!\left(Q \sqrt{z_2\, x_{20}^2}\right)}{\textrm{K}_1^2\!\left(QX_2\right)} \quad \frac{z_2}{2\, z_1\, (1\!-\!z_1)\, x_{20}^2}
\, .\label{ImpFact_NLO_T_low_z_2_strong_recoil}
\end{eqnarray}

The main conclusion is that for both the longitudinal and the transverse impact factors at low $z_2$, the whole part of the integration domain in $\mathbf{x}_{2}$ such that $z_1\, (1\!-\!z_1)\, {x}_{10}^2\lesssim z_2\, {x}_{20}^2 \simeq z_2\, {x}_{21}^2$ should not contribute to leading logs, because of the exponential suppression provided by the modified Bessel functions. This property is not taken into account in the standard resummation of high-energy LL's exposed in the section \ref{sec:std_subtr_LL}.


\subsection{Mellin space analysis of the real NLO impact factors and LL counter-term\label{sec:Mellin_NLO_IF}}

In order to investigate in more details the problems of the standard subtraction of high energy LL's described in the section \eqref{sec:std_subtr_LL}, it is very convenient to compare the approximate Mellin representation of the real NLO corrections to the DIS cross sections \eqref{sigma_TL_z_min} and the one of the counter-term \eqref{LL_ct_real_NLO-like}, supposed to be their LL approximation.
In the dilute regime (or BFKL approximation, or linear regime, see section \ref{sec:evolEqs}), where the Mellin representation is most useful, one has
\begin{eqnarray}
\frac{2\, C_F}{N_c}\: \Big[1- \left\langle {\mathbf S}_{012} \right\rangle_{Y^+}\Big]&=&  \left(1\!-\!\frac{1}{N_c^2}\right) -  \left\langle{\mathbf S}_{02}\,{\mathbf S}_{21}\!-\!\frac{1}{N_c^2}{\mathbf S}_{01} \right\rangle_{Y^+}
\simeq  \left\langle {\textbf N}_{02} \right\rangle_{Y^+} + \left\langle {\textbf N}_{21} \right\rangle_{Y^+} - \frac{1}{N_c^2}\, \left\langle{\textbf N}_{01} \right\rangle_{Y^+}\label{operator_NLO_real}\, .
\end{eqnarray}
As we have discussed previously, the standard resummation of LL's is correct when one daughter dipole is much smaller than the other and the parent, or when the three dipoles are of the same order, when using the factorization scheme in $k^+$. Problems might arise only if the parent dipole is much smaller than the the two daughters, \emph{i.e.} when ${x}_{10}^2\ll {x}_{20}^2 \simeq {x}_{21}^2$.

Due to color transparency, one should have
\begin{equation}
\left\langle{\textbf N}_{ij} \right\rangle_{Y^+} \propto x_{ij}^2 \label{color_transparency}
\end{equation}
up to logarithmic factors, in the limit $x_{ij}\rightarrow 0$ at fixed $Y^+$. Due to quantum evolution effects, the dipole target amplitude $\left\langle{\textbf N}_{ij} \right\rangle_{Y^+}$ typically acquires some anomalous dimension, modifying the behavior \eqref{color_transparency}. But $\left\langle{\textbf N}_{ij} \right\rangle_{Y^+}$ should still behave roughly as a positive power of $x_{ij}$ in all the linear regime. Hence, for ${x}_{10}^2\ll {x}_{20}^2 \simeq {x}_{21}^2$ one has
\begin{equation}
\left\langle {\textbf N}_{02} \right\rangle_{Y^+} \simeq \left\langle {\textbf N}_{21} \right\rangle_{Y^+} \gg \left\langle{\textbf N}_{01} \right\rangle_{Y^+}\, ,\label{coll_ord_dipole_ampl}
\end{equation}
so that the expression \eqref{operator_NLO_real} appearing in the real NLO corrections in \eqref{sigma_TL_z_min} and in the naive counter-term \eqref{LL_ct_real} reduces to
\begin{eqnarray}
\frac{2\, C_F}{N_c}\: \Big[1- \left\langle {\mathbf S}_{012} \right\rangle_{Y^+}\Big]
&\simeq & 2 \left\langle {\textbf N}_{02} \right\rangle_{Y^+}\label{operator_NLO_real_coll}\, .
\end{eqnarray}
in the part of the linear regime satisfying ${x}_{10}^2\ll {x}_{20}^2 \simeq {x}_{21}^2$, and there, the (unknown) virtual NLO corrections to the DIS cross section and the corresponding counter-term \eqref{LL_ct_virt} are power suppressed compared to the real corrections, so that we can ignore them.


As mentioned in the end of the section \ref{sec:std_subtr_LL}, $Y_2^+$ and $Y_f^+$ are the two natural guesses for the scale 
at which one should take the expectation value of the operator $\left\langle {\mathbf S}_{012} \right\rangle$ in the counter-term \eqref{LL_ct_real} and/or in the real NLO correction. The case of $Y_2^+$ is considered here, whereas the case $Y_f^+$ is treated in the appendix \ref{App:Yfplus}. It is shown in that appendix that in the $Y_f^+$ prescription, the Regge limit and the collinear limit do not commute, making the collinear DLL regime ambiguous and quite pathological. Hence, the $Y_2^+$ prescription is more appropriate than the $Y_f^+$ prescription for the expectation value of the operator $\left\langle {\mathbf S}_{012} \right\rangle$ in the real NLO correction and in the counter-term \eqref{LL_ct_real}.

With the $Y_2^+$ prescription, the contribution to the counter-term \eqref{LL_ct_real} from the domain ${x}_{10}^2\ll {x}_{20}^2 \simeq {x}_{21}^2$ writes in the dilute regime
\begin{eqnarray}
\delta\!\left\langle {\mathbf S}_{01} \right\rangle_{Y_f^+}\bigg|_{\textrm{LL, real}}^{{x}_{10}^2\ll {x}_{20}^2 \simeq {x}_{21}^2}
&=&
\bar{\alpha} \int_{0}^{Y_f^+} \!\!\!\!\dd Y_2^+
\int_{{x}_{10}^2\ll {x}_{20}^2} \frac{\textrm{d}^2\mathbf{x}_{2}}{2\pi}\; \textbf{K}_{012}\: \frac{2\, C_F}{N_c}\: \Big[1- \left\langle {\mathbf S}_{012} \right\rangle_{Y_2^+}\Big]
\nonumber\\
&\simeq & \bar{\alpha} \int_{0}^{Y_f^+}\!\!\textrm{d}Y_2^+\; \int_{{x}_{01}^2}^{+\infty} \frac{\textrm{d}({x}_{02}^2)}{2}\;\; \frac{{x}_{01}^2}{{x}_{02}^4}\; \; 2 \left\langle {\textbf N}_{02} \right\rangle_{Y_2^+}\nonumber\\
& \simeq & \bar{\alpha} \int_{1/2-i\infty}^{1/2+i\infty} \frac{\dd \g}{2\pi i}\, \left(\frac{x_{01}^2 \, Q_0^2}{4}\right)^\g\;  \int_{0}^{Y_f^+}\!\!\textrm{d}Y_2^+\;{\cal N}(\g,Y_2^+)\;\;
\int_{{x}_{01}^2}^{+\infty} \frac{\textrm{d}({x}_{02}^2)}{{x}_{01}^2}\;\; \left(\frac{{x}_{02}^2}{{x}_{01}^2}\right)^{\g-2}\nonumber\\
& \simeq & \int_{1/2-i\infty}^{1/2+i\infty} \frac{\dd \g}{2\pi i}\, \left(\frac{x_{01}^2 \, Q_0^2}{4}\right)^\g\;  \int_{\om_0-i\infty}^{\om_0+i\infty} \frac{\dd \om}{2\pi i}\;
e^{\om\, Y_f^+}\; \hat{\cal N}(\g,\om) \;\;  \frac{\bar{\alpha}}{\om (1\!-\! \g)}\, ,
\label{Mellin_naive_subtract}
\end{eqnarray}
where the relation \eqref{Laplace_int} has been used in the last step of the calculation. Large logs manifest themselves in the Mellin representation as poles, and more precisely we have here the correspondence
\begin{equation}
\frac{\bar{\alpha}}{\om (1\!-\! \g)} \quad \leftrightarrow \quad \bar{\alpha}\, Y_f^+\, \log\left(\frac{4}{x_{01}^2 \, Q_0^2}\right)\, ,
\end{equation}
following the discussion in the section \ref{sec:Mellin_BFKL_BK_LL_NLL}. That DLL contribution is not the correct collinear DLL contribution compatible with DGLAP physics, which would contain $Y_f^-$ instead of $Y_f^+$.

When calculating the analogous approximate Mellin representation of the real NLO correction, it is convenient to change the upper bound of the $z_2$ integration from $1\!-\! z_1$ to $z_1(1\!-\! z_1)$ and make the choice of factorization scale $z_f\equiv z_1(1\!-\! z_1)$. All of this does not affect the pattern of singularities in the Mellin representation, or equivalently the presence of large logs. The integrand is taken in the dilute approximation and assuming $z_2 \ll z_f$. In order to facilitate the comparison with the expression \eqref{Mellin_naive_subtract}, one divides the real NLO correction by the LO impact factor $\mathcal{I}_{T,L}^{LO}$, and only the potentially problematic domain ${x}_{10}^2\ll {x}_{20}^2 \simeq {x}_{21}^2$ is considered.

Calculating first the contribution from the region $z_f\, {x}_{10}^2\gg z_2\, {x}_{20}^2 \simeq z_2\, {x}_{21}^2$, in which the factorized approximation \eqref{Fact_IF_NLO_zero_z2} is valid, one finds
\begin{eqnarray}
& &\left.\bar{\alpha} \int_{z_{\min}}^{z_f}\frac{\textrm{d}z_2}{z_2}\; \int \frac{\textrm{d}^2\mathbf{x}_{2}}{2\pi}\;
\frac{\mathcal{I}_{T,L}^{NLO}}{\mathcal{I}_{T,L}^{LO}} \;\frac{2\, C_F}{N_c}\: \Big[1- \left\langle {\mathbf S}_{012} \right\rangle_{Y^+_2}\Big]\right|_{{x}_{10}^2\ll {x}_{20}^2 \simeq {x}_{21}^2 \textrm{ and }z_f\, {x}_{10}^2\gg z_2\, {x}_{20}^2}\nonumber\\
& &\quad \simeq  \bar{\alpha} \int_{0}^{Y_f^+}\!\!\textrm{d}Y_2^+\; \int_{{x}_{01}^2}^{{x}_{01}^2 \exp(Y^+_f\!-\!Y^+_2)} \frac{\textrm{d}({x}_{02}^2)}{2}\;\; \frac{{x}_{01}^2}{{x}_{02}^4}\; \; 2 \left\langle {\textbf N}_{02} \right\rangle_{Y_2^+}\nonumber\\
& & \quad \simeq \int_{1/2-i\infty}^{1/2+i\infty} \frac{\dd \g}{2\pi i}\, \left(\frac{x_{01}^2 \, Q_0^2}{4}\right)^\g\;    \frac{\bar{\alpha}}{(1\!-\! \g)}  \int_{0}^{Y_f^+}\!\!\textrm{d}Y_2^+\;
\left(1\!-\!e^{-(1\!-\!\g)(Y^+_f\!-\!Y^+_2)} \right)\;{\cal N}(\g,Y_2^+)\nonumber\\
& &\quad \simeq  \int_{1/2-i\infty}^{1/2+i\infty} \frac{\dd \g}{2\pi i}\, \left(\frac{x_{01}^2 \, Q_0^2}{4}\right)^\g\;  \int_{\om_0-i\infty}^{\om_0+i\infty} \frac{\dd \om}{2\pi i}\;
e^{\om\, Y_f^+}\; \hat{\cal N}(\g,\om) \;\;  \frac{\bar{\alpha}}{\om (1\!-\! \g\!+\!\om)}\, .
\label{Mellin_NLO real_kc_reg}
\end{eqnarray}
In the last step of the calculation \eqref{Mellin_NLO real_kc_reg}, one writes the inverse Laplace transform of the Laplace transform
\begin{eqnarray}
\int_{0}^{+\infty}\!\!\textrm{d}Y_f^+\; e^{-\om\, Y_f^+}
\int_{0}^{Y_f^+}\!\!\textrm{d}Y_2^+\;
\left(1\!-\!e^{-(1\!-\!\g)(Y^+_f\!-\!Y^+_2)} \right)\;{\cal N}(\g,Y_2^+)
&=& \left[\frac{1}{\om}\!-\!\frac{1}{1\!-\!\g\!+\!\om}\right]\; \hat{\cal N}(\g,\om)\nonumber\\
&=& \frac{(1\!-\!\g)}{\om(1\!-\!\g\!+\!\om)}\; \hat{\cal N}(\g,\om)\, ,
\end{eqnarray}
calculated by interchanging the order of the integrations. The only difference between the results \eqref{Mellin_naive_subtract} and \eqref{Mellin_NLO real_kc_reg} is the shift of the pole in $\g$ from $\g=1$ to $\g=1+\om$. As shown in the section \ref{sec:coll_Mellin}, that shift is induced by the change of variables from $Y_f^-$ to $Y_f^+$, and one has the correspondence
\begin{equation}
\frac{\bar{\alpha}}{\om (1\!-\! \g\!+\!\om)} \quad \leftrightarrow \quad \bar{\alpha}\, \bigg[Y_f^+ - \log\left(\frac{4}{x_{01}^2 \, Q_0^2}\right)\bigg]\, \log\left(\frac{4}{x_{01}^2 \, Q_0^2}\right)\simeq\bar{\alpha}\, Y_f^-\, \log\left(\frac{4}{x_{01}^2 \, Q_0^2}\right)
 \, .
\end{equation}

Hence, the real NLO corrections to the DIS structure functions indeed provide the correct collinear DLL limit, by contrast to the standard counter-term \eqref{LL_ct_real}, see \eqref{Mellin_naive_subtract}.
However, only the contribution from the region $z_f\, {x}_{10}^2\gg z_2\, {x}_{20}^2 \simeq z_2\, {x}_{21}^2$ has been calculated so far. Hence, it remains to show that, in the real NLO corrections, the contributions from the region $z_f\, {x}_{10}^2\ll z_2\, {x}_{20}^2 \simeq z_2\, {x}_{21}^2$ are subleading in the collinear DLL regime. For that purpose, one should consider the transverse and longitudinal photon cases separately, and use the various approximations for $\mathcal{I}_{L}^{NLO}$ and $\mathcal{I}_{T}^{NLO}$ found in the section \ref{sec:NLO_IF_approx}.

Then, one gets in the longitudinal case
\begin{eqnarray}
& &\left.\bar{\alpha} \int_{z_{\min}}^{z_f}\frac{\textrm{d}z_2}{z_2}\; \int \frac{\textrm{d}^2\mathbf{x}_{2}}{2\pi}\;
\frac{\mathcal{I}_{L}^{NLO}}{\mathcal{I}_{L}^{LO}} \;\frac{2\, C_F}{N_c}\: \Big[1- \left\langle {\mathbf S}_{012} \right\rangle_{Y^+_2}\Big]\right|_{z_f\, {x}_{01}^2\ll z_2\, {x}_{02}^2\simeq z_2\, {x}_{21}^2 }\nonumber\\
& &\quad \simeq  \bar{\alpha} \int_{z_{\min}}^{z_f}\frac{\textrm{d}z_2}{z_2}\; \; \int^{+\infty}_{{x}_{01}^2 z_f/z_2} \frac{\textrm{d}({x}_{02}^2)}{2}\;\; \frac{{x}_{01}^2}{{x}_{02}^4}\;\; \frac{\textrm{K}_0^2\!\left(Q\sqrt{z_2\, {x}_{02}^2}\right)}{\textrm{K}_0^2\!\left(Q\sqrt{z_f\, {x}_{01}^2}\right)} \; \; 2 \left\langle {\textbf N}_{02} \right\rangle_{Y_2^+}\nonumber\\
& & \quad \simeq \bar{\alpha} \int_{1/2-i\infty}^{1/2+i\infty} \frac{\dd \g}{2\pi i}\, \left(\frac{x_{01}^2 \, Q_0^2}{4}\right)^\g\;  f_0(\g\!-\!2\, ,z_f {x}_{01}^2 Q^2)\;
\int_{0}^{Y_f^+}\!\!\textrm{d}Y_2^+\; e^{-(1\!-\!\g)(Y^+_f\!-\!Y^+_2)}\; {\cal N}(\g,Y_2^+)\nonumber\\
& &\quad \simeq  \int_{1/2-i\infty}^{1/2+i\infty} \frac{\dd \g}{2\pi i}\, \left(\frac{x_{01}^2 \, Q_0^2}{4}\right)^\g\;  \int_{\om_0-i\infty}^{\om_0+i\infty} \frac{\dd \om}{2\pi i}\;
e^{\om\, Y_f^+}\; \hat{\cal N}(\g,\om) \;\;  f_0(\g\!-\!2\, ,z_f {x}_{01}^2 Q^2)\; \; \frac{\bar{\alpha}}{(1\!-\! \g\!+\!\om)}
\, ,\label{Mellin_NLO real_L_non-kc}
\end{eqnarray}
where we have introduced the notation
\begin{eqnarray}
f_{\beta}(\xi,\tau^2) &=& \int_{1}^{+\infty} \textrm{d}u\;  u^{\xi}\;   \frac{\textrm{K}_{\beta}^2\!\left(\tau \sqrt{u}\right)}{\textrm{K}_{\beta}^2\!\left(\tau\right)}
\, .\label{f_beta}
\end{eqnarray}
Note that, for $\tau>0$, the exponential decay of the modified Bessel function $\textrm{K}_{\beta}$ implies that the integral in $u$ converges no matter what is the value of $\xi$. Therefore, $f_{\beta}(\xi,\tau^2)$ is holomorphic in $\xi$ for any $\tau>0$.
The only singularity in Mellin space in the expression \eqref{Mellin_NLO real_L_non-kc} is thus the pole at $\g=1+\om$. It corresponds to a collinear single log, and there is no collinear DLL in the contribution \eqref{Mellin_NLO real_L_non-kc}, as expected.

In the transverse photon case the intermediate region $z_f {x}_{10}^2/z_2 \ll {x}_{20}^2 \simeq {x}_{21}^2 \ll z_f^2 {x}_{10}^2/z_2^2$, where transverse recoil effects are still negligible, gives the contribution
\begin{eqnarray}
& &\left.\bar{\alpha} \int_{z_{\min}}^{z_f}\frac{\textrm{d}z_2}{z_2}\; \int \frac{\textrm{d}^2\mathbf{x}_{2}}{2\pi}\;
\frac{\mathcal{I}_{T}^{NLO}}{\mathcal{I}_{T}^{LO}} \;\frac{2\, C_F}{N_c}\: \Big[1- \left\langle {\mathbf S}_{012} \right\rangle_{Y^+_2}\Big]\right|_{z_f {x}_{10}^2/z_2 \ll {x}_{20}^2 \simeq {x}_{21}^2 \ll z_f^2 {x}_{10}^2/z_2^2}\nonumber\\
& &\quad \simeq  \bar{\alpha} \int_{z_{\min}}^{z_f}\frac{\textrm{d}z_2}{z_2}\; \; \int^{{x}_{01}^2 z_f^2/z_2^2}_{{x}_{01}^2 z_f/z_2} \frac{\textrm{d}({x}_{02}^2)}{2}\;\; \frac{{x}_{01}^2}{{x}_{02}^4}\;\; \frac{z_f\, {x}_{01}^2}{z_2\, {x}_{02}^2}\;\; \frac{\textrm{K}_1^2\!\left(Q\sqrt{z_2\, {x}_{02}^2}\right)}{\textrm{K}_1^2\!\left(Q\sqrt{z_f\, {x}_{01}^2}\right)} \; \; 2 \left\langle {\textbf N}_{02} \right\rangle_{Y_2^+}\nonumber\\
& & \quad \simeq \bar{\alpha} \int_{1/2-i\infty}^{1/2+i\infty} \frac{\dd \g}{2\pi i}\, \left(\frac{x_{01}^2 \, Q_0^2}{4}\right)^\g\; \int_{0}^{Y_f^+}\!\!\textrm{d}Y_2^+\; e^{-(1\!-\!\g)(Y^+_f\!-\!Y^+_2)}\; {\cal N}(\g,Y_2^+)
\int_{1}^{e^{(Y^+_f\!-\!Y^+_2)}} \!\!\!\!\!\!\!\! \!\!\!\!\!\!\!\!\textrm{d}u\;\;\;  u^{\g-3}\;   \frac{\textrm{K}_{1}^2\!\left(Q\sqrt{z_f\, {x}_{01}^2} \sqrt{u}\right)}{\textrm{K}_{1}^2\!\left(Q\sqrt{z_f\, {x}_{01}^2}\right)}
  \nonumber\\
& &\quad \simeq  \int_{1/2-i\infty}^{1/2+i\infty} \frac{\dd \g}{2\pi i}\, \left(\frac{x_{01}^2 \, Q_0^2}{4}\right)^\g\;  \int_{\om_0-i\infty}^{\om_0+i\infty} \frac{\dd \om}{2\pi i}\;
e^{\om\, Y_f^+}\; \hat{\cal N}(\g,\om) \;\;  f_1(2\g\!-\!\om\!-\!4\, ,z_f {x}_{01}^2 Q^2)\; \; \frac{\bar{\alpha}}{(1\!-\! \g\!+\!\om)}
\, ,\label{Mellin_NLO real_T_non-kc_recoilless}
\end{eqnarray}
This expression has a single pole in Mellin space corresponding to a collinear single log. It differs from the result \eqref{Mellin_NLO real_L_non-kc} found in the longitudinal photon case only by the precise value of the factor in front of the single pole or single log.

Finally, in the extreme region $z_f^2 {x}_{10}^2 \ll z_2^2 {x}_{20}^2 \simeq z_2^2 {x}_{21}^2$ where transverse recoil effects are important, one finds
\begin{eqnarray}
& &\left.\bar{\alpha} \int_{z_{\min}}^{z_f}\frac{\textrm{d}z_2}{z_2}\; \int \frac{\textrm{d}^2\mathbf{x}_{2}}{2\pi}\;
\frac{\mathcal{I}_{T}^{NLO}}{\mathcal{I}_{T}^{LO}} \;\frac{2\, C_F}{N_c}\: \Big[1- \left\langle {\mathbf S}_{012} \right\rangle_{Y^+_2}\Big]\right|_{z_f^2\, {x}_{01}^2\ll z_2^2\, {x}_{02}^2\simeq z_2^2\, {x}_{21}^2}\nonumber\\
& &\quad \simeq  \bar{\alpha} \int_{z_{\min}}^{z_f}\frac{\textrm{d}z_2}{z_2}\; \; \int_{{x}_{01}^2 z_f^2/z_2^2}^{+\infty} \frac{\textrm{d}({x}_{02}^2)}{2}\;\;  \frac{z_2}{2\, z_f\, {x}_{02}^2}\;\; \frac{\textrm{K}_1^2\!\left(Q\sqrt{z_2\, {x}_{02}^2}\right)}{\textrm{K}_1^2\!\left(Q\sqrt{z_f\, {x}_{01}^2}\right)} \; \; 2 \left\langle {\textbf N}_{02} \right\rangle_{Y_2^+}\nonumber\\
& & \quad \simeq \frac{\bar{\alpha}}{2} \int_{1/2-i\infty}^{1/2+i\infty} \frac{\dd \g}{2\pi i}\, \left(\frac{x_{01}^2 \, Q_0^2}{4}\right)^\g\; \int_{0}^{Y_f^+}\!\!\textrm{d}Y_2^+\; e^{-(1\!-\!\g)(Y^+_f\!-\!Y^+_2)}\; {\cal N}(\g,Y_2^+)
\int^{+\infty}_{\exp{(Y^+_f\!-\!Y^+_2)}} \!\!\!\!\!\!\!\! \!\!\!\!\!\!\!\!\textrm{d}u\;\;\;  u^{\g-1}\;   \frac{\textrm{K}_{1}^2\!\left(Q\sqrt{z_f\, {x}_{01}^2} \sqrt{u}\right)}{\textrm{K}_{1}^2\!\left(Q\sqrt{z_f\, {x}_{01}^2}\right)}
  \nonumber\\
& &\quad \simeq  \int_{1/2-i\infty}^{1/2+i\infty} \frac{\dd \g}{2\pi i}\, \left(\frac{x_{01}^2 \, Q_0^2}{4}\right)^\g\;  \int_{\om_0-i\infty}^{\om_0+i\infty} \frac{\dd \om}{2\pi i}\;
e^{\om\, Y_f^+}\; \hat{\cal N}(\g,\om)\nonumber\\
& &\quad \quad\quad \; \times \; \frac{\bar{\alpha}}{2(1\!-\! \g\!+\!\om)} \;\;  \Big[f_1(\g\!-\!1\, ,z_f {x}_{01}^2 Q^2)\!-\! f_1(2\g\!-\!\om\!-\!2\, ,z_f {x}_{01}^2 Q^2)  \Big]
\, ,\label{Mellin_NLO real_T_deep_recoil}
\end{eqnarray}
The two terms cancel each other when $1\!-\! \g\!+\!\om=0$, so that the singularity at $1\!-\! \g\!+\!\om=0$ is actually removable. Hence, there is no true Mellin space singularity in the expression \eqref{Mellin_NLO real_T_deep_recoil}, and thus the region $z_f^2 {x}_{10}^2 \ll z_2^2 {x}_{20}^2 \simeq z_2^2 {x}_{21}^2$ contributes neither to the high-energy LL's nor to the collinear LL's. It may be quite counter-intuitive that this region of the $\mathbf{x}_{2}$ plane does not contribute to DGLAP collinear logs whereas the intermediate region does. But this is only due the choice of the variable $Y^+$, which is not the most appropriate when discussing the collinear limit as explained in the section \ref{sec:coll_Mellin}.
As expected, neither the intermediate region $z_f {x}_{10}^2/z_2 \ll {x}_{20}^2 \simeq {x}_{21}^2 \ll z_f^2 {x}_{10}^2/z_2^2$ nor the extreme region $z_f^2 {x}_{10}^2 \ll z_2^2 {x}_{20}^2 \simeq z_2^2 {x}_{21}^2$ contribute to the high-energy LL's in the transverse photon case.


\subsection{Summary of the analysis of the real NLO corrections to DIS}

The study of the real NLO corrections to DIS in Mellin space done in the previous section \ref{sec:Mellin_NLO_IF} confirms the hints found directly in mixed space. In particular, the emission of a gluon at parametrically large distance in the transverse plane, such that $z_f {x}_{10}^2 \ll z_2 {x}_{20}^2 \simeq z_2 {x}_{21}^2$, does not contribute to high-energy LL's. This constraint is the analog in mixed-space of the $k^-$ ordering in momentum space discussed in the section \ref{sec:kin_mom_space} and of the shift of the collinear pole in Mellin space from $\g=1$ to $\g=1+\om$ discussed in the section \ref{sec:coll_Mellin}. That kinematical constraint, necessary to reproduce the correct DLL limit compatible with the DGLAP evolution of the target, is not included in the standard version of the high-energy evolution equations, BFKL, BK or B-JIMWLK. Hence, those equations lead to slightly overestimate the high-energy LL contributions arising at higher orders in fixed order calculations, and thus do not allow to resum the LL's correctly, following the method presented in the section \ref{sec:std_subtr_LL}. After such an incorrect resummation, the leftover NLO (and higher order) corrections become large and negative when approaching the collinear regime, leading to a breakdown of the naively resummed perturbative expansion.

Partonic Fock states in the photon wave-function which have a formation time larger than the virtual photon lifetime give only exponentially suppressed contributions to the DIS cross section, within fixed order perturbative calculations.
In order to maintain that physically correct property when performing the resummation of high-energy LL's, it is necessary to use a high-energy evolution equation including the kinematical constraint.

\section{Kinematical constraint for LL evolution equations in mixed space\label{sec:kcBK}}


\subsection{Constraint for the real emission kernel}

In the standard LL approximation without kinematical constraint, the probability density for the initial-state emission of a gluon with momentum fraction $z_2$ at position $\mathbf{x}_{2}$ from a single $q\bar{q}$ dipole $(\mathbf{x}_{0},\mathbf{x}_{1})$ writes \cite{Mueller:1993rr}
\begin{equation}
\frac{\alpha_s\, C_F}{\pi^2}\, \frac{\dd z_2}{z_2}\, \dd^2 \mathbf{x}_{2}\,\frac{x_{01}^2}{x_{02}^2\, x_{21}^2}\equiv\abar\, \frac{2 C_F}{N_c}\, \frac{\dd z_2}{z_2}\, \frac{\dd^2 \mathbf{x}_{2}}{2\pi}\, \textbf{K}_{012}\, ,
 \label{Proba_dipole_split_LL_naive}
\end{equation}
where the gluon is assumed to be much softer than the quark and the anti-quark, \emph{i.e.} $z_2\ll z_1$ and $z_2\ll 1\!-\!z_1$.

As shown in the previous section, such a LL gluon emission can actually occur only in some bounded domain of the $\mathbf{x}_{2}$-plane. The probability density \eqref{Proba_dipole_split_LL_naive} should then be kinematically constrained as
\begin{equation}
\abar\, \frac{2 C_F}{N_c}\, \frac{\dd z_2}{z_2}\, \frac{\dd^2 \mathbf{x}_{2}}{2\pi}\, \textbf{K}_{012}\;
\theta\!\Big(z_1 (1\!-\!z_1)\, {x}_{01}^2 \!-\! z_2\, l_{012}^2\Big)\, .
 \label{Proba_parent_dipole_split_LL_kc}
\end{equation}
The quantity $l_{ijk}$ introduced in the equation \eqref{Proba_parent_dipole_split_LL_kc} should satisfy
\begin{equation}
l_{ijk}\simeq {x}_{ik}\simeq {x}_{jk} \quad \textrm{in the regime} \quad {x}_{ij}\ll {x}_{ik}\simeq {x}_{jk}\, ,\label{lijk_large}
\end{equation}
in order to implement the precise kinematical restriction found in the section \ref{sec:NLO_IF_analysis}. On the other hand, the theta function introduced in \eqref{Proba_parent_dipole_split_LL_kc} should not have a significant effect in the rest of the $\mathbf{x}_{2}$-plane, and thus
\begin{equation}
l_{ijk}\lesssim {x}_{ij} 
\quad \textrm{in the regimes} \quad {x}_{ik}\ll {x}_{ij} \simeq {x}_{jk}\; , \;\; {x}_{jk}\ll {x}_{ij} \simeq {x}_{ik}\;\; \textrm{and }\;  {x}_{ij}\simeq {x}_{ik}\simeq {x}_{jk}\label{lijk_short}
\, .
\end{equation}
Apart from the requirements \eqref{lijk_large} and \eqref{lijk_short}, the precise expression of $l_{ijk}$ is essentially arbitrary. Any choice leads in the end to self-consistent kinematically improved BK or BFKL equations. This should be understood as a resummation scheme ambiguity associated with the kinematical constraint.

Most of the calculations in the rest of this paper will be done for arbitrary $l_{ijk}$ obeying the conditions \eqref{lijk_large} and \eqref{lijk_short}. However, for practical applications, one can use the explicit expression
\begin{equation}
l_{ijk}= \min ({x}_{ik}, {x}_{jk})\, ,\label{lijk_min_explicit}
\end{equation}
which is symmetric in the parent dipole legs $i$ and $j$, and minimizes the impact of the theta function when $z_2$ is not too small.

Having the probability density \eqref{Proba_parent_dipole_split_LL_kc} for the emission of a gluon from a single dipole, the next step is to study how multiple gluon emissions iterate in the case of a full initial-state parton (or dipole) cascade.

Most of QCD evolution equations describing parton cascades, like the DGLAP, BFKL or BK equations are local: after several steps, further emission in one branch of the parton cascade is independent of what happens in other branches. Hence, the information about subsequent evolution in other branches can be thrown away, which allows to write those evolution equations in closed form as simple integro-differential equations. The main counter-example is the B-JIMWLK evolution, in which the information about the full cascade has to be kept. This is the reason why B-JIMWLK can be written as a functional equation or as an infinite hierarchy of equations, but not as a closed integro-differential equation.

It has been shown in the section \ref{sec:kin_mom_space} that, in Light-Front perturbation theory, the kinematical constraint is obtained from a careful study of energy denominators, leading to a simultaneous $k^+$ and $k^-$ ordering of successive emissions. However, each energy denominator involve the momenta of all the partons present in the current intermediate Fock state. Hence, it seems that the $k^+$ and $k^-$ orderings are global, \emph{i.e.} with the momentum of a new radiated gluon restricted by the momenta of all of the partons already radiated, including in other branches of the cascade, preventing one to include the kinematical constraint as a simple modification of the BFKL and BK integro-differential equations. This issue has been noticed in Ref.~\cite{Motyka:2009gi}, where the authors have then considered for simplicity a local $k^+$ and $k^-$ ordering - with respect to the legs of the emitting dipole only - with the hope that the mismatch between local and global orderings would not be essential.
However, a more thorough study of those issues, presented in the appendix \ref{App:locality_kc}, shows that Light-Front perturbation theory actually lead to a local instead of a global $k^+$ and $k^-$ ordering. More precisely, there is a global constraint at the level of individual graphs, which however becomes a local one when summing over graphs differing just by the order of gluon emissions by different color dipoles\footnote{I thank Heribert Weigert for an enlightening discussion on a closely related problem.}, up to corrections of NLL order.

Hence, in the case of the emission of a gluon $k$ by a generic dipole $ij$ within a full parton cascade, with the parton $j$ radiated after the parton $i$, one should have the orderings
\begin{eqnarray}
&& k^+_{i}\gg k^+_{j}\gg k^+_{k}\\
&& k^-_{i}\ll k^-_{j}\ll k^-_{k}
\end{eqnarray}
in full momentum space. Then, one can write the probability density for that gluon emission in mixed-space as
\begin{equation}
\abar\, \frac{2 C_F}{N_c}\, \frac{\dd z_k}{z_k}\, \frac{\dd^2 \mathbf{x}_{k}}{2\pi}\, \textbf{K}_{ijk}\;
\theta\!\Big(z_j\, {x}_{ij}^2 \!-\! z_k\, l_{ijk}^2\Big)\, ,
 \label{Proba_generic_dipole_split_LL_kc}
\end{equation}
where the theta function effectively enforces the condition $k^-_{j}< k^-_{k}$.

At this stage, it is clear that one can write an evolution equation in $k^+$ in order to generate the dipole cascade at LL accuracy with the kinematical constraint, involving the probability densities for gluon emission \eqref{Proba_parent_dipole_split_LL_kc} and \eqref{Proba_generic_dipole_split_LL_kc}. Let us consider a dipole ${x}_{ij}$, associated with some factorization scale $k^+_f=z_f\, q^+$. Then, this dipole can emit a gluon with momentum $k^+_{k}\ll k^+_f$ and position $\mathbf{x}_{k}$ with the probability density
\begin{equation}
\abar\, \frac{2 C_F}{N_c}\, \frac{\dd k^+_{k}}{k^+_{k}}\, \frac{\dd^2 \mathbf{x}_{k}}{2\pi}\, \textbf{K}_{ijk}\;
\theta\!\Big(k^+_f\, {x}_{ij}^2 \!-\! k^+_{k}\, l_{ijk}^2\Big)\, ,
 \label{Proba_dipole_split_LL_kc_kfplus}
\end{equation}
and one obtains two dipoles $(\mathbf{x}_{i},\mathbf{x}_{k})$, and $(\mathbf{x}_{k},\mathbf{x}_{j})$, both considered at the new factorization scale $k^+_{k}$. One can then iterate further this evolution for each daughter dipole. In this way, the factorization scale associated to each dipole in the cascade is the $k^+$ of its softest leg, except for the primordial dipole initiating the cascade, for which the factorization scale can be taken as $k^+_f\equiv z_1 (1\!-\!z_1) q^+$. In this way, the evolution equation in $k^+$ reproduces the expression \eqref{Proba_parent_dipole_split_LL_kc} for the first gluon emission and the expression \eqref{Proba_generic_dipole_split_LL_kc} for subsequent gluon emissions in the dipole cascade, as it should. It only remains to include virtual corrections in a consistent way in order to write down explicitly the evolution equation.



\subsection{Calculating the virtual corrections}

\subsubsection{Probability conserving evolution}

The virtual terms in the evolution equation are not directly sensitive to the kinematical issues discussed previously. Indeed, virtual corrections were irrelevant in the discussion of the collinear regime for DIS at NLO in the section \ref{sec:NLO_IF_analysis}. Because of this, there is \emph{a priori} some freedom about the treatment of the virtual terms in the resummation associated with the kinematical constraint: in the evolution equation one can move virtual terms from the higher order contributions to the resummed leading order \emph{a priori} without restriction. Of course, one has to make sure that in the strict Regge limit, the evolution equation reduces to the standard unresummed one, but this still leaves a lot of freedom.

In the previous study of the kinematical constraint in mixed space \cite{Motyka:2009gi},
the focus was clearly on the real emission kernel, and no specific expression for the virtual corrections has been proposed. Those results were used in numerical studies in ref. \cite{Berger:2010sh}, with a particular choice of the implementation of virtual terms, for which no motivation is provided.

However, there is a very natural way to pin down the virtual corrections. It consists in requiring the probabilistic interpretation of the parton cascade \cite{Mueller:1993rr} to be preserved by the kinematical resummation. Indeed, probability conservation is automatically guaranteed when using the real emission kernel \eqref{Proba_dipole_split_LL_kc_kfplus} to write a Bethe-Salpeter-like integral equation for the dipole-target S-matrix, resuming the LL contributions between the scale $k^+_f$ of the projectile dipole and the scale $k^+_{\min}$ of the target (assuming obviously $k^+_f>k^+_{\min}$),
\begin{eqnarray}
\left\langle {\mathbf S}_{01}\right\rangle_{\log (k^+_f/k^+_{\min})}&=& {\mathbf D}_{01}(k^+_f,k^+_{\min})\:  \left\langle {\mathbf S}_{01}\right\rangle_{0}\nonumber\\
& &+  \abar\, \frac{2 C_F}{N_c}\int_{k^+_{\min}}^{k^+_f}\!\! \frac{\dd k^+_2}{k^+_2}\,  {\mathbf D}_{01}(k^+_f,k^+_2)\int \frac{\dd^2 \mathbf{x}_{2}}{2\pi}\, \textbf{K}_{012}\;
\theta\!\left(k^+_f\, {x}_{01}^2 \!-\! k^+_2\, l_{012}^2\right)\,
\left\langle {\mathbf S}_{012}\right\rangle_{\log (k^+_2/k^+_{\min})}\, ,\label{BetheSalp_kplus}
\end{eqnarray}
with the $q\bar{q}g$ tripole operator ${\mathbf S}_{012}$ defined by the equation \eqref{def_tripole}.

In the equation \eqref{BetheSalp_kplus},
the factor ${\mathbf D}_{01}(k^+_f,k^+)$ is the probability that the dipole $01$ doesn't split when evolved from the factorization scale $k^+_f$ down to $k^+<k^+_f$. It should obviously satisfy the initial condition
\begin{equation}
{\mathbf D}_{01}(k^+,k^+)=1\label{init_cond_D01}
\end{equation}
for any positive $k^+$. The first term in the right hand side of the equation \eqref{BetheSalp_kplus} is the contribution of the case when the parent dipole $01$ does not split when evolved from $k^+_f$ all the way down to $k^+_{\min}$, leaving no room for evolution of the target. By contrast, the second term is the contribution of the case when dipole splittings occur, and only the first splitting, on the projectile side, is described explicitly using the real emission kernel \eqref{Proba_dipole_split_LL_kc_kfplus}.

In the equation \eqref{BetheSalp_kplus}, there are UV divergences for $\mathbf{x}_{2}\rightarrow \mathbf{x}_{0}$ and $\mathbf{x}_{2}\rightarrow \mathbf{x}_{1}$. Hence, one should regularize the transverse integration in the equation \eqref{BetheSalp_kplus}, for example like in ref. \cite{Mueller:1993rr} by restricting it to the domain such that ${x}_{02}>\rho$ and ${x}_{12}>\rho$, where $\rho$ is a given short distance cut-off. However, in order to simplify notations, the regularization is kept implicit in equation \eqref{BetheSalp_kplus} and in the following.

The next step is to calculate explicitly the function ${\mathbf D}_{01}(k^+_f,k^+)$ consistent with probability conservation. The evolution equation \eqref{BetheSalp_kplus} is independent of the nature of the target, and should even be valid in the absence of any target. In that case, $\left\langle {\mathbf S}_{01}\right\rangle_{Y^+}\equiv 1$ and $\left\langle {\mathbf S}_{012}\right\rangle_{Y^+}\equiv 1$ for any $Y^+$, and thus the equation \eqref{BetheSalp_kplus} reduces to
\begin{eqnarray}
1&=& {\mathbf D}_{01}(k^+_f,k^+_{\min})+  \abar\, \frac{2 C_F}{N_c}\int_{k^+_{\min}}^{k^+_f}\!\! \frac{\dd k^+_2}{k^+_2}\,  {\mathbf D}_{01}(k^+_f,k^+_2)\int \frac{\dd^2 \mathbf{x}_{2}}{2\pi}\, \textbf{K}_{012}\;
\theta\!\left(k^+_f\, {x}_{01}^2 \!-\! k^+_2\, l_{012}^2\right)\,
\, .\label{BetheSalp_Vac}
\end{eqnarray}
Taking the derivative of that relation with respect to $k^+_{\min}$, one finds
\begin{eqnarray}
k^+_{\min}\, \d_{k^+_{\min}}\,  {\mathbf D}_{01}(k^+_f,k^+_{\min})&=&
  \abar\, \frac{2 C_F}{N_c} {\mathbf D}_{01}(k^+_f,k^+_{\min}) \int \frac{\dd^2 \mathbf{x}_{2}}{2\pi}\, \textbf{K}_{012}\;
\theta\!\left(k^+_f\, {x}_{01}^2 \!-\! k^+_{\min}\, l_{012}^2\right)\, ,
\end{eqnarray}
which is trivially solved by
\begin{eqnarray}
{\mathbf D}_{01}(k^+_f,k^+_{\min})=\exp \bigg[-\abar\, \frac{2 C_F}{N_c}  \int_{k^+_{\min}}^{k^+_f}\!\! \frac{\dd k^+}{k^+}\, \int \frac{\dd^2 \mathbf{x}_{2}}{2\pi}\, \textbf{K}_{012}\;
\theta\!\left(k^+_f\, {x}_{01}^2 \!-\! k^+\, l_{012}^2\right) \bigg]\, ,\label{D01_sol_1}
\end{eqnarray}
where the regularization of the transverse integration is again kept implicit. The integration over $k^+$ can be done explicitly, which gives
\begin{eqnarray}
{\mathbf D}_{01}(k^+_f,k^+_{\min})=\exp \left[-\abar\, \frac{2 C_F}{N_c}  \int \frac{\dd^2 \mathbf{x}_{2}}{2\pi}\, \textbf{K}_{012}\; \left(\log \left(\frac{k^+_f}{k^+_{\min}}\right)-\Delta_{012}\right)\; \; \;
\theta\!\left(\log \left(\frac{k^+_f}{k^+_{\min}}\right)-\Delta_{012}\right) \right]\, ,\label{D01_sol_2}
\end{eqnarray}
with the notation
\begin{equation}
\Delta_{012}= \max \left\{0,\, \log\left(\frac{l_{012}^2}{{x}_{01}^2}\right)  \right\}\, .\label{Delta012}
\end{equation}
The generic behavior of the shift $\Delta_{012}$ is then
\begin{eqnarray}
\Delta_{012}&=& 0 \qquad \textrm{for} \quad {x}_{02}^2\ll {x}_{01}^2 \quad \textrm{or} \quad {x}_{21}^2\ll {x}_{01}^2\nonumber\\
\Delta_{012}&\sim& \log \left(\frac{{x}_{02}^2}{{x}_{01}^2}\right) \: \sim \:\:  \log \left(\frac{{x}_{21}^2}{{x}_{01}^2}\right) \qquad \textrm{for} \quad {x}_{01}^2 \ll  {x}_{02}^2 \sim  {x}_{21}^2\, ,
\end{eqnarray}
and its precise value outside of those limits depends on the choice of $l_{012}$, \emph{i.e.} on the choice of resummation scheme.

With the result \eqref{D01_sol_2}, the integral evolution equation \eqref{BetheSalp_kplus} is fully specified, and is a kinematically improved version of the B-JIMWLK evolution equation \eqref{B_JIMWLK_dipole} for the dipole, obeying the kinematical constraint for the real emission kernel but still preserving the probabilistic interpretation of the parton cascade exactly, by construction.

Thanks to their structure, the equations \eqref{BetheSalp_kplus} and  \eqref{D01_sol_2} can be rewritten in terms of logarithmic variables $Y^+$ instead of the $k^+$'s, as
\begin{eqnarray}
\left\langle {\mathbf S}_{01}\right\rangle_{Y^+_f}&=& {\mathbf D}_{01}(Y^+_f)\:  \left\langle {\mathbf S}_{01}\right\rangle_{0}+  \abar\, \frac{2 C_F}{N_c}\int \frac{\dd^2 \mathbf{x}_{2}}{2\pi}\, \textbf{K}_{012}\;
\theta\!\left(Y^+_f\!-\!  \Delta_{012}\right)\,
 \int_{0}^{Y^+_f\!-\!  \Delta_{012}}\!\! \dd Y^+_2\,  {\mathbf D}_{01}(Y^+_f\!-\!Y^+_2)
\left\langle {\mathbf S}_{012}\right\rangle_{Y^+_2}\label{BetheSalp_Yplus}
\end{eqnarray}
and\footnote{In the equation \eqref{D01_sol_3}, the integration variable has been relabeled from $\mathbf{x}_{2}$ to $\mathbf{x}_{v}$ in order to avoid confusions in later stages of the calculation, in particular in the equation \eqref{B_JIMWLK_kc_untrunc}.}
\begin{eqnarray}
{\mathbf D}_{01}(Y^+_f)=\exp \left[-\abar\, \frac{2 C_F}{N_c}  \int \frac{\dd^2 \mathbf{x}_{v}}{2\pi}\, \textbf{K}_{01v}\; \left(Y^+_f\!-\!\Delta_{01v}\right)\; \; \;
\theta\!\left(Y^+_f\!-\!\Delta_{01v}\right) \right]\, ,\label{D01_sol_3}
\end{eqnarray}
where $Y^+_f>0$ has been assumed in both equations.

Rather than an integral equation like \eqref{BetheSalp_Yplus}, it is often more convenient to have an integro-differential equation. One obtains the latter from the equation \eqref{BetheSalp_Yplus} by dividing by ${\mathbf D}_{01}(Y^+_f)$, taking the derivative with respect to $Y^+_f$, and multiplying again by ${\mathbf D}_{01}(Y^+_f)$. In that way, one obtains
\begin{eqnarray}
\d_{Y_f^+} \left\langle {\mathbf S}_{01}\right\rangle_{Y_f^+}&=&   \abar\, \frac{2 C_F}{N_c}\int \frac{\dd^2 \mathbf{x}_{2}}{2\pi}\, \textbf{K}_{012}\;
\theta\!\left(Y_f^+\!-\! \Delta_{012}\right)\,  \Bigg\{{\mathbf D}_{01}(\Delta_{012})\; \left\langle {\mathbf S}_{012}\right\rangle_{Y_f^+\!-\!\Delta_{012}} -  \left\langle {\mathbf S}_{01}\right\rangle_{Y_f^+} \nonumber\\
& & +\int_{0}^{Y_f^+\!-\! \Delta_{012}}\!\! \dd Y_2^+\,  {\mathbf D}_{01}(Y_f^+\!-\!Y_2^+)
\left\langle {\mathbf S}_{012}\right\rangle_{Y_2^+}\;  \d_{Y_f^+} \log \left(\frac{{\mathbf D}_{01}(Y_f^+\!-\!Y_2^+)}{{\mathbf D}_{01}(Y_f^+)}\right) \Bigg\}\, .\label{B_JIMWLK_kc_blah}
\end{eqnarray}
As a cross-check, one can formally recover from that equation the strict LL equation \eqref{B_JIMWLK_dipole} by setting $\Delta_{012}$ to $0$ and also $\Delta_{01v}$ to $0$ in the expression \eqref{D01_sol_3}. Indeed, in that case, the ratio ${\mathbf D}_{01}(Y_f^+\!-\!Y_2^+)/{\mathbf D}_{01}(Y_f^+)$ becomes independent of $Y_f^+$, so that the second line of the equation \eqref{B_JIMWLK_kc_blah} does not contribute in the standard Regge limit.

Calculating explicitly the last term in the equation \eqref{B_JIMWLK_kc_blah} thanks to the expression \eqref{D01_sol_3}, one finally gets
\begin{eqnarray}
\d_{Y^+} \left\langle {\mathbf S}_{01}\right\rangle_{Y^+}&=&\abar\, \frac{2 C_F}{N_c}\int \frac{\dd^2 \mathbf{x}_{2}}{2\pi}\, \textbf{K}_{012}\; \Bigg\{
\theta\!\left(Y^+\!-\! \Delta_{012}\right)\, \bigg[{\mathbf D}_{01}(\Delta_{012})\; \left\langle {\mathbf S}_{012}\right\rangle_{Y^+\!-\!\Delta_{012}} -  \left\langle {\mathbf S}_{01}\right\rangle_{Y^+} \bigg]\nonumber\\
& & \!\!\!\!\!\!+ \abar\, \frac{2 C_F}{N_c}\int \frac{\dd^2 \mathbf{x}_{3}}{2\pi}\, \textbf{K}_{013}\; \theta\!\left(Y^+\!-\! \Delta_{013}\right)\, \theta\!\left(\Delta_{013}\!-\! \Delta_{012}\right)\,
\int_{Y^+\!-\! \Delta_{013}}^{Y^+\!-\! \Delta_{012}}\!\! \dd Y_2^+\,  {\mathbf D}_{01}(Y^+\!-\!Y_2^+)
\left\langle {\mathbf S}_{012}\right\rangle_{Y_2^+}\!\!\! \Bigg\}\, .\label{B_JIMWLK_kc_untrunc}
\end{eqnarray}

\subsubsection{Discarding explicitly NLL terms}

The aim of this study is to perform a resummation of contributions of higher logarithmic order in the strict Regge limit, in order to provide an improved version of evolution equations at LL accuracy.
In the equation \eqref{B_JIMWLK_kc_untrunc}, such a resummation of higher order contribution into a modification of LL
terms appears most notably as the shift $Y^+ \mapsto Y^+\!-\!\Delta_{012}$ in the first term. However, the equation \eqref{B_JIMWLK_kc_untrunc} also contains contributions which are explicitly of order NLL or higher, for example all the second line, which is a contribution of order ${\cal O}(\abar^2)$. Since we are not including the full NLL BFKL \cite{Fadin:1998py,Ciafaloni:1998gs} or NLL BK \cite{Balitsky:2008zz} kernel, it is not really consistent to keep those terms, which appear only because \emph{exact} probability conservation in the dipole cascade has been required. However, it makes presumably more sense to require only probability conservation up to terms of order ${\cal O}(\abar^2)$ in the equation \eqref{B_JIMWLK_kc_untrunc}, and discard all the explicitly NLL terms, \emph{i.e.} the second line and the higher order terms in the expansion of ${\mathbf D}_{01}(\Delta_{012})$ in $\abar$.
One then obtains the truncated equation
\begin{eqnarray}
\partial_{Y^+}  \left\langle {\mathbf S}_{01} \right\rangle_{Y^+}&=&   \bar{\alpha}\, \frac{2 C_F}{N_c}
\int \frac{\textrm{d}^2\mathbf{x}_{2}}{2\pi} \textbf{K}_{012}\: \theta\!\left(Y^+\!-\! \Delta_{012}\right)\,
\Big[\left\langle {\mathbf S}_{012}\right\rangle_{Y^+\!-\!\Delta_{012}} -  \left\langle {\mathbf S}_{01}\right\rangle_{Y^+} \Big]\, .\label{B_JIMWLK_kc_trunc}
\end{eqnarray}

An additional attractive feature of the equation \eqref{B_JIMWLK_kc_trunc} is that one can safely remove the regulator $\rho$ of the transverse integration, like in the unresummed equation \eqref{B_JIMWLK_dipole}, whereas the regulator $\rho$ is necessary in the equation \eqref{B_JIMWLK_kc_untrunc}.

Physically, $\left\langle {\mathbf S}_{01} \right\rangle_{Y^+}$ and $\left\langle {\mathbf S}_{012} \right\rangle_{Y^+}$ have to take values in the range $[0,1]$, and are decreasing functions of $Y^+$. Indeed, increasing ${Y^+}$ amounts to increase the density of gluons in the target, making the interaction with any projectile stronger. Moreover, one has $\left\langle {\mathbf S}_{012} \right\rangle_{Y^+}\leq \left\langle {\mathbf S}_{01} \right\rangle_{Y^+}$ because the tripole $012$ decoheres in color more easily than the dipole $01$ by interaction with the same target. Hence, in the standard B-JIMWLK dipole evolution equation \eqref{B_JIMWLK_dipole}, the virtual term is driving the decrease of $\left\langle {\mathbf S}_{01} \right\rangle_{Y^+}$ whereas the real term is slowing down that decrease. In the square bracket in the kinematically improved equation \eqref{B_JIMWLK_kc_trunc}, the real term is enhanced because $Y^+\!-\!\Delta_{012}\leq Y^+$ implies $\left\langle {\mathbf S}_{012} \right\rangle_{Y^+\!-\!\Delta_{012}}\geq \left\langle {\mathbf S}_{012} \right\rangle_{Y^+}$, whereas the virtual term is unchanged, which make the evolution of $\left\langle {\mathbf S}_{01} \right\rangle_{Y^+}$ according to the kinematically improved equation \eqref{B_JIMWLK_kc_trunc} slower than according to the standard equation \eqref{B_JIMWLK_dipole}. Obviously, the presence of the theta function in the equation \eqref{B_JIMWLK_kc_trunc} further slows down the evolution of $\left\langle {\mathbf S}_{01} \right\rangle_{Y^+}$.

In order to get some insight into the effect of the truncation of explicitly NLL contributions on probability conservation, one can rewrite the equation \eqref{B_JIMWLK_kc_trunc} in integral form as
\begin{eqnarray}
\left\langle {\mathbf S}_{01}\right\rangle_{Y^+_f}&=& {\mathbf D}_{01}(Y^+_f)\:  \left\langle {\mathbf S}_{01}\right\rangle_{0}+  \abar\, \frac{2 C_F}{N_c}\int \frac{\dd^2 \mathbf{x}_{2}}{2\pi}\, \textbf{K}_{012}\;
\theta\!\left(Y^+_f\!-\!  \Delta_{012}\right)\,
 \int_{0}^{Y^+_f\!-\!  \Delta_{012}}\!\!\!\!\!\!\!\!\!\!\!\!\!\! \dd Y^+_2\,  \frac{{\mathbf D}_{01}(Y^+_f)}{{\mathbf D}_{01}(Y^+_2\!\!+\!\! \Delta_{012})}
\left\langle {\mathbf S}_{012}\right\rangle_{Y^+_2}\, .\label{B_JIMWLK_kc_trunc_integ_1}
\end{eqnarray}
By comparison with the original equation \eqref{BetheSalp_Yplus}, one can see that the truncation of the explicitly NLL contributions amounts to write inaccurately the probability of no splitting before the first splitting, as ${\mathbf D}_{01}(Y^+_f)/{\mathbf D}_{01}(Y^+_2\!\!+\!\! \Delta_{012})$ instead of ${\mathbf D}_{01}(Y^+_f\!-\!Y^+_2)$. Those two expression become equivalent under the replacement $\Delta_{012}\mapsto 0$, as expected because no such truncation is needed for the standard versions of the BFKL and BK equations.

\subsection{Counter-terms for NLO observables and for the NLL evolution equations}


\subsubsection{Kinematically constrained LL counter-term for observables at NLO}

Armed with the kinematically improved evolution equation \eqref{B_JIMWLK_kc_trunc}, one can revisit the resummation of LL's in observables known beyond LO order as outlined in the section \ref{sec:std_subtr_LL}. For observables involving at LO only the dipole operator ${\mathbf S}_{01}$, like DIS structure functions or forward single inclusive particle production in pA collisions, one replaces in the LO term $\left\langle {\mathbf S}_{01} \right\rangle_{0}$ by $\left\langle {\mathbf S}_{01} \right\rangle_{Y_f^+}$ evolved according to the kinematically constrained LL equation  \eqref{B_JIMWLK_kc_trunc}, up to a counterterm, \emph{i.e.}
\begin{equation}
\left\langle {\mathbf S}_{01} \right\rangle_{0}=\left\langle {\mathbf S}_{01} \right\rangle_{Y_f^+}\bigg|_{\textrm{kcLL}} +\delta\!\left\langle {\mathbf S}_{01} \right\rangle_{Y_f^+}\bigg|_{\textrm{kcLL}}\, .\label{def_kcLL_ct}
\end{equation}
From the relation \eqref{def_kcLL_ct}, one gets the expression of the counter-term by integration of the evolution equation \eqref{B_JIMWLK_kc_trunc}, which reads\footnote{If one decides to use the kinematically improved evolution with exact probability conservation \eqref{B_JIMWLK_kc_untrunc}, one can still use the counter-term \eqref{kcLL_ct} in order to remove LL's from NLO corrections to inclusive observables. Indeed, the two improved evolution equations \eqref{B_JIMWLK_kc_untrunc} and \eqref{B_JIMWLK_kc_trunc} differ only by terms of order ${\cal O}(\abar^2)$, which would matter when removing LL's from NNLO corrections instead.}
\begin{equation}
\delta\!\left\langle {\mathbf S}_{01} \right\rangle_{Y_f^+}\bigg|_{\textrm{kcLL}}= -\, \bar{\alpha}\: \frac{2\, C_F}{N_c} \int_{0}^{Y_f^+} \!\!\!\!\dd Y_2^+
\int \frac{\textrm{d}^2\mathbf{x}_{2}}{2\pi}\; \textbf{K}_{012}\: \theta\!\left(Y_2^+\!-\! \Delta_{012}\right)\,
\Big[\left\langle {\mathbf S}_{012}\right\rangle_{Y_2^+\!-\!\Delta_{012}} -  \left\langle {\mathbf S}_{01}\right\rangle_{Y_2^+} \Big]\label{kcLL_ct}
\, .
\end{equation}
As in the section \ref{sec:std_subtr_LL}, it is convenient to split that counter-term into a counter-term for the real NLO correction and a counter-term for the virtual NLO correction, as
\begin{eqnarray}
\delta\!\left\langle {\mathbf S}_{01} \right\rangle_{Y_f^+}\bigg|_{\textrm{kcLL, real}}&=&
\bar{\alpha}\: \frac{2\, C_F}{N_c} \int_{0}^{Y_f^+} \!\!\!\!\dd Y_2^+
\int \frac{\textrm{d}^2\mathbf{x}_{2}}{2\pi}\; \textbf{K}_{012}\: \theta\!\left(Y_2^+\!-\! \Delta_{012}\right)\: \Big[1- \left\langle {\mathbf S}_{012} \right\rangle_{Y_2^+\!-\! \Delta_{012}}\Big]\label{kcLL_ct_real}\\
\delta\!\left\langle {\mathbf S}_{01} \right\rangle_{Y_f^+}\bigg|_{\textrm{kcLL, virt}}&=&
-\, \bar{\alpha}\: \frac{2\, C_F}{N_c} \int_{0}^{Y_f^+} \!\!\!\!\dd Y_2^+ \: \Big[1- \left\langle {\mathbf S}_{01} \right\rangle_{Y_2^+}\Big]
\int \frac{\textrm{d}^2\mathbf{x}_{2}}{2\pi}\; \textbf{K}_{012}\: \theta\!\left(Y_2^+\!-\! \Delta_{012}\right)\label{kcLL_ct_virt}\, .
\end{eqnarray}
The real counter-term \eqref{kcLL_ct_real} can be rewritten as
\begin{equation}
\delta\!\left\langle {\mathbf S}_{01} \right\rangle_{Y_f^+}\bigg|_{\textrm{kcLL, real}}=
\bar{\alpha}\: \frac{2\, C_F}{N_c} \int_{0}^{Y_f^+} \!\!\!\!\dd Y_2^+
\int \frac{\textrm{d}^2\mathbf{x}_{2}}{2\pi}\; \textbf{K}_{012}\: \theta\!\left(Y_f^+ \!-\! Y_2^+\!-\! \Delta_{012}\right)\: \Big[1- \left\langle {\mathbf S}_{012} \right\rangle_{Y_2^+}\Big]\label{kcLL_ct_real_2}\, ,
\end{equation}
or, in $k^+$ variables, as
\begin{equation}
\delta\!\left\langle {\mathbf S}_{01} \right\rangle_{\log(k_f^+/k^+_{\min})}\bigg|_{\textrm{kcLL, real}}=
\bar{\alpha}\: \frac{2\, C_F}{N_c} \int_{k^+_{\min}}^{k^+_f}\frac{\textrm{d}k^+_2}{k^+_2}\;
\int \frac{\textrm{d}^2\mathbf{x}_{2}}{2\pi}\; \textbf{K}_{012}\: \theta\!\left(k^+_f\, {x}_{01}^2 \!-\! k^+_2\, l_{012}^2\right)\: \Big[1- \left\langle {\mathbf S}_{012} \right\rangle_{\log(k_2^+/k^+_{\min})}\Big]
\label{kcLL_ct_real_3}\, .
\end{equation}
The only difference between that kinematically improved counter-term and its analog \eqref{LL_ct_real} associated with the naive LL evolution is the presence of the theta function. Thanks to the general behavior \eqref{lijk_large} and \eqref{lijk_short} of $l_{012}$, that theta function cuts precisely the regime which, in the section \ref{sec:NLO_IF_analysis}, has been found not to contribute to high-energy LL's within the real NLO corrections to DIS. Hence, that counter-term allows to subtract only the high-energy LL contributions actually present in the real NLO corrections to DIS and presumably to other observables, provided high-energy factorization is valid. In particular, when evaluating in the dilute approximation and in Mellin representation the contribution to the real counter-term \eqref{kcLL_ct_real_2} coming from the regime $z_2 \ll z_f$ and ${x}_{10}^2\ll {x}_{20}^2 \simeq {x}_{21}^2$, one obtains again the expression \eqref{Mellin_NLO real_kc_reg}.

Concerning the counter-term for the virtual NLO corrections, the only change between the kinematically improved counter-term \eqref{kcLL_ct_virt} and the naive one \eqref{LL_ct_virt} is also the presence of a theta function. In order to understand the impact of this change, it is convenient to study the difference between these two counterterms
\begin{eqnarray}
\!\!\!\!\!\delta\!\left\langle {\mathbf S}_{01} \right\rangle_{Y_f^+}\bigg|_{\textrm{kcLL, virt}}-\delta\!\left\langle {\mathbf S}_{01} \right\rangle_{Y_f^+}\bigg|_{\textrm{LL, virt}}
&=&\, \bar{\alpha}\: \frac{2\, C_F}{N_c} \int_{0}^{Y_f^+} \!\!\!\!\dd Y_2^+ \: \Big[1- \left\langle {\mathbf S}_{01} \right\rangle_{Y_2^+}\Big]
\int \frac{\textrm{d}^2\mathbf{x}_{2}}{2\pi}\; \textbf{K}_{012}\: \theta\!\left(\Delta_{012}\!-\! Y_2^+\right)\nonumber\\
&=&\, \bar{\alpha}\: \frac{2\, C_F}{N_c} \int_{0}^{Y_f^+} \!\!\!\!\dd Y_2^+ \: \Big[1- \left\langle {\mathbf S}_{01} \right\rangle_{Y_2^+}\Big]
\int \frac{\textrm{d}^2\mathbf{x}_{2}}{2\pi}\; \textbf{K}_{012}\: \theta\!\left(l_{012}^2-{x}_{01}^2\, e^{Y^+_2}
\right)
\label{kcLL-LL_ct_virt} .
\end{eqnarray}

In the expression \eqref{kcLL-LL_ct_virt},  the theta function is completely cutting the anti-collinear regimes ${x}_{20}^2\ll {x}_{10}^2 \simeq {x}_{21}^2$ and ${x}_{21}^2\ll {x}_{10}^2 \simeq {x}_{20}^2$, thus guarantying that the UV divergences present in the counter-term \eqref{kcLL_ct_virt} are identical as in the naive one  \eqref{LL_ct_virt}. As the anti-collinear regimes are also untouched by the kinematical constraint in the case of the real conterterm \eqref{kcLL_ct_real_2}, it is clear that the UV divergences still cancel exactly between the real and virtual counter-terms, as they should.

Moreover, for $Y^+_2$ not too small, the integral in $\mathbf{x}_{2}$ receives contributions only from the collinear regime ${x}_{10}^2\ll {x}_{20}^2 \simeq {x}_{21}^2$, so that it can be calculated approximately as
\begin{equation}
\int \frac{\textrm{d}^2\mathbf{x}_{2}}{2\pi}\; \textbf{K}_{012}\: \theta\!\left(l_{012}^2-{x}_{01}^2\, e^{Y^+_2}
\right) \simeq \int_{{x}_{01}^2\,  e^{Y^+_2}}^{+\infty} \frac{\dd ({x}_{02}^2)}{2} \frac{{x}_{01}^2}{{x}_{02}^4} = \frac{1}{2}\,  e^{-Y^+_2} \qquad \textrm{for $Y^+_2$ not too small,}\label{kc_kernel_virt}
\end{equation}
and thus
\begin{eqnarray}
\!\!\!\!\!\delta\!\left\langle {\mathbf S}_{01} \right\rangle_{Y_f^+}\bigg|_{\textrm{kcLL, virt}}-\delta\!\left\langle {\mathbf S}_{01} \right\rangle_{Y_f^+}\bigg|_{\textrm{LL, virt}}
& \simeq &\, \bar{\alpha}\: \frac{C_F}{N_c} \int_{0}^{Y_f^+} \!\!\!\!\dd Y_2^+ \:  e^{-Y^+_2} \:\Big[1- \left\langle {\mathbf S}_{01} \right\rangle_{Y_2^+}\Big]\nonumber\\
& \rightarrow & \, \bar{\alpha}\: \frac{C_F}{N_c} \int_{0}^{+\infty} \!\!\!\!\dd Y_2^+ \:  e^{-Y^+_2} \:\Big[1- \left\langle {\mathbf S}_{01} \right\rangle_{Y_2^+}\Big]\quad \textrm{for}\quad Y_f^+\rightarrow +\infty
\label{kcLL-LL_ct_virt2}\, .
\end{eqnarray}
Indeed, the convergence of the integral in $Y_2^+$ is guarantied by the exponential decay \eqref{kc_kernel_virt} because $0\leq \left\langle {\mathbf S}_{01} \right\rangle_{Y^+} \leq 1$. Hence, the difference \eqref{kcLL-LL_ct_virt2} between the two virtual counter-term is a finite ${\cal O}(\abar)$ contribution, with no large logs. On the other hand, each of these two counter-terms contain LL contributions of order ${\cal O}(\abar Y_f^+)$ at large $Y_f^+$, with a UV divergent coefficient. Hence, the kinematical constraint does not modify the LL terms subtracted by the counter-term for the virtual NLO correction, by contrast to the case of the counter-term for the real NLO corrections.


\subsubsection{Subtracting kinematical spurious singularities from NLL evolution equations}

In the naive version of the Regge limit, one can calculate the $Y^+$ evolution of the expectation value of multipole operators systematically in powers of $\abar$. For example, for the dipole operator ${\mathbf S}_{01}$, one has
\begin{equation}
\partial_{Y^+}  \left\langle {\mathbf S}_{01} \right\rangle_{Y^+} = \partial_{Y^+}  \left\langle {\mathbf S}_{01} \right\rangle_{Y^+}\bigg|_{LL} + \partial_{Y^+}  \left\langle {\mathbf S}_{01} \right\rangle_{Y^+}\bigg|_{NLL} +{\cal O}(\abar^3)\, .\label{series_evol_naive}
\end{equation}
The first term in the right hand side of \eqref{series_evol_naive}, proportional to $\abar$, is given by the equation \eqref{B_JIMWLK_dipole} (with $\eta=Y^+$), whereas the second term, proportional to $\abar^2$, has been calculated in ref. \cite{Balitsky:2008zz}.

As discussed at length in this paper, that perturbative expansion breaks down due to the appearance of larger and larger higher order corrections, requiring a resummation. The largest corrections at each order are resummed by taking the kinematically improved equation \eqref{B_JIMWLK_kc_trunc} instead of \eqref{B_JIMWLK_dipole} as the first order in the expansion, \emph{i.e.}
\begin{equation}
\partial_{Y^+}  \left\langle {\mathbf S}_{01} \right\rangle_{Y^+} = \partial_{Y^+}  \left\langle {\mathbf S}_{01} \right\rangle_{Y^+}\bigg|_{kcLL} + \partial_{Y^+}  \left\langle {\mathbf S}_{01} \right\rangle_{Y^+}\bigg|_{NLL} -\partial_{Y^+}  \left\langle {\mathbf S}_{01} \right\rangle_{Y^+}\bigg|_{NLL, c.t.} +{\cal O}(\abar^3)\, ,\label{series_evol_kc}
\end{equation}
where the third term is a counter-term accounting for the difference between the LL evolution equations \eqref{B_JIMWLK_kc_trunc} and \eqref{B_JIMWLK_dipole}, which should remove the most pathological parts of the naive NLL contribution to the evolution, corresponding in Mellin space in the dilute regime to the collinear triple pole at $\g=1$, see section \ref{sec:spurious_sing}.

That counter-term can be calculated by re-expanding the kinematically improved equation \eqref{B_JIMWLK_kc_trunc} in the naive Regge limit, and then collecting the terms of order ${\cal O}(\abar^2)$. Formally, it amounts to perform a Taylor expansion around $\Delta_{012}=0$ in \eqref{B_JIMWLK_kc_trunc} as
\begin{eqnarray}
\partial_{Y^+}  \left\langle {\mathbf S}_{01} \right\rangle_{Y^+}\bigg|_{kcLL}&=&   \bar{\alpha}\, \frac{2 C_F}{N_c}
\int \frac{\textrm{d}^2\mathbf{x}_{2}}{2\pi} \textbf{K}_{012}\: \theta\!\left(Y^+\!-\! \Delta_{012}\right)\,
\Big[\left\langle {\mathbf S}_{012}\right\rangle_{Y^+\!-\!\Delta_{012}} -  \left\langle {\mathbf S}_{01}\right\rangle_{Y^+} \Big]\nonumber\\
&=& \bar{\alpha}\, \frac{2 C_F}{N_c}
\int \frac{\textrm{d}^2\mathbf{x}_{2}}{2\pi} \textbf{K}_{012}\:
\Big[\left\langle {\mathbf S}_{012}\right\rangle_{Y^+}  -  \left\langle {\mathbf S}_{01}\right\rangle_{Y^+}
- \Delta_{012}\: \partial_{Y^+} \left\langle {\mathbf S}_{012}\right\rangle_{Y^+} \nonumber\\
& &\qquad\qquad\qquad\qquad\qquad\qquad + {\cal O}\left(\Delta_{012}^2 \partial^2_{Y^+} \left\langle {\mathbf S}_{012}\right\rangle_{Y^+}\right)
 \Big]
\, .\label{B_JIMWLK_kc_trunc_reexpand}
\end{eqnarray}
The theta function completely disappears when doing that Taylor expansion at $Y^+>0$. In the right hand side of the equation \eqref{B_JIMWLK_kc_trunc_reexpand}, the $\partial_{Y^+}$ derivatives correspond to the evolution following the standard LL B-JIMWLK equations, without kinematical constraint, and each $\partial_{Y^+}$ gives a power of $\abar$. Hence, one can read off from \eqref{B_JIMWLK_kc_trunc_reexpand} the expression of the counter-term for the evolution
\begin{eqnarray}
\partial_{Y^+}  \left\langle {\mathbf S}_{01} \right\rangle_{Y^+}\bigg|_{NLL, c.t.} &=&
 -\bar{\alpha}\, \frac{2 C_F}{N_c}
\int \frac{\textrm{d}^2\mathbf{x}_{2}}{2\pi} \textbf{K}_{012}\:
 \Delta_{012}\: \bigg\{   \partial_{Y^+} \left\langle {\mathbf S}_{012}\right\rangle_{Y^+}\bigg|_{LL}
 \bigg\}\nonumber\\
 &=&
 -\bar{\alpha}
\int \frac{\textrm{d}^2\mathbf{x}_{2}}{2\pi} \textbf{K}_{012}\:
 \Delta_{012}\: \bigg\{   \partial_{Y^+} \left\langle {\mathbf S}_{02}\,{\mathbf S}_{21}\right\rangle_{Y^+}\bigg|_{LL}
 -\frac{1}{N_c^2}\;  \partial_{Y^+} \left\langle{\mathbf S}_{01}\right\rangle_{Y^+}\bigg|_{LL}
 \bigg\}\, .\label{NLL_ct_1}
\end{eqnarray}

In the first term, one needs the second equation in Balitsky's hierarchy \cite{Balitsky:1995ub} at LL accuracy in the naive Regge limit, which writes (see \emph{e.g.} eq.(152) in ref. \cite{Balitsky:2001gj})
\begin{eqnarray}
\partial_{Y^+}   \left\langle {\mathbf S}_{02}\,{\mathbf S}_{21}\right\rangle_{Y^+}
&=&   \bar{\alpha} \int \frac{\textrm{d}^2\mathbf{x}_{3}}{2\pi}\;
\bigg\{ \textbf{K}_{023}\: \left\langle \left[{\mathbf S}_{03} {\mathbf S}_{32} \!-\! {\mathbf S}_{02}\right]  {\mathbf S}_{21} \right\rangle_{Y^+}
+\textbf{K}_{213}\: \left\langle  {\mathbf S}_{02} \left[{\mathbf S}_{23} {\mathbf S}_{31} \!-\! {\mathbf S}_{21}\right]\right\rangle_{Y^+}\nonumber\\
& & \qquad\qquad\qquad -\frac{1}{2\, N_c^2}\: \Big[ \textbf{K}_{023}\!+\!\textbf{K}_{213}\!-\!\textbf{K}_{013}\Big]\,
\left\langle{\mathbf S}_{023123} \!+\! {\mathbf S}_{032132} \!-\! 2\, {\mathbf S}_{01} \right\rangle_{Y^+}
\bigg\}\, ,\label{B_JIMWLK_2nd_Eq}
\end{eqnarray}
where we have introduced the fundamental sextupole operator
\begin{equation}
{\mathbf S}_{012345}=\frac{1}{N_c} \textrm{Tr} \left(U_{\mathbf{x}_{0}}\, U_{\mathbf{x}_{1}}^\dag U_{\mathbf{x}_{2}}\, U_{\mathbf{x}_{3}}^\dag U_{\mathbf{x}_{4}}\, U_{\mathbf{x}_{5}}^\dag \right)
\, .\label{def_sextupole}
\end{equation}
Using the first two equations \eqref{B_JIMWLK_dipole} and \eqref{B_JIMWLK_2nd_Eq} of Balitsky's hierarchy, one obtains the final expression for the counter-term \eqref{NLL_ct_1} for the NLL evolution equation
\begin{eqnarray}
\partial_{Y^+}  \left\langle {\mathbf S}_{01} \right\rangle_{Y^+}\bigg|_{NLL, c.t.}
 &\!\!\! =&\!\!\! -\bar{\alpha}^2 \int \frac{\textrm{d}^2\mathbf{x}_{2}}{2\pi} \textbf{K}_{012}\, \Delta_{012}\,  \int \frac{\textrm{d}^2\mathbf{x}_{3}}{2\pi}
 \bigg\{    \textbf{K}_{023}\, \left\langle \left[{\mathbf S}_{03} {\mathbf S}_{32} \!-\! {\mathbf S}_{02}\right]  {\mathbf S}_{21} \right\rangle_{Y^+}
+\textbf{K}_{213}\, \left\langle  {\mathbf S}_{02} \left[{\mathbf S}_{23} {\mathbf S}_{31} \!-\! {\mathbf S}_{21}\right]\right\rangle_{Y^+}\bigg\}\nonumber\\
 & & + \frac{\bar{\alpha}^2}{2\, N_c^2} \int \frac{\textrm{d}^2\mathbf{x}_{2}}{2\pi} \textbf{K}_{012}\, \Delta_{012}\,  \int \frac{\textrm{d}^2\mathbf{x}_{3}}{2\pi} \Big[ \textbf{K}_{023}\!+\!\textbf{K}_{213}\!-\!\textbf{K}_{013}\Big]\,
\left\langle{\mathbf S}_{023123} \!+\! {\mathbf S}_{032132} \!-\! 2\, {\mathbf S}_{01} \right\rangle_{Y^+}\nonumber\\
 & & + \frac{\bar{\alpha}^2}{N_c^2} \left[\int \frac{\textrm{d}^2\mathbf{x}_{2}}{2\pi} \textbf{K}_{012}\, \Delta_{012}\right]\,  \int \frac{\textrm{d}^2\mathbf{x}_{3}}{2\pi}\, \textbf{K}_{013}\,
\left\langle {\mathbf S}_{03} {\mathbf S}_{31} \!-\! {\mathbf S}_{01}\right\rangle_{Y^+}
 \, .\label{NLL_ct_2}
\end{eqnarray}
That counter-term is supposed to cancel the largest and most pathological contributions appearing in the B-JIMWLK evolution equation for the dipole at NLL accuracy \cite{Balitsky:2008zz} in the naive Regge limit. Such contributions should behave as triple logs in the collinear limit \cite{Salam:1998tj}. However, it is rather difficult to track down those triple logs in the NLL evolution equation or in the counter-term \eqref{NLL_ct_2} without further approximation. One can nevertheless notice that all the multipole operators appearing in the counter-term \eqref{NLL_ct_2} also appear in the NLL evolution equation \cite{Balitsky:2008zz} for the dipole $ \left\langle {\mathbf S}_{01} \right\rangle_{Y^+}$, as expected for consistency.

In order to analyse further the counter-term \eqref{NLL_ct_2}, it is convenient to consider the dilute target case, and take accordingly the $2$ gluons exchange approximation for the expectation value of all the operators.
Expanding the Wilson lines in the sextupole operator \eqref{def_sextupole}, collecting the terms of order ${\cal O}(g^2)$ and comparing the result with the similar expansion for the dipole operator ${\mathbf S}_{ij}$, one finds that
\begin{eqnarray}
1-  \left\langle{\mathbf S}_{012345}\right\rangle_{Y^+} &\simeq &  \left\langle
{\textbf N}_{01} +{\textbf N}_{03}+{\textbf N}_{05}+{\textbf N}_{21}+{\textbf N}_{23}+{\textbf N}_{25}+{\textbf N}_{41}+{\textbf N}_{43}+{\textbf N}_{45}\right\rangle_{Y^+}\nonumber\\
& &- \left\langle{\textbf N}_{02}+{\textbf N}_{04}+{\textbf N}_{24}+{\textbf N}_{13}+{\textbf N}_{15}+{\textbf N}_{35}\right\rangle_{Y^+}\, ,
\end{eqnarray}
so that
\begin{eqnarray}
1-  \left\langle{\mathbf S}_{023123}\right\rangle_{Y^+} & \simeq &  \left\langle
{\textbf N}_{01} \right\rangle_{Y^+}\, .
\end{eqnarray}
Hence, in the $2$ gluons exchange approximation,
\begin{equation}
\left\langle{\mathbf S}_{023123} \!+\! {\mathbf S}_{032132} \!-\! 2\, {\mathbf S}_{01} \right\rangle_{Y^+} \simeq 0\, ,
\end{equation}
 and thus the term in the second line of the expression \eqref{NLL_ct_2} disappears completely, and the counter-term reduces to
\begin{eqnarray}
\!\!\!\!\!\!\!\!\!\!\!\!\partial_{Y^+}  \left\langle {\mathbf S}_{01} \right\rangle_{Y^+}\bigg|_{NLL, c.t.; \textrm{ dilute}}
 &\!\!\! = &\!\!\! \bar{\alpha}^2 \int \frac{\textrm{d}^2\mathbf{x}_{2}}{2\pi} \textbf{K}_{012}\, \Delta_{012}\, \bigg[    \int \frac{\textrm{d}^2\mathbf{x}_{3}}{2\pi}
 \textbf{K}_{023}\, \left\langle {\mathbf N}_{03}\!+\!{\mathbf N}_{32} \!-\! {\mathbf N}_{02} \right\rangle_{Y^+}
\nonumber\\
 & & + \int \frac{\textrm{d}^2\mathbf{x}_{3}}{2\pi}\textbf{K}_{213}\, \left\langle {\mathbf N}_{23}\!+\! {\mathbf N}_{31} \!-\! {\mathbf N}_{21}\right\rangle_{Y^+}
  - \frac{1}{N_c^2} \,  \int \frac{\textrm{d}^2\mathbf{x}_{3}}{2\pi}\, \textbf{K}_{013}\,
\left\langle {\mathbf N}_{03}\!+\! {\mathbf N}_{31} \!-\! {\mathbf N}_{01}\right\rangle_{Y^+}\bigg]
 \, .\label{NLL_ct_2glue}
\end{eqnarray}
Each of the three terms in the bracket in the expression \eqref{NLL_ct_2glue} corresponds to the right hand side of the BFKL equation \eqref{BFKL_dipole}. Therefore, introducing the Mellin representation \eqref{Mellin_rep} for the dipole amplitude $\left\langle {\mathbf N}_{ij}\right\rangle_{Y^+}$ and using the characteristic function $\chi(\g)$ of the BFKL kernel \eqref{BFKL_eigenvalue}, one gets
\begin{eqnarray}
\!\!\!\!\!\!\!\!\!\!\!\!\partial_{Y^+}  \left\langle {\mathbf S}_{01} \right\rangle_{Y^+}\bigg|_{NLL, c.t.; \textrm{ dilute}}
 &\!\!\! = &\!\!\!  \int_{1/2-i\infty}^{1/2+i\infty} \frac{\dd \g}{2\pi i}\;   {\cal N}(\g,Y^+)\:   \bar{\alpha}^2\, \chi(\g)
 \int \frac{\textrm{d}^2\mathbf{x}_{2}}{2\pi} \textbf{K}_{012}\, \Delta_{012}\,
  \nonumber\\
 & & \qquad \times
 \left[  \left(\frac{x_{02}^2 \, Q_0^2}{4}\right)^\g
 + \left(\frac{x_{21}^2 \, Q_0^2}{4}\right)^\g \;
  - \frac{1}{N_c^2} \, \left(\frac{x_{01}^2 \, Q_0^2}{4}\right)^\g\right]  \nonumber\\
 & \!\!\! = & \!\!\! \int_{1/2-i\infty}^{1/2+i\infty} \frac{\dd \g}{2\pi i}\;  \left(\frac{x_{01}^2 \, Q_0^2}{4}\right)^\g\, {\cal N}(\g,Y^+)\;   \bar{\alpha}^2\, \chi(\g)\, \left[
{\cal F}_{\Delta}(\g)
- \frac{1}{2\, N_c^2}\, {\cal F}_{\Delta}(0)
 \right]
 \, ,\label{NLL_ct_2glue_2}
\end{eqnarray}
with the notation
\begin{equation}
{\cal F}_{\Delta}(\g) = \int \frac{\textrm{d}^2\mathbf{x}_{2}}{2\pi} \textbf{K}_{012}\, \Delta_{012}\, \left[\left(\frac{x_{02}^2}{x_{01}^2}\right)^\g+ \left(\frac{x_{21}^2}{x_{01}^2}\right)^\g\: \right]\, .
\end{equation}
Thanks to the definition \eqref{Delta012} of the shift $\Delta_{012}$, the regimes $x_{02}\ll x_{21} \simeq x_{01}$ and $x_{02}\ll x_{21} \simeq x_{01}$ are explicitly cut-off. Hence, potential problems for the convergence of the $\mathbf{x}_{2}$ integration can come only from the large $\mathbf{x}_{2}$ limit, \emph{i.e.} the regime $x_{01}\ll x_{02} \simeq x_{21}$. For that reason, ${\cal F}_{\Delta}(\g)$ can not have singularities on the left of the line $\textrm{Re}(\g)=1/2$ but just on the right, and ${\cal F}_{\Delta}(0)$ is a well-defined constant. The first singularity of ${\cal F}_{\Delta}(\g)$ on the right of the line $\textrm{Re}(\g)=1/2$ is obtained by taking the integrand in the limit $x_{01}\ll x_{02} \simeq x_{21}$, as
\begin{equation}
{\cal F}_{\Delta}(\g)\bigg|_{1\textrm{st sing.}} = \int_{x_{01}^2}^{+\infty} \frac{\textrm{d}(x_{02}^2)}{2}\: \frac{x_{01}^2}{x_{02}^4}\, \log\left(\frac{x_{02}^2}{x_{01}^2}\right)\; 2   \left(\frac{x_{02}^2}{x_{01}^2}\right)^\g= \frac{1}{(1\!-\!\g)^2}
\, .
\end{equation}
The next singularities of ${\cal F}_{\Delta}(\g)$, located at $\g=2$ and higher integers, are irrelevant for our purposes.
Finally, the counter-term \eqref{NLL_ct_2glue_2} in the 2 gluons approximation and in Mellin representation is driven in the collinear limit $(x_{01}^2 \, Q_0^2) \rightarrow 0$  by a triple pole at $\g=1$
\begin{equation}
\abar^2 \chi(\g)\, \left[{\cal F}_{\Delta}(\g)- \frac{1}{2\, N_c^2}\, {\cal F}_{\Delta}(0)\right]=  \abar^2 \left[
\frac{1}{(1\!-\!\g)^3} + {\cal O}\left(\frac{1}{(1\!-\!\g)}\right)
\right] \qquad \textrm{for } \g\rightarrow 1 \label{kernel_NLL_ct_coll}
\end{equation}
and in the anti-collinear limit by a simple pole at $\g=0$
\begin{equation}
\abar^2 \chi(\g)\, \left[{\cal F}_{\Delta}(\g)- \frac{1}{2\, N_c^2}\, {\cal F}_{\Delta}(0)\right]=  \abar^2 \left[
\frac{1}{\g}\; {\cal F}_{\Delta}(0) \: \left(1\!-\!\frac{1}{2\, N_c^2}\right)  + {\cal O}\left(\g^2\right)
\right] \qquad \textrm{for } \g\rightarrow 0\, . \label{kernel_NLL_ct_anticoll}
\end{equation}

The triple pole \eqref{kernel_NLL_ct_coll} is the same\footnote{The different sign is trivially due to the fact that the equation \eqref{BFKL_NLL_dipole_Laplace} is written as an equation for $\left\langle {\mathbf N}_{01} \right\rangle_{Y^+}$ whereas the counter-term \eqref{NLL_ct_2glue_2} applies to the equation for $\left\langle {\mathbf S}_{01} \right\rangle_{Y^+}$.} as the one appearing in the characteristic function $\chi_1(\g)$ of the naive NLL evolution equation \eqref{BFKL_NLL_dipole_Laplace}, see equations \eqref{double_poles_0} and \eqref{double_and_triple_poles_1}. Hence, in the dilute regime, the counter-term \eqref{NLL_ct_2glue} cancels exactly the spurious collinear triple logs appearing in the NLL evolution equation obtained in the naive Regge limit, which manifest themselves as a triple pole at $\g=1$ in Mellin space. Apart from this, the counter-term \eqref{NLL_ct_2glue} is modifying contributions to the NLL kernel behaving at most as single logs in the collinear and anti-collinear regimes, which cannot overcome the LL contributions.

As announced in the section \ref{sec:spurious_sing}, the kinematical constraint allows to deal with the triple pole at $\g=1$ in $\chi_1(\g)$ by correcting the kinematics in the collinear limit, but leaves the double poles at $\g=1$ and $\g=0$ unaffected. Those would require distinct resummations.


\section{Conclusion and Discussion\label{sec:Discussion}}

In momentum space, LL's arise in initial-state parton or dipole cascades from configurations strictly ordered in $k^+$ and $k^-$ simultaneously. That fact is overlooked in standard derivations of the high-energy LL evolution equations, where only one ordering is strictly imposed. In momentum space, one can ensure that both orderings are satisfied by introducing a kinematical constraint in the kernel of the BFKL equation.

In this paper, the translation of the kinematical constraint from momentum space to mixed space has been studied, because mixed-space is the most suitable for high-energy evolution equations with gluon saturation, like BK, JIMWLK, and Balitsky's hierarchy. The mixed-space version of the kinematical constraint has been understood in the section \ref{sec:NLO_IF_analysis}, by extracting LL contributions from the explicit expressions of the real NLO corrections to DIS structure functions, known in mixed-space.

The result of that analysis has been used in the section \ref{sec:kcBK} to write down kinematically-improved high-energy LL evolution equations in mixed-space, with the form of the virtual corrections fixed by the requirement of probability conservation along the dipole cascade. More precisely, two equations have been proposed. The first one, eq. \eqref{B_JIMWLK_kc_untrunc}, satisfies exact probability conservation, but includes also a tower of higher order corrections, which depend on a UV cut-off. The second one, eq. \eqref{B_JIMWLK_kc_trunc}, is UV finite and contains only terms of (improved-) LL accuracy, but obeys probability conservation only up to NLL corrections. Presumably, the equation \eqref{B_JIMWLK_kc_trunc} should be preferred for all practical purposes, except for Monte Carlo simulation of dipole cascades, requiring exact probability conservation. The equations \eqref{B_JIMWLK_kc_untrunc} and \eqref{B_JIMWLK_kc_trunc} are written as generalizations of the first equation of Balitsky's hierarchy \eqref{B_JIMWLK_dipole}. But of course, performing the usual mean-field or dilute approximations, one obtains kinematically constrained versions of the BK or BFKL equations in mixed-space. By contrast, constructing a kinematically constrained version of the full hierarchy of Balitsky or of the JIMWLK equation seems to be a complicated task, and is left for further studies.

As a remark, note that the kinematical improvement of the high-energy evolution equations in mixed space has been discussed and performed only within the factorization scheme with a cut-off in $k^+$. In other schemes, the kinematical constraint should look completely different. Its form in the factorization schemes in $k^-$ or in rapidity $y$ mentioned in the section \ref{sec:evol_variables} could be guessed, but not cross-checked, because no explicit higher order calculation has been done in those schemes in mixed space. In particular, in the $k^-$ scheme, the kinematical constraint should affect the small daughter dipole regimes instead of the large daughter dipoles regime.

As a side result of the present paper, it has been found in the appendix \ref{App:locality_kc} that despite naive expectations based on light-front perturbation theory, and by contrast to the claim made in ref. \cite{Motyka:2009gi}, the kinematical constraint is local and not global in a cascade, \emph{i.e.} each emitted gluon has its $k^+$ and $k^-$ constrained by the ones of the two partons forming the parent color dipole, but not by the $k^+$'s and $k^-$'s of the other partons present in the cascade, up to NLL corrections.

In phenomenological studies, one uses the fixed order perturbative results, at LO or NLO, for the considered observable, together with the resummation of large high-energy logs, at LL or NLL accuracy. Accordingly, the kinematically consistent BK equation (kcBK) obtained from eq. \eqref{B_JIMWLK_kc_trunc} is useful in practice at LO+LL accuracy, NLO+LL accuracy, or NLO+NLL accuracy. By consistency, it should be used to evolve the dipole amplitude over the appropriate range $Y^+_f$, as discussed in the section \ref{sec:evol_variables}, like the expression \eqref{Yfplus_final} in the case of DIS.

First, as a finite-energy correction, the kinematical constraint is an improvement of the theoretical framework at LO+LL accuracy, conceptually analog to the one provided by the inclusion of running coupling effects. Both can be understood as a resummation of terms of all logarithmic orders in the naive perturbative expansion. The higher order terms associated with running coupling effects or kinematical constraint effects are independent of each other, and their resummation affects different parts of the LL evolution equations. Therefore, it is straightforward to take both the kinematical constraint and the running coupling into account: one should just replace the factor $\abar \textbf{K}_{012}$ in the equation  \eqref{B_JIMWLK_kc_trunc} by the kernel with the chosen running coupling prescription, for example Balitsky's prescription \cite{Balitsky:2006wa}. In practice, the evolution according to the BK equation with both the kinematical constraint and the running coupling (kcrcBK) should be significantly slower than the evolution according to the BK equation with just running coupling (rcBK), especially in the beginning. Hence, going from rcBK to kcrcBK should improve further the agreement between phenomenological studies at LO+LL accuracy and the DIS data  \cite{Albacete:2010sy,Kuokkanen:2011je}.

Second, the kinematical constraint is absolutely necessary for studies at NLO+LL accuracy. The LL evolution equations with kinematical constraint allow, by construction, to resum properly the LL's and to remove them exactly from the NLO corrections thanks to the counter-term \eqref{kcLL_ct}. By contrast, as discussed in the section \ref{sec:NLO_IF_analysis}, the standard LL evolution equations without kinematical constraint overestimate the LL's present in the NLO corrections, and thus fail at properly resumming them.
In that case, the leftover NLO correction after the failed resummation is negative and overcomes the LO term in the collinear regime, although that effect disappears progressively in the limit of high energy (or large interval $Y^+_f$), in which the kinematical constraint would become weaker and weaker. This phenomenon has indeed been observed for single inclusive hadron production at forward rapidity in pp or pA collisions in ref. \cite{Stasto:2013cha}, which is the only phenomenological study performed so far at NLO+LL accuracy with gluon saturation, and it does not include the kinematical constraint\footnote{It is clear that the lack of kinematical constraint does produce at NLO+LL accuracy large corrections with the systematics observed in ref. \cite{Stasto:2013cha}. However, it is not excluded that the large and negative correction observed in ref. \cite{Stasto:2013cha} is the cumulated effect of the lack of kinematical constraint plus another yet unknown problem. In order to clarify that issue, one should study the LL's in the NLO corrections to single inclusive hadron production \cite{Chirilli:2011km,Chirilli:2012jd}, with the method of section \ref{sec:NLO_IF_analysis}. The same kinematical constraint should occur in a quite different way for that observable than for the dipole cascades relevant for DIS analyzed in the present paper, due to the crossing of Wilson lines from the complex conjugate amplitude into the amplitude. However, it was demonstrated in ref. \cite{Mueller:2012bn} that this crossing does not modify the evolution equation at NLL accuracy. Therefore, the same kcBK equation \eqref{B_JIMWLK_kc_trunc} has to be valid both for DIS structure functions and single inclusive hadron production.}.

Third, the kinematical constraint is a crucial building block for studies at NLO+NLL accuracy. As discussed in the section \ref{sec:spurious_sing}, the NLL evolution equations contain large corrections in the collinear and anti-collinear regimes making them useless without further collinear resummations. The kinematical constraint corresponds to the resummation of the parametrically largest of those corrections. Hence, one should take the kinematically improved equation \eqref{B_JIMWLK_kc_trunc} for the LL term, and use the counter-term \eqref{NLL_ct_2} to remove the contributions induced by the kinematical constraint from the naive NLL terms in the evolution equations. The large corrections associated with the running coupling can be dealt with in a similar way, by taking the LL term with Balitsky's running coupling prescription \cite{Balitsky:2006wa}, or with the smallest dipole prescription. On the other hand, there are also large NLL contributions induced the DGLAP evolution in the collinear and anti-collinear regimes, whose resummation in mixed-space is left for further studies.



\begin{acknowledgments}
I thank Emil Avsar, Ian Balitsky, Giovanni Chirilli, Anna Sta\'sto, Heribert Weigert and Bo-Wen Xiao for helpful discussions at various stages of this project.
This work is funded by European Research Council grant HotLHC ERC-2011-StG-279579; by Ministerio de Ciencia e Innovaci\'on of Spain under project FPA2011-22776; by Xunta de Galicia (Conseller\'{\i}a de Educaci\'on); by the Spanish Consolider-Ingenio 2010 Programme CPAN and by FEDER.
This project has been started when I was a research associate at Brookhaven National Laboratory, working under the Contract No. \#DE-AC02-98CH10886 with the U.S. Department of Energy.
\end{acknowledgments}


\appendix


\section{Locality of the kinematical constraint in Light-Front perturbation theory\label{App:locality_kc}}

In this appendix, the discussion of the kinematical constraint based on Light-Front perturbation theory presented in section \ref{sec:kinematical_approx} is extended, in order to show that the constraint is local in the parton or dipole cascade. More precisely, the aim is to show that gluon emission by a dipole is insensitive, at LL accuracy, to gluon emission by another dipole in the same dipole cascade.

\subsection{Diagrams without gluon splittings}

\begin{figure}
\setbox1\hbox to 10cm{
 \includegraphics{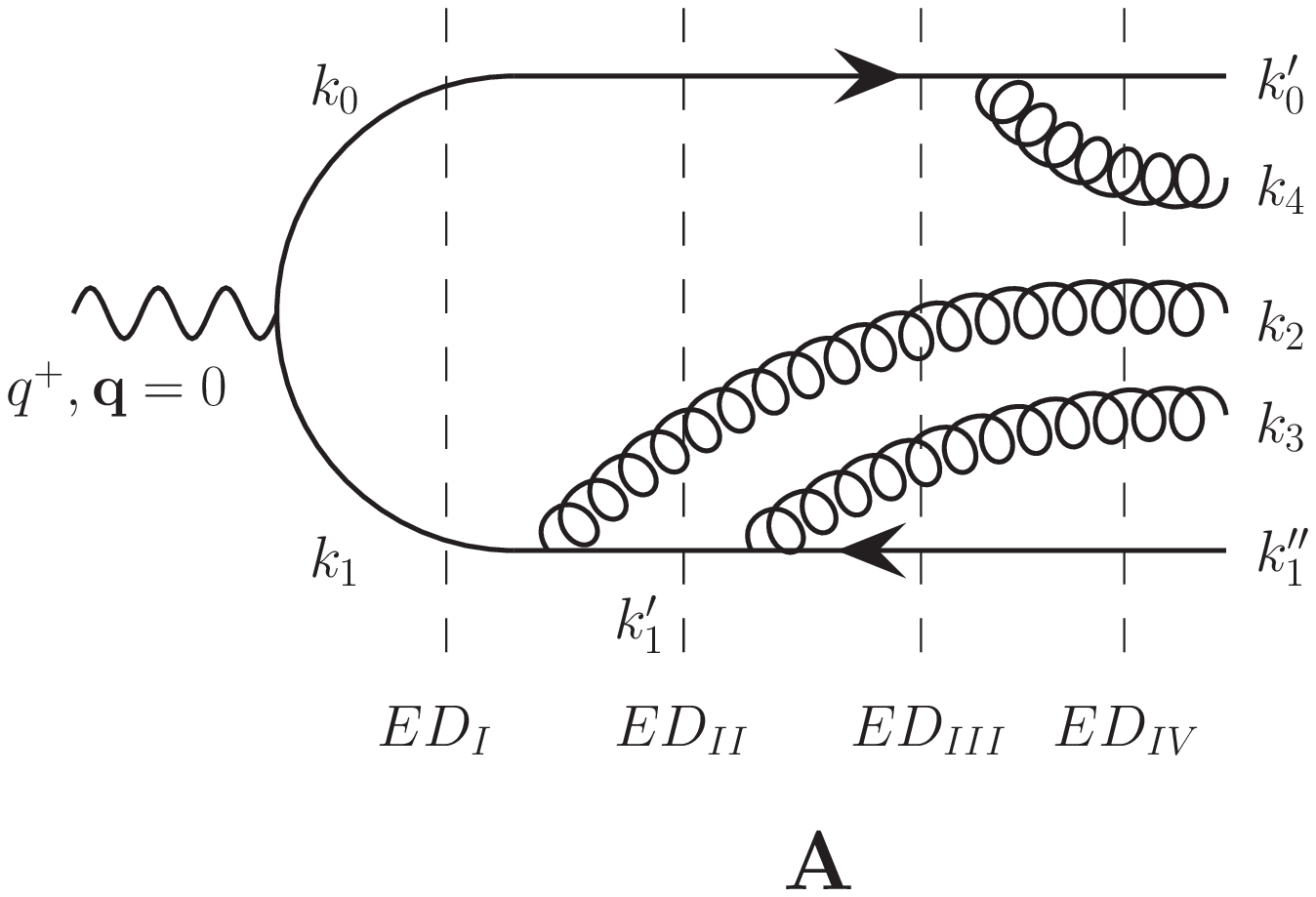}
}
\setbox2\hbox to 10cm{
 \includegraphics{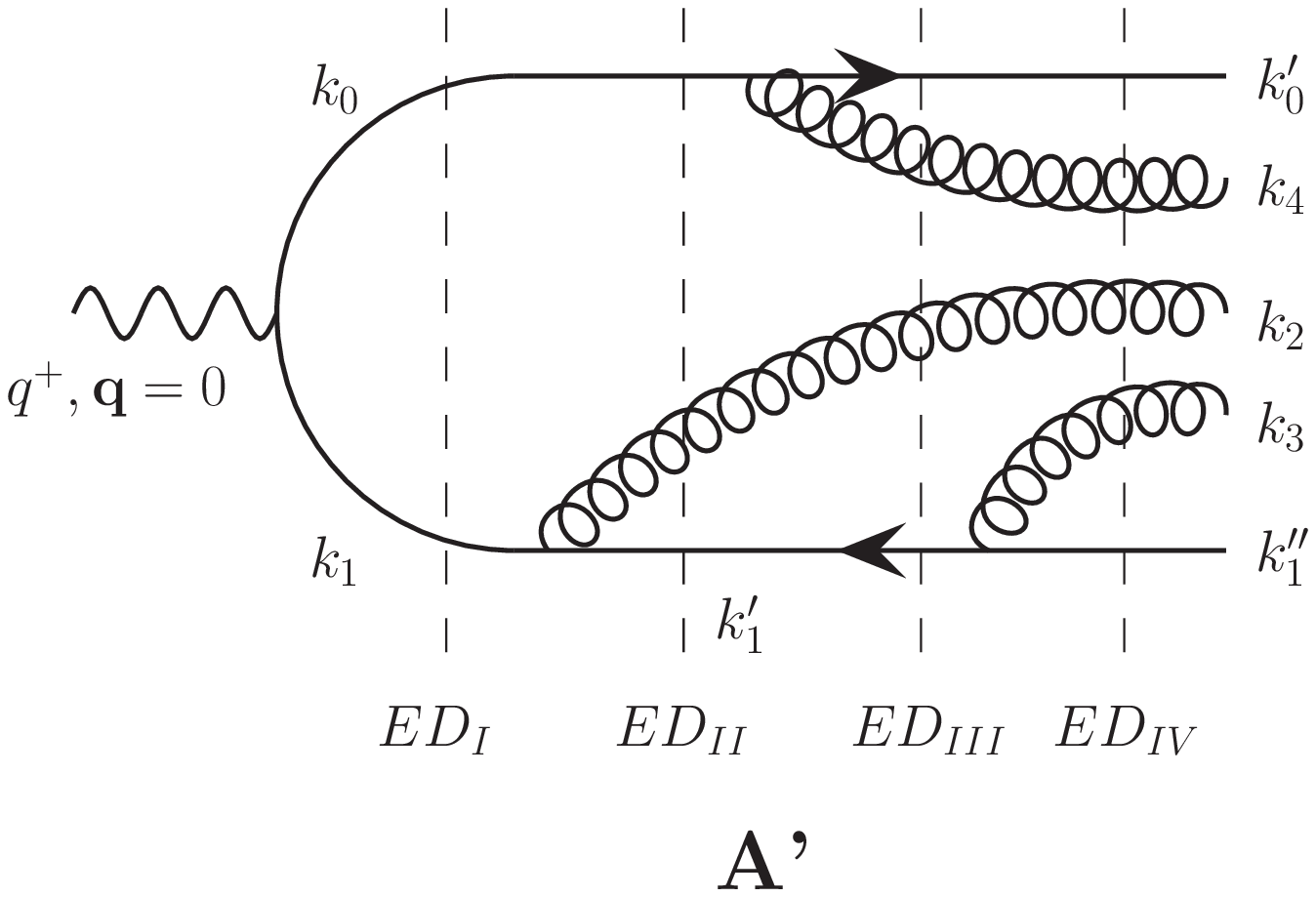}
}
\begin{center}
\hspace{-4cm}
\resizebox*{14cm}{!}{\mbox{\box1 \hspace{5cm} \box2}}
\caption{\label{Fig:qqbarggg_A} Examples of light-front perturbation theory diagrams without gluon splitting contributing to the $q\bar{q}ggg$ Fock component of a photon wave-function.}
\end{center}
\end{figure}

The diagrams \textbf{A} and \textbf{A'} shown in Fig. \ref{Fig:qqbarggg_A} are some of the most simple ones for which the question of local \emph{versus} global kinematical constraint is relevant. In both of them, the parent dipole $01$ has split into two dipoles $02$ and $21$ by emission of the gluon $2$. Then, each of the two dipoles emits another gluon. The light-front diagrams \textbf{A} and \textbf{A'} differ only by the order of emission of the last two gluons, called $3$ and $4$.
The diagrams \textbf{A} and \textbf{A'} have the same color flow and the same momentum flow, and thus their vertices have identical expressions. Moreover, the energy denominators $I$, $II$ and $IV$ are the same for the two graphs, and write
\begin{eqnarray}
& & ED_{I}^{\textbf{A}}= ED_{I}^{\textbf{A'}}= -\frac{Q^2}{2 q^+}- k^-_0 - k^-_1 +i \epsilon \label{ED_I_A}\\
& & ED_{II}^{\textbf{A}}= ED_{II}^{\textbf{A'}}= -\frac{Q^2}{2 q^+}- k^-_0 - {k_{1}'}^- - k^-_2 +i \epsilon  \label{ED_II_A}\\
& & ED_{IV}^{\textbf{A}}= ED_{IV}^{\textbf{A'}}= -\frac{Q^2}{2 q^+}- {k_{0}'}^- - {k_{1}''}^- - k^-_2- k^-_3- k^-_4 +i \epsilon\, . \label{ED_IV_A}
\end{eqnarray}
Hence, the only difference between the expressions of the diagrams \textbf{A} and \textbf{A'} comes from the energy denominator $III$, which writes
\begin{eqnarray}
& & ED_{III}^{\textbf{A}}= -\frac{Q^2}{2 q^+}- k^-_0 - {k_{1}''}^- - k^-_2 - k^-_3 +i \epsilon  \label{ED_III_A}\\
& & ED_{III}^{\textbf{A'}}= -\frac{Q^2}{2 q^+}- {k_{0}'}^- - {k_{1}'}^- - k^-_2 - k^-_4 +i \epsilon\, . \label{ED_III_A_prime}
\end{eqnarray}

Then, the sum of the diagrams \textbf{A} and \textbf{A'} is of the form
\begin{eqnarray}
\textbf{A} + \textbf{A'}&=& \textrm{vertices} \times \frac{1}{ED_{I}^{\textbf{A}}}\; \frac{1}{ED_{II}^{\textbf{A}}}\; \left[\frac{1}{ED_{III}^{\textbf{A}}} +\frac{1}{ED_{III}^{\textbf{A'}}}  \right]\; \frac{1}{ED_{IV}^{\textbf{A}}}\nonumber\\
&=& \textrm{vertices} \times \frac{1}{ED_{I}^{\textbf{A}}}\; \left[\frac{1}{ED_{II}^{\textbf{A}}}+\frac{1}{ED_{IV}^{\textbf{A}}}\right]\; \frac{1}{ED_{III}^{\textbf{A}}}\; \frac{1}{ED_{III}^{\textbf{A'}}}\label{A+A_prime}
\, ,
\end{eqnarray}
using the identity
\begin{equation}
ED_{III}^{\textbf{A}}+ED_{III}^{\textbf{A'}}= ED_{II}^{\textbf{A}}+ED_{IV}^{\textbf{A}}
\end{equation}
satisfied by the energy denominators \eqref{ED_II_A}, \eqref{ED_IV_A}, \eqref{ED_III_A} and \eqref{ED_III_A_prime}.

The aim is now to understand sufficient kinematical conditions for each gluon emission to come with a large soft log.
Following the reasoning introduced in section \ref{sec:kinematical_approx}, one finds that the conditions \begin{eqnarray}
& & k^+_3,\: k^+_4 \ll k^+_2 \ll k^+_0,\: k^+_1 \leq q^+ \label{kplus_ord_A}\\
& & k^-_3,\: k^-_4 \gg k^-_2 \gg \frac{Q^2}{2 q^+}+ k^-_0 + k^-_1\label{kminus_ord_A}
\, ,
\end{eqnarray}
with no constraint on the relative size of $k^+_3$ and $k^+_4$ and of $k^-_3$ and $k^-_4$, are sufficient in order to justify the approximations
\begin{eqnarray}
ED_{II}^{\textbf{A}} & \simeq & - k^-_2 +i \epsilon \\
ED_{III}^{\textbf{A}} & \simeq & - k^-_3 +i \epsilon \\
ED_{III}^{\textbf{A'}} & \simeq & - k^-_4 +i \epsilon\, .
\end{eqnarray}
Assuming only the $k^+$ ordering \eqref{kplus_ord_A}, the energy denominator $IV$ simplifies as
\begin{eqnarray}
ED_{IV}^{\textbf{A}} & \simeq & -\frac{Q^2}{2 q^+}-   \frac{(\mathbf{k}_0\!-\!\mathbf{k}_4)^2}{2\, {k_0}^+}    - \frac{(\mathbf{k}_1\!-\!\mathbf{k}_2\!-\!\mathbf{k}_3)^2}{2\, {k_1}^+}
 - \frac{\mathbf{k}_2^2}{2\, {k_2}^+}
 - \frac{\mathbf{k}_3^2}{2\, {k_3}^+}- \frac{\mathbf{k}_4^2}{2\, {k_4}^+} +i \epsilon \, .\label{ED_IV_A_approx1}
\end{eqnarray}
For generic values of the transverse momenta, the expression \eqref{ED_IV_A_approx1} is dominated by the terms with ${k_3}^+$ or ${k_4}^+$ in the denominator.
Other terms are then relevant only in the regime where both $\mathbf{k}_3$ and $\mathbf{k}_4$ are parametrically small, where typically the term with ${k_2}^+$ in the denominator is the dominant one, unless $\mathbf{k}_2$ is also parametrically small. Hence, thanks to the $k^+$ ordering \eqref{kplus_ord_A}, one can drop $\mathbf{k}_2$, $\mathbf{k}_3$ and $\mathbf{k}_4$ from the second and third terms in the right-hand side of the equation \eqref{ED_IV_A_approx1} no matter what is the relative size of all the transverse momenta, and obtain
\begin{eqnarray}
ED_{IV}^{\textbf{A}} & \simeq & ED_{I}^{\textbf{A}} - k^-_2- k^-_3- k^-_4 +i \epsilon \, ,\label{ED_IV_A_approx2}
\end{eqnarray}
under the assumption \eqref{kplus_ord_A} only.
Therefore, when assuming both the $k^+$ ordering \eqref{kplus_ord_A} and the $k^-$ ordering \eqref{kminus_ord_A},
one has
\begin{equation}
\left| ED_{IV}^{\textbf{A}} \right| \simeq k^-_3 + k^-_4 \gg \left| ED_{II}^{\textbf{A}} \right| \simeq k^-_2\, .
\end{equation}
All in all, the sum \eqref{A+A_prime} of the diagrams \textbf{A} and \textbf{A'} simplifies to
\begin{eqnarray}
\textbf{A} + \textbf{A'}&\simeq & \textrm{vertices} \times \frac{1}{ED_{I}^{\textbf{A}}}\; \frac{1}{[- k^-_2 +i \epsilon]}\; \frac{1}{[- k^-_3 +i \epsilon]}\; \frac{1}{[- k^-_4 +i \epsilon]}\label{A+A_prime_approx}
\, ,
\end{eqnarray}
in the regime where all the conditions  \eqref{kplus_ord_A} and \eqref{kminus_ord_A} are satisfied. As usual, those conditions also allow one to neglect the non-eikonal terms in the vertices, as well as the transverse and longitudinal recoil effects.
Hence, the conditions  \eqref{kplus_ord_A} and \eqref{kminus_ord_A} are sufficient for the sum of the diagrams \textbf{A} and \textbf{A'} to exhibit the factorized form which will give, upon Fourier transform to mixed space and squaring of the wave-function, one large soft log for each gluon emission.

The important point is that the relative size of $k^+_3$ and $k^+_4$ and of $k^-_3$ and $k^-_4$ becomes irrelevant to the appearance of high-energy LL's, when considering the sum \textbf{A}$+$\textbf{A'}.

Obviously, this observation generalizes to a large class of graphs contributing to the $\gamma\rightarrow q + \bar{q} + n\, g$ sector of the photon wave-function at tree level, namely the graphs with all the $n$ gluons emitted directly from the quark or the anti-quark, without any gluon splitting. The configurations contributing to LL's for those graphs are the ones where each gluon has a smaller $k^+$ and a larger $k^-$ than the gluon previously emitted by the same parent (quark or anti-quark). That previous gluon plays the role of the other leg of the emitting dipole. On the other hand, there is no sensitivity to possible gluon emission on the other side of the cascade, when taking the sum over graphs with identical color and momentum flow, but different $x^+$ ordering of gluon emission vertices.


\subsection{Diagrams with gluon splittings}

\begin{figure}
\setbox1\hbox to 10cm{
 \includegraphics{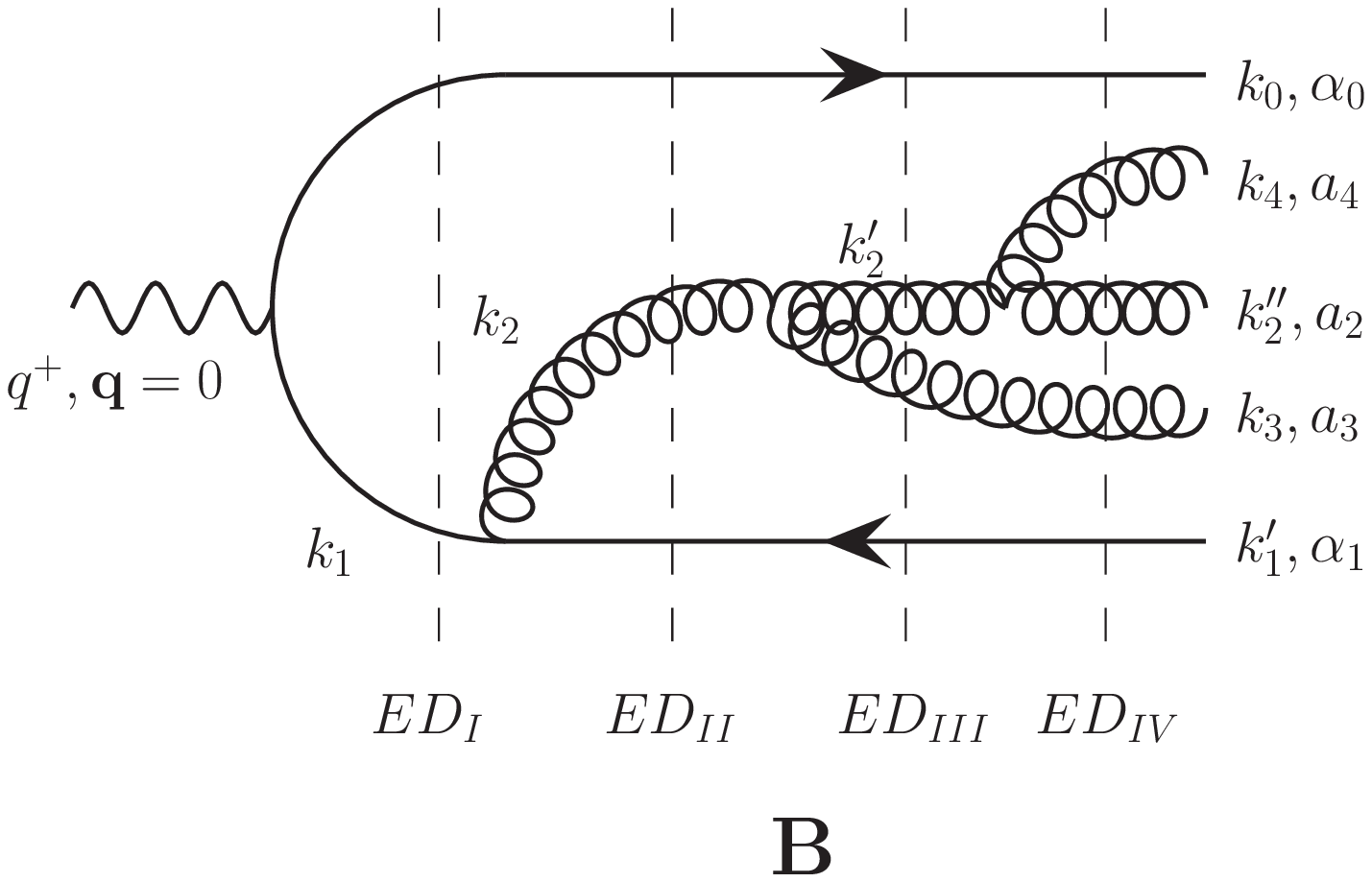}
}
\setbox2\hbox to 10cm{
 \includegraphics{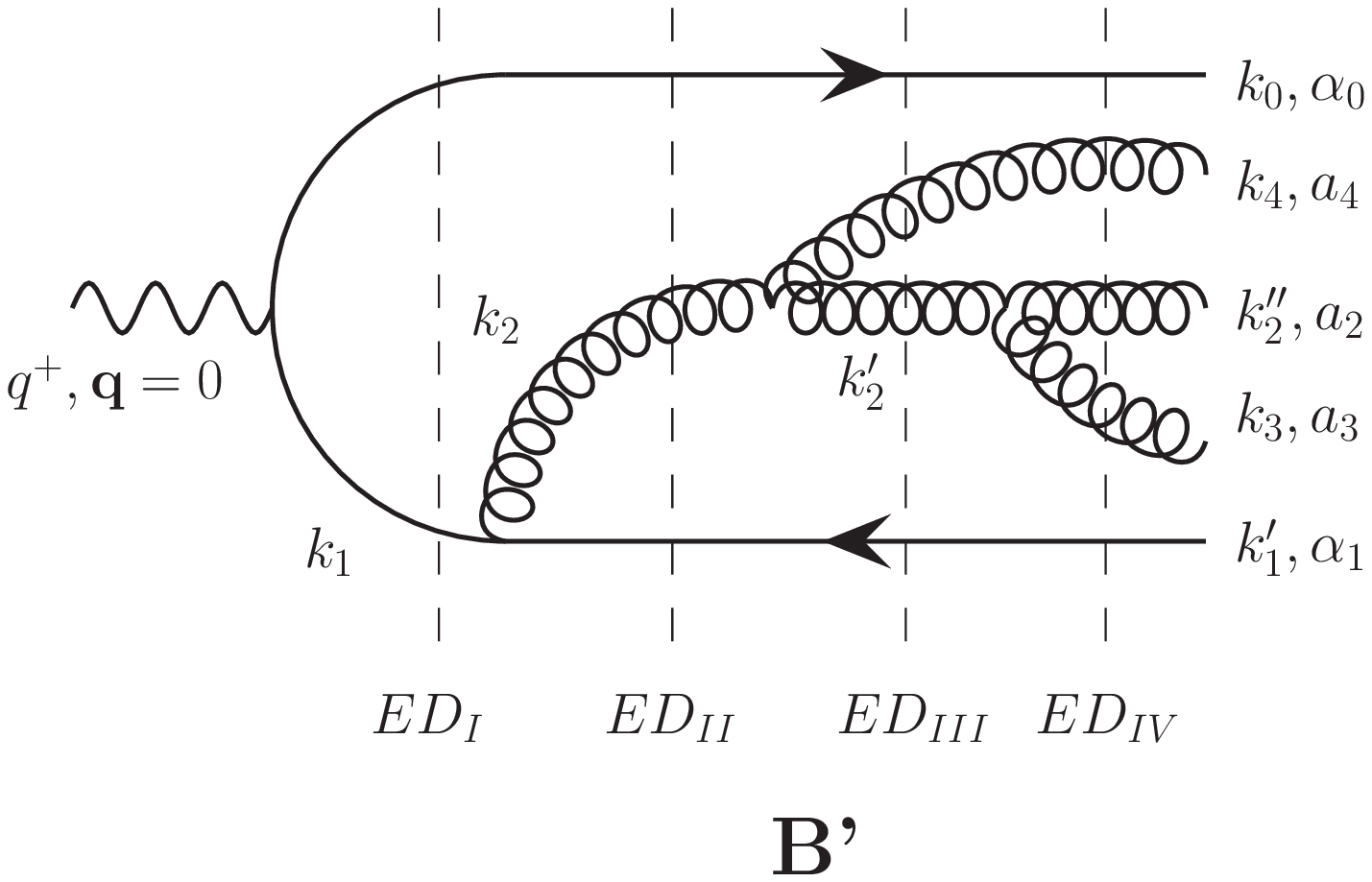}
}
\begin{center}
\hspace{-4cm}
\resizebox*{14cm}{!}{\mbox{\box1 \hspace{5cm} \box2}}
\caption{\label{Fig:qqbarggg_B} Examples of light-front perturbation theory diagrams with gluon splittings contributing to the $q\bar{q}ggg$ Fock component of a photon wave-function. These diagrams can be interpreted in the standard way, or as color-ordered contributions to the photon wave-function, with the color factor $[t^{a_4}t^{a_2}t^{a_3}]_{\alpha_0 \alpha_1}$.}
\end{center}
\end{figure}

Diagrams which include $3$-gluons vertices, like the diagrams \textbf{B} and \textbf{B'} shown on Fig. \ref{Fig:qqbarggg_B}, bring an extra complication. Indeed, we would like to consider each gluon as emitted from a color singlet dipole. However, using the usual expression for the $3$-gluons vertices, there is an ambiguity in associating the gluon emission to one or the other of the dipoles delimitated by the parent gluon. In the original construction of the dipole model \cite{Mueller:1993rr}, that ambiguity disappears when considering the squared wave-function and taking the large $N_c$ limit.

By contrast, one can also resolve that ambiguity before taking the square, by splitting the wave-functions into their color-ordered components. This is the analog for the wave-functions of the color-ordered amplitudes \cite{Berends:1987cv,Mangano:1987xk,Mangano:1988kk} (for a recent pedagogical introduction, see ref. \cite{Dixon:2013uaa}). The idea is the following: using the commutation relation
\begin{equation}
[t^a,t^b]=i f^{abc} t^c\, ,
\end{equation}
one can rewrite the color factor of the diagram \textbf{B} as
\begin{equation}
(-i) f^{a_4 a_2 b}\, (-i) f^{a_3 b c}  [t^c]_{\alpha_0 \alpha_1}
=-[t^{a_4}t^{a_2}t^{a_3}]_{\alpha_0 \alpha_1} -[t^{a_3}t^{a_2}t^{a_4}]_{\alpha_0 \alpha_1}
+[t^{a_3}t^{a_4}t^{a_2}]_{\alpha_0 \alpha_1}+[t^{a_2}t^{a_4}t^{a_3}]_{\alpha_0 \alpha_1}
\label{color_fact_B}
\end{equation}
and the one of the diagram \textbf{B'} as
\begin{equation}
(-i) f^{a_3 a_2 b}\, (-i) f^{a_4 b c}  [t^c]_{\alpha_0 \alpha_1}
=-[t^{a_4}t^{a_2}t^{a_3}]_{\alpha_0 \alpha_1} -[t^{a_3}t^{a_2}t^{a_4}]_{\alpha_0 \alpha_1}
+[t^{a_4}t^{a_3}t^{a_2}]_{\alpha_0 \alpha_1}+[t^{a_2}t^{a_3}t^{a_4}]_{\alpha_0 \alpha_1}
\, .\label{color_fact_B_prime}
\end{equation}
The first term in the expressions \eqref{color_fact_B} and \eqref{color_fact_B_prime}, $[t^{a_4}t^{a_2}t^{a_3}]_{\alpha_0 \alpha_1}$, is the same as the color factor of the diagrams \textbf{A} and \textbf{A'}. Only that term in the diagrams \textbf{B} and \textbf{B'} corresponds to the emission of the gluon $4$ by the dipole $02$ and of the gluon $3$ by the dipole $21$. Instead, the second term in the expressions \eqref{color_fact_B} and \eqref{color_fact_B_prime} is associated with the emission of the gluon $4$ by the dipole $21$ and of the gluon $3$ by the dipole $02$, and the other terms in the expressions \eqref{color_fact_B} and \eqref{color_fact_B_prime}, differing between the diagrams \textbf{B} and \textbf{B'}, correspond to the emission of both the gluons $3$ and $4$ on the same side. Hence, only the term $[t^{a_4}t^{a_2}t^{a_3}]_{\alpha_0 \alpha_1}$ is relevant for our purposes, whereas the other ones, contributing to distinct dipole cascades, can be discarded.

Concerning the kinematics, the only difference in the momentum flow of the diagrams \textbf{B} and \textbf{B'} is the momentum ${k_{2}'}$, which is constrained to take different values. The energy denominators $I$, $II$ and $IV$ are the same for the diagrams \textbf{B} and \textbf{B'}, whereas
\begin{eqnarray}
& & ED_{III}^{\textbf{B}}= -\frac{Q^2}{2 q^+}- k^-_0 - {k_{1}'}^- - \frac{(\mathbf{k}_2\!-\!\mathbf{k}_3)^2}{2\, ({k_2}^+\!-\!{k_3}^+)} - k^-_3 +i \epsilon  \label{ED_III_B}\\
& & ED_{III}^{\textbf{B'}}= -\frac{Q^2}{2 q^+}- {k_{0}}^- - {k_{1}'}^- - \frac{(\mathbf{k}_2\!-\!\mathbf{k}_4)^2}{2\, ({k_2}^+\!-\!{k_4}^+)} - k^-_4 +i \epsilon\, . \label{ED_III_B_prime}
\end{eqnarray}
Under the conditions \eqref{kplus_ord_A} and \eqref{kminus_ord_A}, the energy denominators simplify to
\begin{eqnarray}
ED_{II}^{\textbf{B}} & \simeq & - k^-_2 +i \epsilon \\
ED_{III}^{\textbf{B}} & \simeq & - k^-_3 +i \epsilon \\
ED_{III}^{\textbf{B'}} & \simeq & - k^-_4 +i \epsilon\\
ED_{IV}^{\textbf{B}} & \simeq & - k^-_3 - k^-_4 +i \epsilon
\, ,
\end{eqnarray}
so that the diagrams \textbf{B} and \textbf{B'} reduce to
\begin{eqnarray}
\textbf{B} &\simeq & \textrm{vertices} \times \frac{1}{ED_{I}^{\textbf{B}}}\; \frac{1}{[- k^-_2 +i \epsilon]}\; \frac{1}{[- k^-_3 +i \epsilon]}\; \frac{1}{[- k^-_3- k^-_4 +i \epsilon]}\label{B_approx}\\
\textbf{B'} &\simeq & \textrm{vertices} \times \frac{1}{ED_{I}^{\textbf{B}}}\; \frac{1}{[- k^-_2 +i \epsilon]}\; \frac{1}{[- k^-_4 +i \epsilon]}\; \frac{1}{[- k^-_3- k^-_4 +i \epsilon]}\label{B_prime_approx}
\, .
\end{eqnarray}
Neglecting longitudinal and transverse recoil effects and non-eikonal contributions due to the assumptions \eqref{kplus_ord_A} and \eqref{kminus_ord_A}, the vertices take identical values in the diagrams \textbf{B} and \textbf{B'}, up to the different color factors \eqref{color_fact_B} and \eqref{color_fact_B_prime} already discussed.

Hence, for the color-ordered versions of \textbf{B} and \textbf{B'} associated with the color-factor $[t^{a_4}t^{a_2}t^{a_3}]_{\alpha_0 \alpha_1}$, one has
\begin{eqnarray}
\bigg(\textbf{B}+\textbf{B'}\bigg)_{\textrm{color-ordered}} &\simeq & \textrm{vertices} \times \frac{1}{ED_{I}^{\textbf{B}}}\; \frac{1}{[- k^-_2 +i \epsilon]}\; \frac{1}{[- k^-_3 +i \epsilon]}\; \frac{1}{[- k^-_4 +i \epsilon]}\label{B+B_prime_approx}
\, .
\end{eqnarray}
This approximated expression for \textbf{B} and \textbf{B'}, leading eventually to one large log for each of the $3$ gluons, is valid under the assumptions \eqref{kplus_ord_A} and \eqref{kminus_ord_A}, no matter what is the relative size of $k^+_3$ and $k^+_4$ and of $k^-_3$ and $k^-_4$.

It is thus clear that, when considering properly color-ordered diagrams, the presence of gluon splitting vertices do not affect our discussion. Therefore, in a generic dipole cascade, the configurations contributing to the high-energy LL's are the ones where each gluon has a smaller $k^+$ and a larger $k^-$ than both partons delimitating the color-singlet dipole emitting that gluon, independently of what happens in the rest of the cascade. The kinematical constraint is then local instead of global, by contrast to the statement made in ref. \cite{Motyka:2009gi}.


\section{Basics of Laplace transform\label{App:Laplace}}

Consider a function ${ F}(Y)$ defined for $Y\in [0,+\infty[$. Its Laplace transform is defined by
\begin{equation}
\hat{F}(\om)= \int_0^{+\infty} \dd Y\;
e^{-\om\, Y}\, {F}(Y) \label{Laplace}\, .
\end{equation}
More precisely, the formula \eqref{Laplace} is being used when $\textrm{Re}(\om)$ is large enough to make the integral convergent, and then $\hat{F}(\om)$ is obtained in the rest of the complex plane by analytical continuation.
The Laplace transform is invertible, and the inverse formula is
\begin{equation}
{F}(Y)= \int_{\om_0\!-\!i\infty}^{\om_0\!+\!i\infty} \frac{\dd \om}{2\pi i}\;
e^{\om\, Y}\;  \hat{F}(\om)\label{Laplace_inv}\, .
\end{equation}
In \eqref{Laplace_inv}, $\om_0$ is a real number large enough so that the integration path passes on the right of all singularities of $\hat{F}(\om)$.

By moving the integration contour to the left in \eqref{Laplace_inv}, one picks progressively contributions from the singularities of $\hat{F}(\om)$, for example $r_s\: e^{\om_s\, Y}$ if $\hat{F}(\om)$ has a simple pole in $\om=\om_s$ with residue $r_s$. Hence, it is clear that the large $Y$ asymptotic behavior of ${F}(Y)$ is determined by the rightmost singularity of $\hat{F}(\om)$.

The properties of the Laplace transform with respect to derivation and integration are also needed in this paper. Let $f$ be the derivative of ${F}$: ${ f}(Y)=\d_Y {F}(Y)$. Then, its Laplace transform is
\begin{equation}
\hat{f}(\om)= \int_0^{+\infty} \dd Y\;
e^{-\om\, Y}\, \d_Y {F}(Y) = \om\, \hat{F}(\om) - {F}(0) \label{Laplace_deriv}\, .
\end{equation}

Let ${G}(Y)$ be the primitive
\begin{equation}
{G}(Y)= \int_0^{Y} \dd y\;
 {F}(y) \label{primitive}\, .
\end{equation}
Then, its Laplace transform is
\begin{equation}
\hat{G}(\om)= \int_0^{+\infty} \dd y\; {F}(y) \int_y^{+\infty} \dd Y\;
e^{-\om\, Y} = \frac{\hat{F}(\om)}{\om}\label{Laplace_int}\, .
\end{equation}


\section{Basics of Mellin representation\label{App:Mellin}}

Consider a dimensionless function ${\mathbf F}_{ij}$, which depends explicitly on the distance $x_{ij}$ between the points ${\mathbf x}_{i}$ and ${\mathbf x}_{j}$ in the transverse plane, but also implicitly on a reference scale $Q_0$. Then, one can use the Mellin representation
\begin{equation}
{\mathbf F}_{ij}= \int_{1/2-i\infty}^{1/2+i\infty} \frac{\dd \g}{2\pi i}\;
\left(\frac{x_{ij}^2 \, Q_0^2}{4}\right)^\g \; {\cal F}(\g)\label{Mellin_rep}\, .
\end{equation}
Since $x_{ij}$ can be both larger or smaller than $2/Q_0$, or equivalently $\log(x_{ij}^2 \, Q_0^2/4)$ can be both positive or negative, there is no inverse formula for \eqref{Mellin_rep} which would be the analog of \eqref{Laplace}, and generically ${\cal F}(\g)$ has singularities on both sides of the integration path.

When $x_{ij}\rightarrow +\infty$, it is convenient to move the integration path to the left in order to make the integrand small. By doing so, one picks progressively contributions from the singularities of ${\cal F}(\g)$ located on the left of the initial integration path. Hence, by analogy with the Laplace transform case, the behavior of ${\mathbf F}_{ij}$ in the limit $x_{ij}\rightarrow +\infty$ is determined by the first singularity on the left of the line $\textrm{Re}(\g)=1/2$.

By symmetry, the behavior of ${\mathbf F}_{ij}$ in the limit $x_{ij}\rightarrow 0$ is determined by the first singularity of ${\cal F}(\g)$ on the right of the line $\textrm{Re}(\g)=1/2$.


\section{Mellin space analysis of the NLO DIS impact factors with the operator evaluated at $Y^+_f$ or at $0$\label{App:Yfplus}}

In the section \ref{sec:Mellin_NLO_IF}, the behavior of the NLO DIS impact factors and of the counter-term for the naive LL evolution in the dilute regime in the domain ${x}_{10}^2\ll {x}_{20}^2 \simeq {x}_{21}^2$ has been studied in Mellin space, evaluating the expectation value $\left\langle {\mathbf S}_{012} \right\rangle$ at the scale $Y_2^+$.
However, it might also be natural to take that expectation value at the scale $Y^+_f$. In the discussion of the naive resummation of high-energy LL's in the section \ref{sec:std_subtr_LL}, the un-evolved expectation value $\left\langle {\mathbf S}_{012} \right\rangle_0$ has also been considered.
Those two other prescriptions deserve further study.
It is easy to deal at once with a whole range of prescriptions for $\left\langle {\mathbf S}_{012} \right\rangle$ including $Y^+_f$ and $0$, \emph{i.e.} choosing a generic positive or zero constant $Y_c^+$, independent of both $Y_2^+$ and $z_2$. In that case, the operator expectation value factors out of the integration in $Y_2^+$ or $z_2$. It is then not helpful to take the Laplace transform from $Y^+$ space to $\om$ space, which would transform an ordinary product into a convolution product. Revisiting the calculations of the section \ref{sec:Mellin_NLO_IF}, but in $(\g,Y^+)$ space and with constant $Y_c^+$, one obtains the following.

The contribution to the counter-term \eqref{LL_ct_real} from the domain ${x}_{10}^2\ll {x}_{20}^2 \simeq {x}_{21}^2$ writes in the dilute regime
\begin{eqnarray}
\delta\!\left\langle {\mathbf S}_{01} \right\rangle_{Y_f^+}\bigg|_{\textrm{LL, real}}^{{x}_{10}^2\ll {x}_{20}^2 \simeq {x}_{21}^2}
&=&
\bar{\alpha} \int_{0}^{Y_f^+} \!\!\!\!\dd Y_2^+
\int_{{x}_{10}^2\ll {x}_{20}^2} \frac{\textrm{d}^2\mathbf{x}_{2}}{2\pi}\; \textbf{K}_{012}\: \frac{2\, C_F}{N_c}\: \Big[1- \left\langle {\mathbf S}_{012} \right\rangle_{Y_c^+}\Big]
\nonumber\\
&\simeq & \int_{1/2-i\infty}^{1/2+i\infty} \frac{\dd \g}{2\pi i}\, \left(\frac{x_{01}^2 \, Q_0^2}{4}\right)^\g\;  {\cal N}(\g,Y_c^+)\;\;
  \frac{\bar{\alpha}\: Y_f^+}{(1\!-\! \g)}\, .
\label{Mellin_naive_subtract_Yc}
\end{eqnarray}
This is indeed the direct analog of the result \eqref{Mellin_naive_subtract}, given the correspondence $Y_f^+ \leftrightarrow 1/\om$. The expression \eqref{Mellin_naive_subtract_Yc} shows the same type of DLL behavior as \eqref{Mellin_naive_subtract}, which is not the correct collinear DLL due to the presence of $Y_f^+$ instead of $Y_f^-$.

The contribution from the region $z_f\, {x}_{10}^2\gg z_2\, {x}_{20}^2 \simeq z_2\, {x}_{21}^2$ to the approximate Mellin representation of the real NLO correction (still restricted to $z_2 \ll z_f$ and ${x}_{10}^2\ll {x}_{20}^2 \simeq {x}_{21}^2$), analog to \eqref{Mellin_NLO real_kc_reg}, writes
\begin{eqnarray}
& &\left.\bar{\alpha} \int_{z_{\min}}^{z_f}\frac{\textrm{d}z_2}{z_2}\; \int \frac{\textrm{d}^2\mathbf{x}_{2}}{2\pi}\;
\frac{\mathcal{I}_{T,L}^{NLO}}{\mathcal{I}_{T,L}^{LO}} \; \frac{2\, C_F}{N_c}\: \Big[1- \left\langle {\mathbf S}_{012} \right\rangle_{Y_c^+}\Big]\right|_{{x}_{10}^2\ll {x}_{20}^2 \simeq {x}_{21}^2 \textrm{ and }z_f\, {x}_{10}^2\gg z_2\, {x}_{20}^2}\nonumber\\
& & \quad \simeq \int_{1/2-i\infty}^{1/2+i\infty} \frac{\dd \g}{2\pi i}\, \left(\frac{x_{01}^2 \, Q_0^2}{4}\right)^\g\;\;{\cal N}(\g,Y_c^+) \;    \frac{\bar{\alpha}}{(1\!-\! \g)}  \int_{0}^{Y_f^+}\!\!\textrm{d}Y_2^+\;
\left(1\!-\!e^{-(1\!-\!\g)(Y^+_f\!-\!Y^+_2)} \right)\nonumber\\
& & \quad \simeq \int_{1/2-i\infty}^{1/2+i\infty} \frac{\dd \g}{2\pi i}\, \left(\frac{x_{01}^2 \, Q_0^2}{4}\right)^\g\;\;{\cal N}(\g,Y_c^+) \;    \frac{\bar{\alpha}}{(1\!-\! \g)^2}  \;
\left[e^{-(1\!-\!\g)Y^+_f}\!-\!1 + (1\!-\!\g)Y^+_f\right]\, .
\label{Mellin_NLO real_kc_reg_Yc}
\end{eqnarray}

In the longitudinal photon case, the contribution from the region $z_f\, {x}_{10}^2\ll z_2\, {x}_{20}^2 \simeq z_2\, {x}_{21}^2$, analog to \eqref{Mellin_NLO real_L_non-kc}, is now
\begin{eqnarray}
& &\left.\bar{\alpha} \int_{z_{\min}}^{z_f}\frac{\textrm{d}z_2}{z_2}\; \int \frac{\textrm{d}^2\mathbf{x}_{2}}{2\pi}\;
\frac{\mathcal{I}_{L}^{NLO}}{\mathcal{I}_{L}^{LO}} \; \frac{2\, C_F}{N_c}\: \Big[1- \left\langle {\mathbf S}_{012} \right\rangle_{Y_c^+}\Big]\right|_{z_f\, {x}_{01}^2\ll z_2\, {x}_{02}^2\simeq z_2\, {x}_{21}^2 }\nonumber\\
& & \quad \simeq \bar{\alpha} \int_{1/2-i\infty}^{1/2+i\infty} \frac{\dd \g}{2\pi i}\, \left(\frac{x_{01}^2 \, Q_0^2}{4}\right)^\g\; {\cal N}(\g,Y_c^+)\;  f_0(\g\!-\!2\, ,z_f {x}_{01}^2 Q^2)\;
\int_{0}^{Y_f^+}\!\!\textrm{d}Y_2^+\; e^{-(1\!-\!\g)(Y^+_f\!-\!Y^+_2)}\nonumber\\
& &\quad \simeq \int_{1/2-i\infty}^{1/2+i\infty} \frac{\dd \g}{2\pi i}\, \left(\frac{x_{01}^2 \, Q_0^2}{4}\right)^\g\;\;{\cal N}(\g,Y_c^+) \;  f_0(\g\!-\!2\, ,z_f {x}_{01}^2 Q^2)\;   \frac{\bar{\alpha}}{(1\!-\! \g)}  \;
\left[1\!-\!e^{-(1\!-\!\g)Y^+_f}\right]
\, .\label{Mellin_NLO real_L_non-kc_Yc}
\end{eqnarray}

In the transverse photon case, the contribution from the intermediate region $z_f {x}_{10}^2/z_2 \ll {x}_{20}^2 \simeq {x}_{21}^2 \ll z_f^2 {x}_{10}^2/z_2^2$, analog to \eqref{Mellin_NLO real_T_non-kc_recoilless}, becomes
\begin{eqnarray}
& &\left.\bar{\alpha} \int_{z_{\min}}^{z_f}\frac{\textrm{d}z_2}{z_2}\; \int \frac{\textrm{d}^2\mathbf{x}_{2}}{2\pi}\;
\frac{\mathcal{I}_{T}^{NLO}}{\mathcal{I}_{T}^{LO}} \; \frac{2\, C_F}{N_c}\: \Big[1- \left\langle {\mathbf S}_{012} \right\rangle_{Y_c^+}\Big]\right|_{z_f {x}_{10}^2/z_2 \ll {x}_{20}^2 \simeq {x}_{21}^2 \ll z_f^2 {x}_{10}^2/z_2^2}\nonumber\\
& & \quad \simeq \bar{\alpha} \int_{1/2-i\infty}^{1/2+i\infty} \frac{\dd \g}{2\pi i}\, \left(\frac{x_{01}^2 \, Q_0^2}{4}\right)^\g\; {\cal N}(\g,Y_c^+) \; \int_{0}^{Y_f^+}\!\!\textrm{d}Y_2^+\; e^{-(1\!-\!\g)(Y^+_f\!-\!Y^+_2)}
\int_{1}^{e^{(Y^+_f\!-\!Y^+_2)}} \!\!\!\!\!\!\!\! \!\!\!\!\!\!\!\!\textrm{d}u\;\;\;  u^{\g-3}\;   \frac{\textrm{K}_{1}^2\!\left(Q\sqrt{z_f\, {x}_{01}^2} \sqrt{u}\right)}{\textrm{K}_{1}^2\!\left(Q\sqrt{z_f\, {x}_{01}^2}\right)}
  \nonumber\\
& & \quad \simeq \int_{1/2-i\infty}^{1/2+i\infty} \frac{\dd \g}{2\pi i}\, \left(\frac{x_{01}^2 \, Q_0^2}{4}\right)^\g {\cal N}(\g,Y_c^+) \; \frac{\bar{\alpha}}{(1\!-\! \g)} \;
\int_{1}^{e^{Y^+_f}}\!\!\!\!\!\!\!\!\textrm{d}u\; u^{\g-3}  \left[u^{-(1\!-\!\g)}\!-\!e^{-(1\!-\!\g)Y^+_f}\right] \frac{\textrm{K}_{1}^2\!\left(Q\sqrt{z_f\, {x}_{01}^2} \sqrt{u}\right)}{\textrm{K}_{1}^2\!\left(Q\sqrt{z_f\, {x}_{01}^2}\right)}
\, .\label{Mellin_NLO real_T_non-kc_recoilless_Yc}
\end{eqnarray}

Finally, the contribution from the extreme region $z_f^2 {x}_{10}^2 \ll z_2^2 {x}_{20}^2 \simeq z_2^2 {x}_{21}^2$ in the transverse photon case, analog to \eqref{Mellin_NLO real_T_deep_recoil}, writes
\begin{eqnarray}
& &\left.\bar{\alpha} \int_{z_{\min}}^{z_f}\frac{\textrm{d}z_2}{z_2}\; \int \frac{\textrm{d}^2\mathbf{x}_{2}}{2\pi}\;
\frac{\mathcal{I}_{T}^{NLO}}{\mathcal{I}_{T}^{LO}} \; \frac{2\, C_F}{N_c}\: \Big[1- \left\langle {\mathbf S}_{012} \right\rangle_{Y_c^+}\Big]\right|_{z_f^2\, {x}_{01}^2\ll z_2^2\, {x}_{02}^2\simeq z_2^2\, {x}_{21}^2}\nonumber\\
& & \quad \simeq \frac{\bar{\alpha}}{2} \int_{1/2-i\infty}^{1/2+i\infty} \frac{\dd \g}{2\pi i}\, \left(\frac{x_{01}^2 \, Q_0^2}{4}\right)^\g\; {\cal N}(\g,Y_c^+)\; \int_{0}^{Y_f^+}\!\!\textrm{d}Y_2^+\; e^{-(1\!-\!\g)(Y^+_f\!-\!Y^+_2)}
\int^{+\infty}_{\exp{(Y^+_f\!-\!Y^+_2)}} \!\!\!\!\!\!\!\! \!\!\!\!\!\!\!\!\textrm{d}u\;\;\;  u^{\g-1}\;   \frac{\textrm{K}_{1}^2\!\left(Q\sqrt{z_f\, {x}_{01}^2} \sqrt{u}\right)}{\textrm{K}_{1}^2\!\left(Q\sqrt{z_f\, {x}_{01}^2}\right)}
  \nonumber\\
& & \quad \simeq \int_{1/2-i\infty}^{1/2+i\infty} \frac{\dd \g}{2\pi i}\, \left(\frac{x_{01}^2 \, Q_0^2}{4}\right)^\g\; {\cal N}(\g,Y_c^+)\;  \Bigg\{
\frac{\bar{\alpha}}{2(1\!-\! \g)} \;
\int_{1}^{e^{Y^+_f}}\!\!\!\!\!\!\!\!\textrm{d}u\; u^{\g-1}  \Big[1\!-\!u^{\g\!-\!1}\Big] \frac{\textrm{K}_{1}^2\!\left(Q\sqrt{z_f\, {x}_{01}^2} \sqrt{u}\right)}{\textrm{K}_{1}^2\!\left(Q\sqrt{z_f\, {x}_{01}^2}\right)}\nonumber\\
& & \qquad \qquad
+\frac{\bar{\alpha}}{2(1\!-\! \g)}\; \left[1\!-\!e^{-(1\!-\!\g)Y^+_f}\right] \; \int^{+\infty}_{e^{Y^+_f}}\!\!\!\textrm{d}u\; u^{\g-1}  \frac{\textrm{K}_{1}^2\!\left(Q\sqrt{z_f\, {x}_{01}^2} \sqrt{u}\right)}{\textrm{K}_{1}^2\!\left(Q\sqrt{z_f\, {x}_{01}^2}\right)} \Bigg\}
\, .\label{Mellin_NLO real_T_deep_recoil_Yc}
\end{eqnarray}

For any finite (possibly large) value of $Y_f^+$, all the potential singularities in $\g$ space appearing in the various contributions \eqref{Mellin_NLO real_kc_reg_Yc}, \eqref{Mellin_NLO real_L_non-kc_Yc}, \eqref{Mellin_NLO real_T_non-kc_recoilless_Yc} and \eqref{Mellin_NLO real_T_deep_recoil_Yc} to the real NLO corrections the the DIS structure functions, all located at $\g=1$, are actually removable. Hence, it seems that the real NLO corrections do not contain any DGLAP-like collinear logs if $\left\langle {\mathbf S}_{012} \right\rangle$ is evaluated at a scale $Y_{c}^+$ independent of the integration variables $Y_2^+$ and $z_2^+$. This is quite unexpected and worrisome.

However, the standard Regge limit is equivalent here to the $Y_f^+\rightarrow +\infty$ limit while keeping $\g$ fixed, in the range $0<\textrm{Re}(\g)<1$. That Regge limit amounts to make the replacements $e^{-(1\!-\!\g)Y^+_f}\rightarrow 0$ and $e^{Y^+_f}\rightarrow +\infty$ in the expressions \eqref{Mellin_NLO real_kc_reg_Yc}, \eqref{Mellin_NLO real_L_non-kc_Yc}, \eqref{Mellin_NLO real_T_non-kc_recoilless_Yc} and \eqref{Mellin_NLO real_T_deep_recoil_Yc}. Then, many of the removable singularities at $\g=1$ are transformed into poles, with a pattern reminiscent of the case of $\left\langle {\mathbf S}_{012} \right\rangle$ evaluated at $Y_2^+$.
\begin{itemize}
\item In the contribution from the domain $z_f\, {x}_{10}^2\gg z_2\, {x}_{20}^2 \simeq z_2\, {x}_{21}^2$ where the factorized approximation \eqref{Fact_IF_NLO_zero_z2} of $\mathcal{I}_{T,L}^{NLO}$ is valid, one obtains both a single and a double pole terms, which can be interpreted together as
\begin{equation}
\bar{\alpha}\Bigg[Y_f^+-\frac{1}{(1\!-\! \g)}\Bigg] \frac{1}{(1\!-\! \g)}\quad \leftrightarrow \quad \bar{\alpha}\, \bigg[Y_f^+ - \log\left(\frac{4}{x_{01}^2 \, Q_0^2}\right)\bigg]\, \log\left(\frac{4}{x_{01}^2 \, Q_0^2}\right)\simeq\bar{\alpha}\, Y_f^-\, \log\left(\frac{4}{x_{01}^2 \, Q_0^2}\right)
 \, ,
\end{equation}
which is the correct collinear DLL contribution.
\item In the transverse photon case, there is still no true Mellin space singularity, and thus no large logs coming from the domain $z_f^2 {x}_{10}^2 \ll z_2^2 {x}_{20}^2 \simeq z_2^2 {x}_{21}^2$ where transverse recoil effects are important.
\item The other contributions \eqref{Mellin_NLO real_L_non-kc_Yc} and \eqref{Mellin_NLO real_T_non-kc_recoilless_Yc} provide a single pole at $\g=1$ with a coefficient independent of $Y_f^+$. These correspond to single collinear logs, with no high-energy logs.
\end{itemize}

Hence, if one takes first the standard Regge limit and then the collinear limit, one obtains the correct collinear DLL contributions from the real NLO corrections to DIS, as with the $Y_2^+$ scale choice. However, the absence of singularities in $\g$ at finite $Y_f^+$ means that if one takes the collinear limit first and then the high-energy limit, one cannot obtain collinear DLL contributions but just high-energy logs. The lack of commutation of the collinear and high-energy limits prevent us to obtain a smooth interpolation between the BFKL/BK regime and the collinear DGLAP regime with the choices of scale $Y^+=0$ or $Y^+=Y_f^+$ for $\left\langle {\mathbf S}_{012} \right\rangle_{Y^+}$. Hence, the choice $Y^+=Y_2^+$ should be done instead in practice.







\bibliography{MaBiblioHEQCD}

\begin{thebibliography}{78}
\expandafter\ifx\csname natexlab\endcsname\relax\def\natexlab#1{#1}\fi
\expandafter\ifx\csname bibnamefont\endcsname\relax
  \def\bibnamefont#1{#1}\fi
\expandafter\ifx\csname bibfnamefont\endcsname\relax
  \def\bibfnamefont#1{#1}\fi
\expandafter\ifx\csname citenamefont\endcsname\relax
  \def\citenamefont#1{#1}\fi
\expandafter\ifx\csname url\endcsname\relax
  \def\url#1{\texttt{#1}}\fi
\expandafter\ifx\csname urlprefix\endcsname\relax\def\urlprefix{URL }\fi
\providecommand{\bibinfo}[2]{#2}
\providecommand{\eprint}[2][]{\url{#2}}

\bibitem[{\citenamefont{Lipatov}(1976)}]{Lipatov:1976zz}
\bibinfo{author}{\bibfnamefont{L.~N.} \bibnamefont{Lipatov}},
  \bibinfo{journal}{Sov. J. Nucl. Phys.} \textbf{\bibinfo{volume}{23}},
  \bibinfo{pages}{338} (\bibinfo{year}{1976}).

\bibitem[{\citenamefont{Kuraev et~al.}(1977)\citenamefont{Kuraev, Lipatov, and
  Fadin}}]{Kuraev:1977fs}
\bibinfo{author}{\bibfnamefont{E.~A.} \bibnamefont{Kuraev}},
  \bibinfo{author}{\bibfnamefont{L.~N.} \bibnamefont{Lipatov}},
  \bibnamefont{and} \bibinfo{author}{\bibfnamefont{V.~S.} \bibnamefont{Fadin}},
  \bibinfo{journal}{Sov. Phys. JETP} \textbf{\bibinfo{volume}{45}},
  \bibinfo{pages}{199} (\bibinfo{year}{1977}).

\bibitem[{\citenamefont{Balitsky and Lipatov}(1978)}]{Balitsky:1978ic}
\bibinfo{author}{\bibfnamefont{I.~I.} \bibnamefont{Balitsky}} \bibnamefont{and}
  \bibinfo{author}{\bibfnamefont{L.~N.} \bibnamefont{Lipatov}},
  \bibinfo{journal}{Sov. J. Nucl. Phys.} \textbf{\bibinfo{volume}{28}},
  \bibinfo{pages}{822} (\bibinfo{year}{1978}).

\bibitem[{\citenamefont{Jalilian-Marian
  et~al.}(1997)\citenamefont{Jalilian-Marian, Kovner, Leonidov, and
  Weigert}}]{Jalilian-Marian:1997jx}
\bibinfo{author}{\bibfnamefont{J.}~\bibnamefont{Jalilian-Marian}},
  \bibinfo{author}{\bibfnamefont{A.}~\bibnamefont{Kovner}},
  \bibinfo{author}{\bibfnamefont{A.}~\bibnamefont{Leonidov}}, \bibnamefont{and}
  \bibinfo{author}{\bibfnamefont{H.}~\bibnamefont{Weigert}},
  \bibinfo{journal}{Nucl. Phys.} \textbf{\bibinfo{volume}{B504}},
  \bibinfo{pages}{415} (\bibinfo{year}{1997}), \eprint{hep-ph/9701284}.

\bibitem[{\citenamefont{Jalilian-Marian
  et~al.}(1998{\natexlab{a}})\citenamefont{Jalilian-Marian, Kovner, Leonidov,
  and Weigert}}]{Jalilian-Marian:1997gr}
\bibinfo{author}{\bibfnamefont{J.}~\bibnamefont{Jalilian-Marian}},
  \bibinfo{author}{\bibfnamefont{A.}~\bibnamefont{Kovner}},
  \bibinfo{author}{\bibfnamefont{A.}~\bibnamefont{Leonidov}}, \bibnamefont{and}
  \bibinfo{author}{\bibfnamefont{H.}~\bibnamefont{Weigert}},
  \bibinfo{journal}{Phys. Rev.} \textbf{\bibinfo{volume}{D59}},
  \bibinfo{pages}{014014} (\bibinfo{year}{1998}{\natexlab{a}}),
  \eprint{hep-ph/9706377}.

\bibitem[{\citenamefont{Jalilian-Marian
  et~al.}(1998{\natexlab{b}})\citenamefont{Jalilian-Marian, Kovner, and
  Weigert}}]{Jalilian-Marian:1997dw}
\bibinfo{author}{\bibfnamefont{J.}~\bibnamefont{Jalilian-Marian}},
  \bibinfo{author}{\bibfnamefont{A.}~\bibnamefont{Kovner}}, \bibnamefont{and}
  \bibinfo{author}{\bibfnamefont{H.}~\bibnamefont{Weigert}},
  \bibinfo{journal}{Phys. Rev.} \textbf{\bibinfo{volume}{D59}},
  \bibinfo{pages}{014015} (\bibinfo{year}{1998}{\natexlab{b}}),
  \eprint{hep-ph/9709432}.

\bibitem[{\citenamefont{Kovner et~al.}(2000)\citenamefont{Kovner, Milhano, and
  Weigert}}]{Kovner:2000pt}
\bibinfo{author}{\bibfnamefont{A.}~\bibnamefont{Kovner}},
  \bibinfo{author}{\bibfnamefont{J.~G.} \bibnamefont{Milhano}},
  \bibnamefont{and} \bibinfo{author}{\bibfnamefont{H.}~\bibnamefont{Weigert}},
  \bibinfo{journal}{Phys. Rev.} \textbf{\bibinfo{volume}{D62}},
  \bibinfo{pages}{114005} (\bibinfo{year}{2000}), \eprint{hep-ph/0004014}.

\bibitem[{\citenamefont{Weigert}(2002)}]{Weigert:2000gi}
\bibinfo{author}{\bibfnamefont{H.}~\bibnamefont{Weigert}},
  \bibinfo{journal}{Nucl. Phys.} \textbf{\bibinfo{volume}{A703}},
  \bibinfo{pages}{823} (\bibinfo{year}{2002}), \eprint{hep-ph/0004044}.

\bibitem[{\citenamefont{Iancu et~al.}(2001{\natexlab{a}})\citenamefont{Iancu,
  Leonidov, and McLerran}}]{Iancu:2000hn}
\bibinfo{author}{\bibfnamefont{E.}~\bibnamefont{Iancu}},
  \bibinfo{author}{\bibfnamefont{A.}~\bibnamefont{Leonidov}}, \bibnamefont{and}
  \bibinfo{author}{\bibfnamefont{L.~D.} \bibnamefont{McLerran}},
  \bibinfo{journal}{Nucl. Phys.} \textbf{\bibinfo{volume}{A692}},
  \bibinfo{pages}{583} (\bibinfo{year}{2001}{\natexlab{a}}),
  \eprint{hep-ph/0011241}.

\bibitem[{\citenamefont{Iancu et~al.}(2001{\natexlab{b}})\citenamefont{Iancu,
  Leonidov, and McLerran}}]{Iancu:2001ad}
\bibinfo{author}{\bibfnamefont{E.}~\bibnamefont{Iancu}},
  \bibinfo{author}{\bibfnamefont{A.}~\bibnamefont{Leonidov}}, \bibnamefont{and}
  \bibinfo{author}{\bibfnamefont{L.~D.} \bibnamefont{McLerran}},
  \bibinfo{journal}{Phys. Lett.} \textbf{\bibinfo{volume}{B510}},
  \bibinfo{pages}{133} (\bibinfo{year}{2001}{\natexlab{b}}),
  \eprint{hep-ph/0102009}.

\bibitem[{\citenamefont{Ferreiro et~al.}(2002)\citenamefont{Ferreiro, Iancu,
  Leonidov, and McLerran}}]{Ferreiro:2001qy}
\bibinfo{author}{\bibfnamefont{E.}~\bibnamefont{Ferreiro}},
  \bibinfo{author}{\bibfnamefont{E.}~\bibnamefont{Iancu}},
  \bibinfo{author}{\bibfnamefont{A.}~\bibnamefont{Leonidov}}, \bibnamefont{and}
  \bibinfo{author}{\bibfnamefont{L.}~\bibnamefont{McLerran}},
  \bibinfo{journal}{Nucl. Phys.} \textbf{\bibinfo{volume}{A703}},
  \bibinfo{pages}{489} (\bibinfo{year}{2002}), \eprint{hep-ph/0109115}.

\bibitem[{\citenamefont{Balitsky}(1996)}]{Balitsky:1995ub}
\bibinfo{author}{\bibfnamefont{I.}~\bibnamefont{Balitsky}},
  \bibinfo{journal}{Nucl. Phys.} \textbf{\bibinfo{volume}{B463}},
  \bibinfo{pages}{99} (\bibinfo{year}{1996}), \eprint{hep-ph/9509348}.

\bibitem[{\citenamefont{Gribov et~al.}(1983)\citenamefont{Gribov, Levin, and
  Ryskin}}]{Gribov:1984tu}
\bibinfo{author}{\bibfnamefont{L.~V.} \bibnamefont{Gribov}},
  \bibinfo{author}{\bibfnamefont{E.~M.} \bibnamefont{Levin}}, \bibnamefont{and}
  \bibinfo{author}{\bibfnamefont{M.~G.} \bibnamefont{Ryskin}},
  \bibinfo{journal}{Phys. Rept.} \textbf{\bibinfo{volume}{100}},
  \bibinfo{pages}{1} (\bibinfo{year}{1983}).

\bibitem[{\citenamefont{Mueller and Qiu}(1986)}]{Mueller:1985wy}
\bibinfo{author}{\bibfnamefont{A.~H.} \bibnamefont{Mueller}} \bibnamefont{and}
  \bibinfo{author}{\bibfnamefont{J.-w.} \bibnamefont{Qiu}},
  \bibinfo{journal}{Nucl. Phys.} \textbf{\bibinfo{volume}{B268}},
  \bibinfo{pages}{427} (\bibinfo{year}{1986}).

\bibitem[{\citenamefont{McLerran and
  Venugopalan}(1994{\natexlab{a}})}]{McLerran:1993ni}
\bibinfo{author}{\bibfnamefont{L.~D.} \bibnamefont{McLerran}} \bibnamefont{and}
  \bibinfo{author}{\bibfnamefont{R.}~\bibnamefont{Venugopalan}},
  \bibinfo{journal}{Phys. Rev.} \textbf{\bibinfo{volume}{D49}},
  \bibinfo{pages}{2233} (\bibinfo{year}{1994}{\natexlab{a}}),
  \eprint{hep-ph/9309289}.

\bibitem[{\citenamefont{McLerran and
  Venugopalan}(1994{\natexlab{b}})}]{McLerran:1993ka}
\bibinfo{author}{\bibfnamefont{L.~D.} \bibnamefont{McLerran}} \bibnamefont{and}
  \bibinfo{author}{\bibfnamefont{R.}~\bibnamefont{Venugopalan}},
  \bibinfo{journal}{Phys. Rev.} \textbf{\bibinfo{volume}{D49}},
  \bibinfo{pages}{3352} (\bibinfo{year}{1994}{\natexlab{b}}),
  \eprint{hep-ph/9311205}.

\bibitem[{\citenamefont{McLerran and
  Venugopalan}(1994{\natexlab{c}})}]{McLerran:1994vd}
\bibinfo{author}{\bibfnamefont{L.~D.} \bibnamefont{McLerran}} \bibnamefont{and}
  \bibinfo{author}{\bibfnamefont{R.}~\bibnamefont{Venugopalan}},
  \bibinfo{journal}{Phys. Rev.} \textbf{\bibinfo{volume}{D50}},
  \bibinfo{pages}{2225} (\bibinfo{year}{1994}{\natexlab{c}}),
  \eprint{hep-ph/9402335}.

\bibitem[{\citenamefont{Gelis et~al.}(2008)\citenamefont{Gelis, Lappi, and
  Venugopalan}}]{Gelis:2008rw}
\bibinfo{author}{\bibfnamefont{F.}~\bibnamefont{Gelis}},
  \bibinfo{author}{\bibfnamefont{T.}~\bibnamefont{Lappi}}, \bibnamefont{and}
  \bibinfo{author}{\bibfnamefont{R.}~\bibnamefont{Venugopalan}},
  \bibinfo{journal}{Phys. Rev.} \textbf{\bibinfo{volume}{D78}},
  \bibinfo{pages}{054019} (\bibinfo{year}{2008}), \eprint{0804.2630}.

\bibitem[{\citenamefont{Kovchegov}(1999)}]{Kovchegov:1999yj}
\bibinfo{author}{\bibfnamefont{Y.~V.} \bibnamefont{Kovchegov}},
  \bibinfo{journal}{Phys. Rev.} \textbf{\bibinfo{volume}{D60}},
  \bibinfo{pages}{034008} (\bibinfo{year}{1999}), \eprint{hep-ph/9901281}.

\bibitem[{\citenamefont{Kovchegov}(2000)}]{Kovchegov:1999ua}
\bibinfo{author}{\bibfnamefont{Y.~V.} \bibnamefont{Kovchegov}},
  \bibinfo{journal}{Phys. Rev.} \textbf{\bibinfo{volume}{D61}},
  \bibinfo{pages}{074018} (\bibinfo{year}{2000}), \eprint{hep-ph/9905214}.

\bibitem[{\citenamefont{Ciafaloni}(1988)}]{Ciafaloni:1987ur}
\bibinfo{author}{\bibfnamefont{M.}~\bibnamefont{Ciafaloni}},
  \bibinfo{journal}{Nucl.Phys.} \textbf{\bibinfo{volume}{B296}},
  \bibinfo{pages}{49} (\bibinfo{year}{1988}).

\bibitem[{\citenamefont{Catani et~al.}(1990{\natexlab{a}})\citenamefont{Catani,
  Fiorani, and Marchesini}}]{Catani:1989sg}
\bibinfo{author}{\bibfnamefont{S.}~\bibnamefont{Catani}},
  \bibinfo{author}{\bibfnamefont{F.}~\bibnamefont{Fiorani}}, \bibnamefont{and}
  \bibinfo{author}{\bibfnamefont{G.}~\bibnamefont{Marchesini}},
  \bibinfo{journal}{Nucl.Phys.} \textbf{\bibinfo{volume}{B336}},
  \bibinfo{pages}{18} (\bibinfo{year}{1990}{\natexlab{a}}).

\bibitem[{\citenamefont{Catani et~al.}(1990{\natexlab{b}})\citenamefont{Catani,
  Fiorani, and Marchesini}}]{Catani:1989yc}
\bibinfo{author}{\bibfnamefont{S.}~\bibnamefont{Catani}},
  \bibinfo{author}{\bibfnamefont{F.}~\bibnamefont{Fiorani}}, \bibnamefont{and}
  \bibinfo{author}{\bibfnamefont{G.}~\bibnamefont{Marchesini}},
  \bibinfo{journal}{Phys.Lett.} \textbf{\bibinfo{volume}{B234}},
  \bibinfo{pages}{339} (\bibinfo{year}{1990}{\natexlab{b}}).

\bibitem[{\citenamefont{Andersson
  et~al.}(1996{\natexlab{a}})\citenamefont{Andersson, Gustafson, Kharraziha,
  and Samuelsson}}]{Andersson:1995jt}
\bibinfo{author}{\bibfnamefont{B.}~\bibnamefont{Andersson}},
  \bibinfo{author}{\bibfnamefont{G.}~\bibnamefont{Gustafson}},
  \bibinfo{author}{\bibfnamefont{H.}~\bibnamefont{Kharraziha}},
  \bibnamefont{and}
  \bibinfo{author}{\bibfnamefont{J.}~\bibnamefont{Samuelsson}},
  \bibinfo{journal}{Z.Phys.} \textbf{\bibinfo{volume}{C71}},
  \bibinfo{pages}{613} (\bibinfo{year}{1996}{\natexlab{a}}).

\bibitem[{\citenamefont{Andersson
  et~al.}(1996{\natexlab{b}})\citenamefont{Andersson, Gustafson, and
  Samuelsson}}]{Andersson:1995ju}
\bibinfo{author}{\bibfnamefont{B.}~\bibnamefont{Andersson}},
  \bibinfo{author}{\bibfnamefont{G.}~\bibnamefont{Gustafson}},
  \bibnamefont{and}
  \bibinfo{author}{\bibfnamefont{J.}~\bibnamefont{Samuelsson}},
  \bibinfo{journal}{Nucl.Phys.} \textbf{\bibinfo{volume}{B467}},
  \bibinfo{pages}{443} (\bibinfo{year}{1996}{\natexlab{b}}).

\bibitem[{\citenamefont{Kwiecinski et~al.}(1996)\citenamefont{Kwiecinski,
  Martin, and Sutton}}]{Kwiecinski:1996td}
\bibinfo{author}{\bibfnamefont{J.}~\bibnamefont{Kwiecinski}},
  \bibinfo{author}{\bibfnamefont{A.~D.} \bibnamefont{Martin}},
  \bibnamefont{and} \bibinfo{author}{\bibfnamefont{P.}~\bibnamefont{Sutton}},
  \bibinfo{journal}{Z.Phys.} \textbf{\bibinfo{volume}{C71}},
  \bibinfo{pages}{585} (\bibinfo{year}{1996}), \eprint{hep-ph/9602320}.

\bibitem[{\citenamefont{Avsar et~al.}(2005)\citenamefont{Avsar, Gustafson, and
  Lonnblad}}]{Avsar:2005iz}
\bibinfo{author}{\bibfnamefont{E.}~\bibnamefont{Avsar}},
  \bibinfo{author}{\bibfnamefont{G.}~\bibnamefont{Gustafson}},
  \bibnamefont{and} \bibinfo{author}{\bibfnamefont{L.}~\bibnamefont{Lonnblad}},
  \bibinfo{journal}{JHEP} \textbf{\bibinfo{volume}{0507}}, \bibinfo{pages}{062}
  (\bibinfo{year}{2005}), \eprint{hep-ph/0503181}.

\bibitem[{\citenamefont{Avsar et~al.}(2007{\natexlab{a}})\citenamefont{Avsar,
  Gustafson, and Lonnblad}}]{Avsar:2006jy}
\bibinfo{author}{\bibfnamefont{E.}~\bibnamefont{Avsar}},
  \bibinfo{author}{\bibfnamefont{G.}~\bibnamefont{Gustafson}},
  \bibnamefont{and} \bibinfo{author}{\bibfnamefont{L.}~\bibnamefont{Lonnblad}},
  \bibinfo{journal}{JHEP} \textbf{\bibinfo{volume}{0701}}, \bibinfo{pages}{012}
  (\bibinfo{year}{2007}{\natexlab{a}}), \eprint{hep-ph/0610157}.

\bibitem[{\citenamefont{Avsar et~al.}(2007{\natexlab{b}})\citenamefont{Avsar,
  Gustafson, and Lonnblad}}]{Avsar:2007xg}
\bibinfo{author}{\bibfnamefont{E.}~\bibnamefont{Avsar}},
  \bibinfo{author}{\bibfnamefont{G.}~\bibnamefont{Gustafson}},
  \bibnamefont{and} \bibinfo{author}{\bibfnamefont{L.}~\bibnamefont{Lonnblad}},
  \bibinfo{journal}{JHEP} \textbf{\bibinfo{volume}{0712}}, \bibinfo{pages}{012}
  (\bibinfo{year}{2007}{\natexlab{b}}), \eprint{0709.1368}.

\bibitem[{\citenamefont{Flensburg et~al.}(2009)\citenamefont{Flensburg,
  Gustafson, and Lonnblad}}]{Flensburg:2008ag}
\bibinfo{author}{\bibfnamefont{C.}~\bibnamefont{Flensburg}},
  \bibinfo{author}{\bibfnamefont{G.}~\bibnamefont{Gustafson}},
  \bibnamefont{and} \bibinfo{author}{\bibfnamefont{L.}~\bibnamefont{Lonnblad}},
  \bibinfo{journal}{Eur.Phys.J.} \textbf{\bibinfo{volume}{C60}},
  \bibinfo{pages}{233} (\bibinfo{year}{2009}), \eprint{0807.0325}.

\bibitem[{\citenamefont{Flensburg and Gustafson}(2010)}]{Flensburg:2010kq}
\bibinfo{author}{\bibfnamefont{C.}~\bibnamefont{Flensburg}} \bibnamefont{and}
  \bibinfo{author}{\bibfnamefont{G.}~\bibnamefont{Gustafson}},
  \bibinfo{journal}{JHEP} \textbf{\bibinfo{volume}{1010}}, \bibinfo{pages}{014}
  (\bibinfo{year}{2010}), \eprint{1004.5502}.

\bibitem[{\citenamefont{Flensburg et~al.}(2011)\citenamefont{Flensburg,
  Gustafson, and Lonnblad}}]{Flensburg:2011kk}
\bibinfo{author}{\bibfnamefont{C.}~\bibnamefont{Flensburg}},
  \bibinfo{author}{\bibfnamefont{G.}~\bibnamefont{Gustafson}},
  \bibnamefont{and} \bibinfo{author}{\bibfnamefont{L.}~\bibnamefont{Lonnblad}},
  \bibinfo{journal}{JHEP} \textbf{\bibinfo{volume}{1108}}, \bibinfo{pages}{103}
  (\bibinfo{year}{2011}), \eprint{1103.4321}.

\bibitem[{\citenamefont{Flensburg et~al.}(2012)\citenamefont{Flensburg,
  Gustafson, and Lönnblad}}]{Flensburg:2012zy}
\bibinfo{author}{\bibfnamefont{C.}~\bibnamefont{Flensburg}},
  \bibinfo{author}{\bibfnamefont{G.}~\bibnamefont{Gustafson}},
  \bibnamefont{and} \bibinfo{author}{\bibfnamefont{L.}~\bibnamefont{Lönnblad}},
  \bibinfo{journal}{JHEP} \textbf{\bibinfo{volume}{1212}}, \bibinfo{pages}{115}
  (\bibinfo{year}{2012}), \eprint{1210.2407}.

\bibitem[{\citenamefont{Fadin and Lipatov}(1998)}]{Fadin:1998py}
\bibinfo{author}{\bibfnamefont{V.~S.} \bibnamefont{Fadin}} \bibnamefont{and}
  \bibinfo{author}{\bibfnamefont{L.~N.} \bibnamefont{Lipatov}},
  \bibinfo{journal}{Phys. Lett.} \textbf{\bibinfo{volume}{B429}},
  \bibinfo{pages}{127} (\bibinfo{year}{1998}), \eprint{hep-ph/9802290}.

\bibitem[{\citenamefont{Ciafaloni and Camici}(1998)}]{Ciafaloni:1998gs}
\bibinfo{author}{\bibfnamefont{M.}~\bibnamefont{Ciafaloni}} \bibnamefont{and}
  \bibinfo{author}{\bibfnamefont{G.}~\bibnamefont{Camici}},
  \bibinfo{journal}{Phys. Lett.} \textbf{\bibinfo{volume}{B430}},
  \bibinfo{pages}{349} (\bibinfo{year}{1998}), \eprint{hep-ph/9803389}.

\bibitem[{\citenamefont{Salam}(1998)}]{Salam:1998tj}
\bibinfo{author}{\bibfnamefont{G.~P.} \bibnamefont{Salam}},
  \bibinfo{journal}{JHEP} \textbf{\bibinfo{volume}{07}}, \bibinfo{pages}{019}
  (\bibinfo{year}{1998}), \eprint{hep-ph/9806482}.

\bibitem[{\citenamefont{Ciafaloni et~al.}(1999)\citenamefont{Ciafaloni,
  Colferai, and Salam}}]{Ciafaloni:1999yw}
\bibinfo{author}{\bibfnamefont{M.}~\bibnamefont{Ciafaloni}},
  \bibinfo{author}{\bibfnamefont{D.}~\bibnamefont{Colferai}}, \bibnamefont{and}
  \bibinfo{author}{\bibfnamefont{G.~P.} \bibnamefont{Salam}},
  \bibinfo{journal}{Phys. Rev.} \textbf{\bibinfo{volume}{D60}},
  \bibinfo{pages}{114036} (\bibinfo{year}{1999}), \eprint{hep-ph/9905566}.

\bibitem[{\citenamefont{Altarelli et~al.}(2000)\citenamefont{Altarelli, Ball,
  and Forte}}]{Altarelli:1999vw}
\bibinfo{author}{\bibfnamefont{G.}~\bibnamefont{Altarelli}},
  \bibinfo{author}{\bibfnamefont{R.~D.} \bibnamefont{Ball}}, \bibnamefont{and}
  \bibinfo{author}{\bibfnamefont{S.}~\bibnamefont{Forte}},
  \bibinfo{journal}{Nucl. Phys.} \textbf{\bibinfo{volume}{B575}},
  \bibinfo{pages}{313} (\bibinfo{year}{2000}), \eprint{hep-ph/9911273}.

\bibitem[{\citenamefont{Ciafaloni et~al.}(2003)\citenamefont{Ciafaloni,
  Colferai, Salam, and Stasto}}]{Ciafaloni:2003rd}
\bibinfo{author}{\bibfnamefont{M.}~\bibnamefont{Ciafaloni}},
  \bibinfo{author}{\bibfnamefont{D.}~\bibnamefont{Colferai}},
  \bibinfo{author}{\bibfnamefont{G.~P.} \bibnamefont{Salam}}, \bibnamefont{and}
  \bibinfo{author}{\bibfnamefont{A.~M.} \bibnamefont{Stasto}},
  \bibinfo{journal}{Phys. Rev.} \textbf{\bibinfo{volume}{D68}},
  \bibinfo{pages}{114003} (\bibinfo{year}{2003}), \eprint{hep-ph/0307188}.

\bibitem[{\citenamefont{Altarelli et~al.}(2006)\citenamefont{Altarelli, Ball,
  and Forte}}]{Altarelli:2005ni}
\bibinfo{author}{\bibfnamefont{G.}~\bibnamefont{Altarelli}},
  \bibinfo{author}{\bibfnamefont{R.~D.} \bibnamefont{Ball}}, \bibnamefont{and}
  \bibinfo{author}{\bibfnamefont{S.}~\bibnamefont{Forte}},
  \bibinfo{journal}{Nucl. Phys.} \textbf{\bibinfo{volume}{B742}},
  \bibinfo{pages}{1} (\bibinfo{year}{2006}), \eprint{hep-ph/0512237}.

\bibitem[{\citenamefont{Ciafaloni et~al.}(2007)\citenamefont{Ciafaloni,
  Colferai, Salam, and Stasto}}]{Ciafaloni:2007gf}
\bibinfo{author}{\bibfnamefont{M.}~\bibnamefont{Ciafaloni}},
  \bibinfo{author}{\bibfnamefont{D.}~\bibnamefont{Colferai}},
  \bibinfo{author}{\bibfnamefont{G.}~\bibnamefont{Salam}}, \bibnamefont{and}
  \bibinfo{author}{\bibfnamefont{A.}~\bibnamefont{Stasto}},
  \bibinfo{journal}{JHEP} \textbf{\bibinfo{volume}{0708}}, \bibinfo{pages}{046}
  (\bibinfo{year}{2007}), \eprint{0707.1453}.

\bibitem[{\citenamefont{Altarelli et~al.}(2008)\citenamefont{Altarelli, Ball,
  and Forte}}]{Altarelli:2008aj}
\bibinfo{author}{\bibfnamefont{G.}~\bibnamefont{Altarelli}},
  \bibinfo{author}{\bibfnamefont{R.~D.} \bibnamefont{Ball}}, \bibnamefont{and}
  \bibinfo{author}{\bibfnamefont{S.}~\bibnamefont{Forte}},
  \bibinfo{journal}{Nucl.Phys.} \textbf{\bibinfo{volume}{B799}},
  \bibinfo{pages}{199} (\bibinfo{year}{2008}), \eprint{0802.0032}.

\bibitem[{\citenamefont{Balitsky and Chirilli}(2008)}]{Balitsky:2008zz}
\bibinfo{author}{\bibfnamefont{I.}~\bibnamefont{Balitsky}} \bibnamefont{and}
  \bibinfo{author}{\bibfnamefont{G.~A.} \bibnamefont{Chirilli}},
  \bibinfo{journal}{Phys. Rev.} \textbf{\bibinfo{volume}{D77}},
  \bibinfo{pages}{014019} (\bibinfo{year}{2008}), \eprint{arXiv:0710.4330
  [hep-ph]}.

\bibitem[{\citenamefont{Balitsky and Chirilli}(2009)}]{Balitsky:2009xg}
\bibinfo{author}{\bibfnamefont{I.}~\bibnamefont{Balitsky}} \bibnamefont{and}
  \bibinfo{author}{\bibfnamefont{G.~A.} \bibnamefont{Chirilli}},
  \bibinfo{journal}{Nucl.Phys.} \textbf{\bibinfo{volume}{B822}},
  \bibinfo{pages}{45} (\bibinfo{year}{2009}), \eprint{arXiv:0903.5326
  [hep-ph]}.

\bibitem[{\citenamefont{Balitsky and Chirilli}(2011)}]{Balitsky:2010ze}
\bibinfo{author}{\bibfnamefont{I.}~\bibnamefont{Balitsky}} \bibnamefont{and}
  \bibinfo{author}{\bibfnamefont{G.~A.} \bibnamefont{Chirilli}},
  \bibinfo{journal}{Phys.Rev.} \textbf{\bibinfo{volume}{D83}},
  \bibinfo{pages}{031502} (\bibinfo{year}{2011}), \eprint{1009.4729}.

\bibitem[{\citenamefont{Beuf}(2012)}]{Beuf:2011xd}
\bibinfo{author}{\bibfnamefont{G.}~\bibnamefont{Beuf}},
  \bibinfo{journal}{Phys.Rev.} \textbf{\bibinfo{volume}{D85}},
  \bibinfo{pages}{034039} (\bibinfo{year}{2012}), \eprint{1112.4501}.

\bibitem[{\citenamefont{Chirilli et~al.}(2011)\citenamefont{Chirilli, Xiao, and
  Yuan}}]{Chirilli:2011km}
\bibinfo{author}{\bibfnamefont{G.~A.} \bibnamefont{Chirilli}},
  \bibinfo{author}{\bibfnamefont{B.-W.} \bibnamefont{Xiao}}, \bibnamefont{and}
  \bibinfo{author}{\bibfnamefont{F.}~\bibnamefont{Yuan}}
  (\bibinfo{year}{2011}), \eprint{1112.1061}.

\bibitem[{\citenamefont{Chirilli et~al.}(2012)\citenamefont{Chirilli, Xiao, and
  Yuan}}]{Chirilli:2012jd}
\bibinfo{author}{\bibfnamefont{G.~A.} \bibnamefont{Chirilli}},
  \bibinfo{author}{\bibfnamefont{B.-W.} \bibnamefont{Xiao}}, \bibnamefont{and}
  \bibinfo{author}{\bibfnamefont{F.}~\bibnamefont{Yuan}},
  \bibinfo{journal}{Phys.Rev.} \textbf{\bibinfo{volume}{D86}},
  \bibinfo{pages}{054005} (\bibinfo{year}{2012}), \eprint{1203.6139}.

\bibitem[{\citenamefont{Balitsky and Chirilli}(2013)}]{Balitsky:2013fea}
\bibinfo{author}{\bibfnamefont{I.}~\bibnamefont{Balitsky}} \bibnamefont{and}
  \bibinfo{author}{\bibfnamefont{G.~A.} \bibnamefont{Chirilli}},
  \bibinfo{journal}{Phys.Rev.} \textbf{\bibinfo{volume}{D88}},
  \bibinfo{pages}{111501} (\bibinfo{year}{2013}), \eprint{1309.7644}.

\bibitem[{\citenamefont{Kovner et~al.}(2013)\citenamefont{Kovner, Lublinsky,
  and Mulian}}]{Kovner:2013ona}
\bibinfo{author}{\bibfnamefont{A.}~\bibnamefont{Kovner}},
  \bibinfo{author}{\bibfnamefont{M.}~\bibnamefont{Lublinsky}},
  \bibnamefont{and} \bibinfo{author}{\bibfnamefont{Y.}~\bibnamefont{Mulian}}
  (\bibinfo{year}{2013}), \eprint{1310.0378}.

\bibitem[{\citenamefont{Avsar et~al.}(2011)\citenamefont{Avsar, Stasto,
  Triantafyllopoulos, and Zaslavsky}}]{Avsar:2011ds}
\bibinfo{author}{\bibfnamefont{E.}~\bibnamefont{Avsar}},
  \bibinfo{author}{\bibfnamefont{A.}~\bibnamefont{Stasto}},
  \bibinfo{author}{\bibfnamefont{D.}~\bibnamefont{Triantafyllopoulos}},
  \bibnamefont{and}
  \bibinfo{author}{\bibfnamefont{D.}~\bibnamefont{Zaslavsky}},
  \bibinfo{journal}{JHEP} \textbf{\bibinfo{volume}{1110}}, \bibinfo{pages}{138}
  (\bibinfo{year}{2011}), \eprint{1107.1252}.

\bibitem[{\citenamefont{Motyka and Stasto}(2009)}]{Motyka:2009gi}
\bibinfo{author}{\bibfnamefont{L.}~\bibnamefont{Motyka}} \bibnamefont{and}
  \bibinfo{author}{\bibfnamefont{A.~M.} \bibnamefont{Stasto}},
  \bibinfo{journal}{Phys.Rev.} \textbf{\bibinfo{volume}{D79}},
  \bibinfo{pages}{085016} (\bibinfo{year}{2009}), \eprint{0901.4949}.

\bibitem[{\citenamefont{Mueller}(1994)}]{Mueller:1993rr}
\bibinfo{author}{\bibfnamefont{A.~H.} \bibnamefont{Mueller}},
  \bibinfo{journal}{Nucl. Phys.} \textbf{\bibinfo{volume}{B415}},
  \bibinfo{pages}{373} (\bibinfo{year}{1994}).

\bibitem[{\citenamefont{Mueller and Patel}(1994)}]{Mueller:1994jq}
\bibinfo{author}{\bibfnamefont{A.~H.} \bibnamefont{Mueller}} \bibnamefont{and}
  \bibinfo{author}{\bibfnamefont{B.}~\bibnamefont{Patel}},
  \bibinfo{journal}{Nucl. Phys.} \textbf{\bibinfo{volume}{B425}},
  \bibinfo{pages}{471} (\bibinfo{year}{1994}), \eprint{hep-ph/9403256}.

\bibitem[{\citenamefont{Jeon}(2013)}]{Jeon:2013zga}
\bibinfo{author}{\bibfnamefont{S.}~\bibnamefont{Jeon}} (\bibinfo{year}{2013}),
  \eprint{1308.0263}.

\bibitem[{\citenamefont{Collins}(2011)}]{Collins_TMD_book}
\bibinfo{author}{\bibfnamefont{J.~C.} \bibnamefont{Collins}},
  \emph{\bibinfo{title}{{Foundations of Perturbative QCD}}}
  (\bibinfo{publisher}{Cambridge University Press, Cambridge},
  \bibinfo{year}{2011}).

\bibitem[{\citenamefont{Kogut and Soper}(1970)}]{Kogut:1969xa}
\bibinfo{author}{\bibfnamefont{J.~B.} \bibnamefont{Kogut}} \bibnamefont{and}
  \bibinfo{author}{\bibfnamefont{D.~E.} \bibnamefont{Soper}},
  \bibinfo{journal}{Phys.Rev.} \textbf{\bibinfo{volume}{D1}},
  \bibinfo{pages}{2901} (\bibinfo{year}{1970}).

\bibitem[{\citenamefont{Bjorken et~al.}(1971)\citenamefont{Bjorken, Kogut, and
  Soper}}]{Bjorken:1970ah}
\bibinfo{author}{\bibfnamefont{J.}~\bibnamefont{Bjorken}},
  \bibinfo{author}{\bibfnamefont{J.~B.} \bibnamefont{Kogut}}, \bibnamefont{and}
  \bibinfo{author}{\bibfnamefont{D.~E.} \bibnamefont{Soper}},
  \bibinfo{journal}{Phys.Rev.} \textbf{\bibinfo{volume}{D3}},
  \bibinfo{pages}{1382} (\bibinfo{year}{1971}).

\bibitem[{\citenamefont{Lipatov}(1986)}]{Lipatov:1985uk}
\bibinfo{author}{\bibfnamefont{L.~N.} \bibnamefont{Lipatov}},
  \bibinfo{journal}{Sov. Phys. JETP} \textbf{\bibinfo{volume}{63}},
  \bibinfo{pages}{904} (\bibinfo{year}{1986}).

\bibitem[{\citenamefont{Chirilli and Kovchegov}(2013)}]{Chirilli:2013kca}
\bibinfo{author}{\bibfnamefont{G.~A.} \bibnamefont{Chirilli}} \bibnamefont{and}
  \bibinfo{author}{\bibfnamefont{Y.~V.} \bibnamefont{Kovchegov}},
  \bibinfo{journal}{JHEP} \textbf{\bibinfo{volume}{1306}}, \bibinfo{pages}{055}
  (\bibinfo{year}{2013}), \eprint{1305.1924}.

\bibitem[{\citenamefont{Kotikov and Lipatov}(2000)}]{Kotikov:2000pm}
\bibinfo{author}{\bibfnamefont{A.}~\bibnamefont{Kotikov}} \bibnamefont{and}
  \bibinfo{author}{\bibfnamefont{L.}~\bibnamefont{Lipatov}},
  \bibinfo{journal}{Nucl.Phys.} \textbf{\bibinfo{volume}{B582}},
  \bibinfo{pages}{19} (\bibinfo{year}{2000}), \eprint{hep-ph/0004008}.

\bibitem[{\citenamefont{Dominguez
  et~al.}(2011{\natexlab{a}})\citenamefont{Dominguez, Xiao, and
  Yuan}}]{Dominguez:2010xd}
\bibinfo{author}{\bibfnamefont{F.}~\bibnamefont{Dominguez}},
  \bibinfo{author}{\bibfnamefont{B.-W.} \bibnamefont{Xiao}}, \bibnamefont{and}
  \bibinfo{author}{\bibfnamefont{F.}~\bibnamefont{Yuan}},
  \bibinfo{journal}{Phys.Rev.Lett.} \textbf{\bibinfo{volume}{106}},
  \bibinfo{pages}{022301} (\bibinfo{year}{2011}{\natexlab{a}}),
  \eprint{1009.2141}.

\bibitem[{\citenamefont{Dominguez
  et~al.}(2011{\natexlab{b}})\citenamefont{Dominguez, Marquet, Xiao, and
  Yuan}}]{Dominguez:2011wm}
\bibinfo{author}{\bibfnamefont{F.}~\bibnamefont{Dominguez}},
  \bibinfo{author}{\bibfnamefont{C.}~\bibnamefont{Marquet}},
  \bibinfo{author}{\bibfnamefont{B.-W.} \bibnamefont{Xiao}}, \bibnamefont{and}
  \bibinfo{author}{\bibfnamefont{F.}~\bibnamefont{Yuan}},
  \bibinfo{journal}{Phys.Rev.} \textbf{\bibinfo{volume}{D83}},
  \bibinfo{pages}{105005} (\bibinfo{year}{2011}{\natexlab{b}}),
  \eprint{1101.0715}.

\bibitem[{\citenamefont{Kovchegov and Weigert}(2007)}]{Kovchegov:2006vj}
\bibinfo{author}{\bibfnamefont{Y.~V.} \bibnamefont{Kovchegov}}
  \bibnamefont{and} \bibinfo{author}{\bibfnamefont{H.}~\bibnamefont{Weigert}},
  \bibinfo{journal}{Nucl. Phys.} \textbf{\bibinfo{volume}{A784}},
  \bibinfo{pages}{188} (\bibinfo{year}{2007}), \eprint{hep-ph/0609090}.

\bibitem[{\citenamefont{Balitsky}(2007)}]{Balitsky:2006wa}
\bibinfo{author}{\bibfnamefont{I.}~\bibnamefont{Balitsky}},
  \bibinfo{journal}{Phys. Rev.} \textbf{\bibinfo{volume}{D75}},
  \bibinfo{pages}{014001} (\bibinfo{year}{2007}), \eprint{hep-ph/0609105}.

\bibitem[{\citenamefont{Berger and Stasto}(2011{\natexlab{a}})}]{Berger:2011ew}
\bibinfo{author}{\bibfnamefont{J.}~\bibnamefont{Berger}} \bibnamefont{and}
  \bibinfo{author}{\bibfnamefont{A.~M.} \bibnamefont{Stasto}},
  \bibinfo{journal}{Phys.Rev.} \textbf{\bibinfo{volume}{D84}},
  \bibinfo{pages}{094022} (\bibinfo{year}{2011}{\natexlab{a}}),
  \eprint{1106.5740}.

\bibitem[{\citenamefont{Nikolaev and Zakharov}(1991)}]{Nikolaev:1990ja}
\bibinfo{author}{\bibfnamefont{N.~N.} \bibnamefont{Nikolaev}} \bibnamefont{and}
  \bibinfo{author}{\bibfnamefont{B.~G.} \bibnamefont{Zakharov}},
  \bibinfo{journal}{Z. Phys.} \textbf{\bibinfo{volume}{C49}},
  \bibinfo{pages}{607} (\bibinfo{year}{1991}).

\bibitem[{\citenamefont{Beuf}()}]{BeufToAppear}
\bibinfo{author}{\bibfnamefont{G.}~\bibnamefont{Beuf}},
  \bibinfo{howpublished}{in preparation}.

\bibitem[{\citenamefont{Berger and Stasto}(2011{\natexlab{b}})}]{Berger:2010sh}
\bibinfo{author}{\bibfnamefont{J.}~\bibnamefont{Berger}} \bibnamefont{and}
  \bibinfo{author}{\bibfnamefont{A.}~\bibnamefont{Stasto}},
  \bibinfo{journal}{Phys.Rev.} \textbf{\bibinfo{volume}{D83}},
  \bibinfo{pages}{034015} (\bibinfo{year}{2011}{\natexlab{b}}),
  \eprint{arXiv:1010.0671 [hep-ph]}.

\bibitem[{\citenamefont{Balitsky}(2001)}]{Balitsky:2001gj}
\bibinfo{author}{\bibfnamefont{I.}~\bibnamefont{Balitsky}}
  (\bibinfo{year}{2001}), \eprint{hep-ph/0101042}.

\bibitem[{\citenamefont{Albacete et~al.}(2011)\citenamefont{Albacete, Armesto,
  Milhano, Quiroga-Arias, and Salgado}}]{Albacete:2010sy}
\bibinfo{author}{\bibfnamefont{J.~L.} \bibnamefont{Albacete}},
  \bibinfo{author}{\bibfnamefont{N.}~\bibnamefont{Armesto}},
  \bibinfo{author}{\bibfnamefont{J.~G.} \bibnamefont{Milhano}},
  \bibinfo{author}{\bibfnamefont{P.}~\bibnamefont{Quiroga-Arias}},
  \bibnamefont{and} \bibinfo{author}{\bibfnamefont{C.~A.}
  \bibnamefont{Salgado}}, \bibinfo{journal}{Eur.Phys.J.}
  \textbf{\bibinfo{volume}{C71}}, \bibinfo{pages}{1705} (\bibinfo{year}{2011}),
  \eprint{1012.4408}.

\bibitem[{\citenamefont{Kuokkanen et~al.}(2011)\citenamefont{Kuokkanen,
  Rummukainen, and Weigert}}]{Kuokkanen:2011je}
\bibinfo{author}{\bibfnamefont{J.}~\bibnamefont{Kuokkanen}},
  \bibinfo{author}{\bibfnamefont{K.}~\bibnamefont{Rummukainen}},
  \bibnamefont{and} \bibinfo{author}{\bibfnamefont{H.}~\bibnamefont{Weigert}}
  (\bibinfo{year}{2011}), \eprint{1108.1867}.

\bibitem[{\citenamefont{Stasto et~al.}(2013)\citenamefont{Stasto, Xiao, and
  Zaslavsky}}]{Stasto:2013cha}
\bibinfo{author}{\bibfnamefont{A.~M.} \bibnamefont{Stasto}},
  \bibinfo{author}{\bibfnamefont{B.-W.} \bibnamefont{Xiao}}, \bibnamefont{and}
  \bibinfo{author}{\bibfnamefont{D.}~\bibnamefont{Zaslavsky}}
  (\bibinfo{year}{2013}), \eprint{1307.4057}.

\bibitem[{\citenamefont{Mueller and Munier}(2012)}]{Mueller:2012bn}
\bibinfo{author}{\bibfnamefont{A.}~\bibnamefont{Mueller}} \bibnamefont{and}
  \bibinfo{author}{\bibfnamefont{S.}~\bibnamefont{Munier}},
  \bibinfo{journal}{Nucl.Phys.} \textbf{\bibinfo{volume}{A893}},
  \bibinfo{pages}{43} (\bibinfo{year}{2012}), \eprint{1206.1333}.

\bibitem[{\citenamefont{Berends and Giele}(1987)}]{Berends:1987cv}
\bibinfo{author}{\bibfnamefont{F.~A.} \bibnamefont{Berends}} \bibnamefont{and}
  \bibinfo{author}{\bibfnamefont{W.}~\bibnamefont{Giele}},
  \bibinfo{journal}{Nucl.Phys.} \textbf{\bibinfo{volume}{B294}},
  \bibinfo{pages}{700} (\bibinfo{year}{1987}).

\bibitem[{\citenamefont{Mangano et~al.}(1988)\citenamefont{Mangano, Parke, and
  Xu}}]{Mangano:1987xk}
\bibinfo{author}{\bibfnamefont{M.~L.} \bibnamefont{Mangano}},
  \bibinfo{author}{\bibfnamefont{S.~J.} \bibnamefont{Parke}}, \bibnamefont{and}
  \bibinfo{author}{\bibfnamefont{Z.}~\bibnamefont{Xu}},
  \bibinfo{journal}{Nucl.Phys.} \textbf{\bibinfo{volume}{B298}},
  \bibinfo{pages}{653} (\bibinfo{year}{1988}).

\bibitem[{\citenamefont{Mangano}(1988)}]{Mangano:1988kk}
\bibinfo{author}{\bibfnamefont{M.~L.} \bibnamefont{Mangano}},
  \bibinfo{journal}{Nucl.Phys.} \textbf{\bibinfo{volume}{B309}},
  \bibinfo{pages}{461} (\bibinfo{year}{1988}).

\bibitem[{\citenamefont{Dixon}(2013)}]{Dixon:2013uaa}
\bibinfo{author}{\bibfnamefont{L.~J.} \bibnamefont{Dixon}}
  (\bibinfo{year}{2013}), \eprint{1310.5353}.

\end{thebibliography}


\end{document}